\documentclass[final,1p,times]{elsarticle}

\usepackage{graphicx}
\usepackage{lscape}
\usepackage{rotating}
\usepackage{graphicx}
\usepackage{epsfig,multirow,booktabs}

\usepackage{amssymb,amsmath}

\usepackage{color}

\def\be{\begin{equation}}
\def\ee{\end{equation}}
\def\beq{\begin{eqnarray}}
\def\eeq{\end{eqnarray}}

\journal{Comptes Rendus de l'Academie des Sciences}

\begin{document}

\begin{frontmatter}



\title{Multiple physical elements to determine the gravitational-wave signatures
 of core-collapse 
supernovae}


\author[naoj,cfca]{Kei Kotake}

\address[naoj]{Division of Theoretical Astronomy, National Astronomical Observatory of Japan, 2-21-1, Osawa, Mitaka, Tokyo, 181-8588, Japan}
\address[cfca]{Center for Computational Astrophysics, National
  Astronomical Observatory of Japan, Mitaka, Tokyo 181-8588, Japan}

\begin{abstract}
We review recent progress in the theoretical predictions of gravitational waves (GWs)
 of core-collapse supernovae. Following a brief summary of the methods in the numerical
 modeling, we summarize multiple physical elements that 
 determine the GW signatures which have been considered to be 
important in extracting the information
 of the long-veiled explosion mechanism
 from the observation of the GWs. We conclude with a summary  of the most urgent tasks
 to make the dream come true.
\end{abstract}


\end{frontmatter}


\section{Introduction}\label{s1}

 Massive stars in the approximate mass range of 8 to 100 solar masses ($M_{\odot}$)
end their lives as core-collapse supernovae. They have long attracted the attention of
astrophysicists because they have many aspects playing important roles
in astrophysics. They are the mother of neutron stars and black holes; they play an 
important role for acceleration of cosmic rays; they influence
galactic dynamics triggering further star formation; they are
 gigantic emitters of neutrinos and gravitational waves.
 They are also a major site for nucleosynthesis, so, naturally, any attempt 
to address human origins may need to start
with an understanding of core-collapse supernovae (SNe). 

Current estimates of core-collapse SN rates in our Galaxy predict one 
  event every $\sim 40 \pm 10$ year \cite{ando_new}. 
When a massive star undergoes a core-collapse SN 
in our Galactic center, copious numbers of neutrinos are produced, 
some of which may be detected on the earth. Such ``supernova neutrinos'' will carry 
valuable information from deep inside the core. 
In fact, the detection of neutrinos from SN1987A 
(albeit in the Large Magellanic Cloud)
 opened up the {\it neutrino astronomy}, which is an alternative to 
conventional astronomy by electromagnetic waves \cite{hirata1987,bionta1987}.
 Even though there were just two dozen neutrino events from SN1987A, these 
events have been studied extensively (yielding $\sim$ 500 papers) and
 have allowed us to 
have a confidence that our basic picture of the supernova 
 physics is correct (e.g., \cite{Sato-and-Suzuki}, see
 \cite{raffelt_review} for a recent review).
 If a supernova occurs in our Galactic center ($\sim 10$ kpc), 
 about 10,000 $\bar{\nu}_e$ events are estimated to be detected by 
 Super-Kamiokande (SK) \cite{totsuka} (e.g., \cite{totani,beacom99}).
 Those successful neutrino detections 
are important not only to study the supernova physics but also to unveil 
 the nature of neutrinos itself such as the neutrino oscillation parameters  
and the mass hierarchy (e.g., \cite{raffelt_2010} for a recent review).

Core-collapse SNe are now about to start even another astronomy,
  {\it Gravitational-Wave Astronomy}.
 Currently long-baseline laser interferometers such as
LIGO (USA)\cite{firstligonew},
VIRGO (Italy)\footnote{http://www.ego-gw.it/},
GEO600 (Germany)\footnote{http://geo600.aei.mpg.de/},
and TAMA300 (Japan) \cite{tamanew} are currently operational and preparing 
 for the first observation (see, e.g., \cite{hough} for a recent review), 
 by which the prediction by Einstein's theory of General Relativity (GR) can be confirmed.
 These instruments are being updated to their {\it Advanced} status, and may start 
 taking data, possibly detecting GWs for the first time, as soon as 2015 (see 
\cite{vandel} for a recent review).
 In fact, {\it Advanced }LIGO/VIRGO, which is an upgrade of the initial LIGO and VIGRO,
 are expected to be completed by 2015 and will increase the observable detection volume
 by a factor of $\sim 1000$ \cite{harry}. The Large-scale Cryogenic Gravitational-wave
 Telescope (LCGT \cite{lcgt}) in Japan was funded in late 2010, 
which is being built under the Kamioka mine and is expected to take its first 
data in 2016. At such a high level of precision, those GW detectors are
 sensitive to many different sources, including chirp, ring-down, and merger phases
 of black-hole and neutron star binaries (e.g., \cite{schutz,faber,duez}), neutron star normal mode 
oscillations (e.g., \cite{nils1}), rotating neutron star mountains 
(e.g., \cite{horowitz}), and core-collapse supernova explosions 
(e.g., \cite{kota06,ott_rev,fryer11} for recent reviews), on the final of which 
we focus in this article.

According to the Einstein's theory
 of general relativity (e.g., \cite{shap83}), no GWs can be emitted if gravitational collapse of the 
supernova core proceeds perfectly spherically symmetric. 
To produce GWs, the gravitational collapse should 
proceed aspherically and dynamically. Observational evidence gathered over
the last few decades has pointed towards core-collapse SNe indeed being
generally aspherical (e.g., \cite{wang01,wang02,maeda08,tanaka} and references 
therein). 
The most unequivocal example is SN1987A. The HST
images of SN1987A are directly showing that the expanding envelope is 
elliptical with the long axis aligned with the rotation axis inferred
from the ring (\cite{wang02}, see however \cite{87a} for a recent counter argument). 
The aspect ratio and position angle of the symmetry axis are 
consistent with those predicted earlier from the observations of speckle and 
linear polarization. What is more, the linear polarization became greater as time 
passed (e.g., \cite{wang01,leonard,leon06}), a fact which has been used to argue that 
the central engine of the explosion is responsible for the
non-sphericity (e.g., \cite{khok99,whee00}).
From a theoretical point of view, clarifying what makes the dynamics of the core
deviate from spherical symmetry is essential in understanding the GW emission
mechanism. Here it is worth mentioning that GWs are primary observables, which 
imprint a live information of the central engine, because they carry the information 
directly to us without being affected in propagating from the stellar center to the 
earth. On the other hand, SN neutrinos, which are affected by the 
well-known Mikheyev-Smirnov-Wolfenstein (MSW) effect
 (e.g., \cite{mikheev1986}) in propagating the stellar envelope,
 may be rather indirect in extracting the information of the central engine compared 
 to GWs.\footnote{However they can be a useful tool like a tomography 
to monitor the density profile in the stellar mantle, which is expected 
 to tell us the evolution of the supernova shock (e.g.,
 \cite{kneller} for a recent review).}

The breaking of the sphericity in the supernova engine 
has been also considered as a most important 
ingredient to understand the explosion mechanism, for 
which supernova theorists have been continuously keeping their efforts 
for the past $\sim$ 40 years. Albeit being a topic of hot debate yet, 
current multi-dimensional (multi-D) simulations based on refined numerical models 
show several promising scenarios. Among the candidates are the neutrino heating 
mechanism aided by convection and hydrodynamic instabilities of the supernova shock
 (e.g., \cite{jank07} for a review), the acoustic mechanism
 \cite{burr_rev}, or the magnetohydrodynamic (MHD) mechanism
 (e.g., \cite{kota06} see references therein).

Putting things together, the multi-dimensionality determines the explosion mechanism, in turn 
we may extract the information that traces the multi-dimensionality by 
 the SN messengers, which would be only possible by a careful analysis primary 
on GWs as well as neutrinos, and photons. In this review article, we hope to 
bring together findings from recent multi-D supernova simulations and the obtained 
predictions of the GWs so far (for GWs from other high-energy 
astrophysical sources such as magnetars, gamma-ray bursts, and coalescing binaries, see 
\cite{eric1,szabi,prad,aso} for recent reviews). Before moving on to the 
 next sections, we first have to draw a caution that the 
 current generation of numerical results that we report in this article and 
 the associated predictions of the GW emission should depend
 on the next-generation calculations by which more sophistication can be made 
not only in determining the efficiency of neutrino-matter coupling (the so-called 
 neutrino transport problem), but also in the treatment of general relativity.
 Therefore we provide here only a snapshot of the moving 
 (long-run) documentary film that records our endeavours for making
 our dream of "GW astronomy of core-collapse supernovae" come true.

\section{On core-collapse supernova simulations}\label{s2}

In order to extract a true astrophysical GW signal from
 the confusing detector noises, we need not only require sensitive
 detectors, but also extensive data analysis of the detector output based on 
 reliable theoretical estimates for the GW signals. 
 In most cases, a detailed numerical modeling is 
 required to make the precise prediction of the dynamics for the GW sources.
As will be summarized briefly in this section, SN modelers have 
 been keeping their efforts to this end for these +45 years.
  
 The abstract in the paper by Colgate and White in 1966 \cite{colgate} 
 who reported the first SN simulation finished with the sentence,{\it "The energy release 
 (at the moment of explosion) corresponds to the change in gravitational potential of the unstable imploding core; the transfer of energy takes place by the emission and deposition of neutrinos"}\footnote{I boldly added several words in (...) for making 
the meaning of the sentence complete.}. As well-known, this is the essence of 
the so-called neutrino-heating mechanism of core-collapse supernovae. The 
 mechanism was first proposed to occur in 
 a prompt manner, but was later reinforced by 
Bethe and Wilson \cite{wils85,bethe85,bethe} to take a currently prevailing delayed form,
 in which a stalled bounce shock is revived by neutrino energy deposition to trigger
  explosions in several hundred milliseconds after bounce. 
The mechanism has been the working hypothesis of supernova theorists 
for these $\sim$45 years (see collective references in Janka et al. (2007)
 \cite{jank07} that was presented in the Bethe Centennial Volume in Physics Report, 
 and also a summary given in section 2 of Nordhaus et al. (2010) \cite{nordhaus}).

 However, one important lesson we have learned from a series of 
  the most up-to-date simulations
\cite{rampp00,lieb01,thom03,sumi05} which implemented the best input physics and 
numerics to date, is that the mechanism fails to blow up 
canonical massive stars\footnote{except for $8-9 M_{\odot}$ stars (e.g., \cite{kitaura}).}
in spherical symmetric (1D) simulations.
Pushed by supernova observations of the blast morphology
\cite[e.g.,][]{wang01,tanaka} mentioned above, it is now almost certain that the
breaking of the spherical symmetry is the key to solve the supernova problem.
 So far a number of multi-D hydrodynamic simulations have been reported, which  
 {\it demonstrated} that hydrodynamic motions associated with convective overturn (e.g., 
\cite{herant,burr95,jankamueller96,frye02,fryer04a}) as well as the 
Standing-Accretion-Shock-Instability (SASI, e.g., 
\cite{blon03,sche04,scheck06,Ohni05,ohnishi07,thierry,murphy08,rodrigo09,rodrigo09_2,
iwakami1,iwakami2,fern} see
 references therein) can help the onset of the neutrino-driven explosion. 

To test the neutrino heating mechanism in the multi-D context, 
it is of crucial importance to solve accurately the neutrino-matter coupling in 
  spatially non-uniform hydrodynamic environments. 
For the purpose, one ultimately needs to solve the 
six-dimensional (6D) neutrino radiation transport problem
(three in space, three in the momentum space of neutrinos), 
thus the supernova simulation has been counted as one of the most challenging tasks 
in numerical astrophysics. In the final sentence of the last paragraph, 
we wrote "demonstrated" because the neutrino heating was given by hand as 
an input parameter in most of the simulations cited above 
(see, however \cite{frye02,fryer04a}). The neutrino heating proceeds 
dominantly via the charged current interactions ($\nu_{\text{e}} + \text{n} \rightleftarrows \text{e}^{-} + \text{p},~ \bar{\nu}_{\text{e}} + \text{p} \rightleftarrows \text{e}^{+} + \text{n}$) in the gain region. The neutrino heating rate in the gain region 
 can be roughly expressed as \cite{jank01} 
$Q^{+} \propto L_{\nu}\langle\mu_{\nu}\rangle^{-1}/r^2$ where $L_{\nu}$ is the neutrino 
luminosity emitted from the surface of neutrino sphere and it determines the 
 amplitude of the neutrino heating as well as cooling, and $r$ and $\langle \mu_{\nu} \rangle$ 
is the distance from the stellar center and the flux factor\footnote{This quantity 
represents the degree of anisotropy in neutrino emission; $\langle\mu_{\nu}\rangle
\sim 0.25$ near at the neutrino sphere, $\langle\mu_{\nu}\rangle=1$ in the free-streaming limit ($r \rightarrow 
\infty$) (e.g., Janka (2001) \cite{jank01}).}, respectively 
(e.g., \cite{jank01}). For example, $L_{\nu}$ is treated as an input parameter
in the so-called `light-bulb'' approach (e.g., \cite{jankamueller96}).
 This is one of the most prevailing approximations in recent
 3D simulations \cite{iwakami1,iwakami2,annop1,nordhaus} because
 it is handy to study multi-D effects on the neutrino heating mechanism 
 (albeit on the qualitative grounds). To go beyond the light-bulb scheme,
  $L_{\nu}$ should be determined in a self-consistent manner. 
 For the purpose, one needs to tackle with 
neutrino transport problem, only by which energy as well as angle 
dependence of the neutrino distribution function can be determined without any 
 assumptions. 
Since the focus of this review is on GWs, a detailed discussion of
 various approximations and numerical techniques taken in the recent 
radiation-hydrodynamic SN simulations cannot be provided. 
 Table 1 is not intended as a comprehensive compilation, but we just want 
 to summarize milestones that have recently reported {\it exploding} models so far.
 The table will be useful in later sections when we discuss 
the GW signatures obtained in some of these simulations.

In Table 1,the first column ("Progenitor") shows the progenitor model employed in each 
simulation. The abbreviation of "NH", "WHW", and "WW" means Nomoto \& Hashimoto (1988)
 \cite{nomo88}, Woosley, Heger, and Weaver (2002) \cite{woos02}, and Woosley and Weaver
 (1995) \cite{WW95}. The second column shows SN groups with the published or 
 submitted year of the corresponding work. "MPA" stands for the core-collapse 
SN group in Max Planck Institute for Astrophysics led by H.T. Janka and E. M\"uller.
 "Princeton+" stands for the group chiefly consisting of
 the staffs in the Princeton University (A. Burrows), Caltech (C.D.Ott), Hebrew University 
 (E. Livne) and their collaborators. The SN group in the Basel university
 is led by M. Liebend\"orfer and F.K. Thielemann. "OakRidge+" stands for the 
SN group mainly 
consisting of the Florida Atlantic University (S. Bruenn) and the Oak Ridge National 
Laboratory (A. Mezzacappa, O.E.B. Messer) and their collaborators. 
NAOJ+ is the SN group chiefly consisting 
 of the staffs in the National Astronomical Observatory of Japan (myself, 
T. Takiwaki), Kyoto university (Y. Suwa), Waseda University (S. Yamada),
 and their collaborators. The third column represents the mechanism of explosions 
 which are basically categorized into two (to date),
 namely by the neutrino-heating mechanism
 (indicated by "$\nu$-driven") or by the acoustic mechanism ("Acoustic").
 "Dim." in the fourth column is the fluid space dimensions which is 
 one-, two-, or three-dimension (1,2,3D). The abbreviation ``N'' stands for `Newtonian,' while ``PN''---for `Post-Newtonian'---stands for some attempt at inclusion of general relativistic effects, and ``GR'' denotes full relativity.
 $t_{\rm exp}$ in the fifth column indicates an
 approximate typical timescale when the explosion initiates and $E_{\rm exp}$ represents
 the explosion energy normalized by Bethe(=$10^{51}$ erg) given at the 
 postbounce time of $t_{\rm pb}$, both of which 
 are attempted to be sought in literatures\footnote{but if we cannot find them, 
we remain them as blank "-".}. In the final column of "$\nu$ transport", 
  "Dim" represents $\nu$ momentum dimensions and the treatment of the 
 velocity dependent term in the transport equations is symbolized by ${\cal O}(v/c)$.
  The definition of the "RBR", "IDSA", and "MGFLD" will be given soon in the following.

\begin{table*}
\begin{center}
\begin{tabular}{lcccccc}
\hline
Progenitor & Group & Mechanism & Dim. & $t_{\rm exp}$ & $E_{\rm exp}$(B) & $\nu$ transport \\ 
           & (Year)  &  & (Hydro) & (ms) & @$t_{\rm pb}$ (ms)& (Dim,  ${\cal O}(v/c)$) \\ 

\hline
\multirow{3}*{8.8 {$M_{\odot}$ }} & MPA\cite{kitaura,janka08} & $\nu$-driven & 1D/2D & $\sim$200 
    & 0.1  &  Boltzmann \\
        & (2006)      & & (PN)   & & ($\sim$800) & 2, ${\cal O}(v/c)$     \\    \cmidrule(l){2-7}
 (NH88\cite{nomo88})   & Princeton+ & $\nu$-driven & 2D & $\lesssim$125 & 0.1 &  MGFLD \\
                       & \cite{burr06}(2006) &  & (N)   & & - &1, (N)     \\    \cmidrule(l){1-7}
10 $M_{\odot}$  & Basel\cite{sage09} & $\nu$+(QCD & 1D &  255 & 0.44 &  Boltzmann \\
(WHW02\cite{woos02})           & (2009)      & transition) & (GR)   & & (350) &2, (GR)     \\    \cmidrule(l){1-7}
11 $M_{\odot}$  & {Princeton+} & Acoustic & 2D & $\gtrsim$550 & $\sim$0.1* &  MGFLD \\
(WW95\cite{WW95})               & \cite{burr06}(2006)  &  & (N)   & & (1000) &1, (N)     \\    \cmidrule(l){1-7}
\multirow{3}*{11.2 {$M_{\odot}$ }} & MPA\cite{buras06} & $\nu$-driven & 2D & $\sim$100 &
 $\sim 0.005$  
      &  "RBR" Boltz- \\
   & (2006)  & & (PN)   & & ($\sim$220) & mann, 2, ${\cal O}(v/c)$     \\    \cmidrule(l){2-7} (WHW02\cite{woos02})   
 & Princeton+ & Acoustic & 2D & $\gtrsim$1100 & $\sim$0.1* &  MGFLD \\
      &  \cite{burrows2} (2007) &  & (N)   & & (1000) &1, (N)     \\    \cmidrule(l){2-7}
      & NAOJ+ & $\nu$-driven  & {\bf 3D} & $\sim$100 & 0.01  &  IDSA \\
    &  \cite{taki11}(2011) &  & (N)   & & (300) &1, (N)     \\    \cmidrule(l){1-7}
12 $M_{\odot}$    & Oak Ridge+ & $\nu$-driven & 2D & $\sim$300 & 0.3 &  "RBR" MGFLD \\
  (WHW02\cite{woos02})  &  \cite{bruenn}(2009) &  & (PN)   & & (1000) &1, ${\cal O}(v/c)$  \\    \cmidrule(l){1-7}
13 $M_{\odot}$    & Princeton+ & Acoustic & 2D & $\gtrsim$1100 & $\sim$0.3* &  MGFLD \\
  (WHW02\cite{woos02})  &  \cite{burrows2}(2007) &  & (N)   & & (1400) &1, (N)     \\    \cmidrule(l){2-7}
(NH88\cite{nomo88})    & NAOJ+ & $\nu$-driven  & 2D & $\sim$200 & 0.1  &  IDSA \\
    &  \cite{suwa}(2010) &  & (N)   & & (500) &1, (N)     \\    \cmidrule(l){1-7}
15 $M_{\odot}$    &  MPA\cite{marek} & $\nu$-driven & 2D & $\sim$600 & 0.025 
    &  Boltzmann \\
(WW95\cite{WW95})  & (2009) & & (PN)   & & ($\sim$700) & 2,${\cal O}(v/c)$     \\    \cmidrule(l){2-7}
(WHW02\cite{woos02})  & Princeton+ & Acoustic & 2D & - & - &  MGFLD \\
    &  \cite{burrows2} &  & (N)   & & (-) &1, (N)     \\    \cmidrule(l){2-7}
                       & OakRidge+ &  $\nu$-driven & 2D & $\sim$300 & $\sim$ 0.3 &  "RBR" MGFLD \\
                    &  \cite{bruenn}(2009) &  & (PN)   & & (600) &1,${\cal O}(v/c)$  \\    \cmidrule(l){1-7}
20 $M_{\odot}$    & Princeton+ & Acoustic & 2D & $\gtrsim$1200 & $\sim$0.7* &  MGFLD \\
  (WHW02\cite{woos02})  &  \cite{burrows2}(2007) &  & (N)   & & (1400) &1, (N)     \\    \cmidrule(l){1-7}
25 $M_{\odot}$    & Princeton+ & Acoustic & 2D & $\gtrsim$1200 & - &  MGFLD \\
  (WHW02\cite{woos02})  &  \cite{burrows2}(2007) &  & (N)   & & (-) &1, (N)     \\    \cmidrule(l){2-7} 
    & Oak Ridge+ & $\nu$-driven & 2D & $\sim$300 & $\sim$ 0.7 &  "RBR" MGFLD \\
                    &  \cite{bruenn}(2009) &  & (PN)   & & (1200) &1, ${\cal O}(v/c)$  \\    \cmidrule(l){1-7}
\end{tabular}
\end{center}
\caption{Selected lists of recent neutrino-radiation hydrodynamic milestones 
 reported by many SN groups around the world ("Group"), which obtained explosions 
by the neutrino-heating mechanism
 (indicated by "$\nu$-driven") or the acoustic mechanism ("Acoustic")
 (See text for more details).
}
\end{table*}

Due to the page limit of this article, we have to start the story only 
 after 2006 (see, e.g., Janka et al. (2007) \cite{jank07} for a grand review, 
and also Cardall (2005) \cite{cardall} for a similar table before 2006). 
The first news of the 
exploding model was reported by the MPA group.
By using their MuDBaTH code which includes one of the best available neutrino transfer 
approximations, they reported 1D and 2D explosions for the 8.8 $M_{\odot}$ star
 by NH88 whose progenitor has a very tenuous outer envelope with steep density 
 gradient, which is a characteristic property of AGB stars. Also in 1D, the Basel+ group
 reported explosions for 10 and 15 $M_{\odot}$ progenitors of WHW02 triggered 
 by the hypothesised first-order QCD phase transition in the protoneutron 
star (PNS). 
To date, these two are the only modern numerical results where the neutrino-driven 
mechanism succeeded in 1D. 
 In the 2D MPA simulations, they obtained explosions 
 for a non-rotating 11.2 $M_\odot$ progenitor of 
WHW02 \cite{buras06}, and then for a $15 M_{\odot}$ progenitor \cite{marek} 
of WW95 with a relatively rapid rotation imposed\footnote{by comparing 
 the precollapse angular velocity to the one predicted in a recent stellar 
 evolution calculation \cite{hege05}.}. They newly 
 brought in the so-called ``ray-by-ray'' approach (indicated by "RBR" in the table),
 in which the neutrino transport is solved 
along a given radial direction assuming that the hydrodynamic medium for the direction 
is spherically symmetric. This method, which reduces the 2D problem partly to 1D, 
fits well with their original 1D Boltzmann solver \cite{rampp00}\footnote{Note that 
 the ray-by-ray approach has an advantage compared to other approximation schemes, such
 that it can fully take into account the available neutrino reactions
(e.g., \cite{buras1} for references therein) 
 and also give us the most accurate solution for a given angular direction.}.
 For 2D hydrodynamic simulations with the ray-by-ray transport, 
one needs to solve the 4D radiation transport problem 
(two in space and two in the neutrino momentum space).
Regarding the explosion energies obtained in the MPA simulations, 
 their values at their final simulation time are typically underpowered by one or 
two orders of magnitudes to explain the canonical supernova kinetic energy 
($\sim 10^{51}$ erg). But the explosion energies presented in their figures 
 are still growing with time, and they could be as high as 1 B if they were able to 
 follow a much longer evolution as discussed in \cite{buras06}.

 More recently, fully 2D multi-angle Boltzmann transport simulations become practicable 
 by the Princeton+ group \cite{ott_multi,brandt}. In this case, one needs to handle 
 the 5D problem for 2D simulations (two in space, and three in the neutrino 
 momentum space). However this scheme is very 
 computationally expensive currently to perform long-term supernova 
simulations. In fact, the most recent 2D work by \cite{brandt} succeeded in 
 following the dynamics until $\sim 400$ ms after bounce for a non-rotating 
 and a rapidly rotating 20 $M_{\odot}$ model of WHW02, but explosions seemingly 
 have not been obtained in such an earlier phase either by the neutrino-heating or 
 the acoustic mechanism.
 
In the table, "MGFLD" stands for the Multi-Group Flux-Limited Diffusion scheme
  which eliminates
 the angular dependence of the neutrino distribution function 
(see, e.g., Bruenn (1985) \cite{bruenn85} for more details).
For 2D simulations, one needs to solve the 3D problem, namely 
two in space, and one in the neutrino momentum space.  By implementing 
 the MGFLD algorithm to the CHIMERA code in a ray-by-ray fashion (e.g., \cite{bruenn}), 
 Bruenn et al. (2009) 
\cite{yakunin} obtained neutrino-driven explosions for non-rotating progenitors 
in a relatively wide range in 12, 15, 20, 25 $M_{\odot}$ of WHW02 (see table). 
These models tend to start exploding at around 300 m after bounce, and 
 the explosion energy for the longest running model of the 
 25 M$_{\odot}$ progenitor is reaching to 1 B at 1.2 s after bounce \cite{bruenn}.  

 On the other hand, the 2D MGFLD simulations implemented 
 in the VULCAN code \cite{burr06} obtained explosions for a variety of progenitors
 of 11, 11.2, 13, 15, 20, and 25 $M_{\odot}$ not by the neutrino-heating mechanism 
but by the acoustic mechanism\footnote{For the 8.8 $M_{\odot}$ progenitor,
  they obtained neutrino-driven explosions (see table 1).}. 
 The acoustic mechanism relies on the revival of the stalled bounce shock by the 
 energy deposition via the acoustic waves that the oscillating protoneutron stars (PNSs) 
would emit in a much delayed phase ($\sim 1$ second) compared to
 the conventional neutrino-heating mechanism ($\sim 300 - 600 $ milliseconds). 
 If the core pulsation energy given in Burrows et al. (2007) 
\cite{burrows2} could be used to measure 
 the explosion energy in the acoustic mechanism, they reach to 1 B after 1000 ms after
 bounce\footnote{Due to the ambiguity of the relation between the pulsation energy 
and the explosion energy, an asterisk is added in the table.}.
 The additional energy input from acoustic waves is very appealing, 
but it may remain a matter of vivid debate and has yet to be confirmed by other groups. 

 By performing 2D simulations in which the spectral neutrino transport
   was solved  by the isotropic diffusion source approximation (IDSA)
   scheme \cite{idsa}, the NAOJ+ group reported explosions for a
   non-rotating and rapidly rotating 13 $M_{\odot}$ progenitor of NH88. 
 They pointed out that a stronger explosion is 
obtained for the rotating model comparing to the corresponding
non-rotating model. The IDSA scheme splits the neutrino distribution
 into two components (namely the streaming and trapped neutrinos),
 both of which are solved using separate numerical techniques
 (see, Liebend\"orfer et al. (2009) \cite{idsa} for more details). 
The approximation level of the IDSA
 scheme is basically the same as the one of the MGFLD.
 The main advantage of the IDSA scheme is that the fluxes in the transparent
 region can be determined by the non-local distribution of sources
 rather than the gradient of the local intensity like in MGFLD.
 A drawback in the current
 version of the IDSA scheme is that heavy lepton neutrinos
 ($\nu_x$, i.e., $\nu_{\mu}$, $\nu_{\tau}$ and their 
anti-particles) as well as the energy-coupling weak interactions 
 have yet to be implemented.
 Extending the 2D modules in Suwa et al. (2009) \cite{suwa} to 3D,
 they recently reported explosions in the 3D models for an 11.2 $M_{\odot}$ 
progenitor of WHW02 (Takiwaki et al.(2011) \cite{taki11}).
 By comparing the convective motions as well as neutrino luminosities and energies between their 2D and 3D models, they pointed out whether 3D effects would help
 explosions or not is sensitive to the employed numerical resolutions.
  They argued that next-generation supercomputers are at least needed
 to draw a robust conclusion of the 3D effects.

 Having summarized a status of the current supernova simulations,
 one might easily see a number of issues that remain to be clarified.
 First of all, the employed progenitors usually rather scatter
 (e.g., Table 1).  Different SN groups seem to have a tendency to 
employ different progenitors, providing different results.
 By climbing over a wall which may have rather separated exchanges 
 among the groups,
  a detailed comparison for a given progenitor
   needs to be done seriously in the multi-D results 
(as have been conducted in the Boltzmann 1D simulations between
 the MPA, Basel+, and Oak Ridge+ groups \cite{lieb05}). 

In addition to the importance of 3D modeling as mentioned above,
a more complete ''realistic'' supernova model should naturally include 
 general relativity (GR) with magnetohydrodynamics (MHD) with multi-D GR
 Boltzmann neutrino transport, in which a microphysical treatment of
 equation of state (EOS) and nuclear-neutrino interactions are
  appropriately implemented. Unfortunately none of the
currently published SN simulations satisfy the "ultimate" requirement.
 In this sense,  all the mentioned studies
 employ some approximations (albeit with different levels of sophistication)
 for multiple physical ingredients (for example, as listed in Table 1\footnote{
 However it is also worth mentioning that a rush of nice work to this end 
has been reported recently including new schemes towards the GR radiation-hydrodynamic 
 simulations \cite{hari,mueller,shibata11} and new sets of supernova EOSs 
\cite{hempel,furusawa,gshen}.}).


In the same way,  theoretical predictions of the GWs that one can obtain 
by analyzing the currently available numerical results, cannot unambiguously 
give us the final answer yet. 
 Again we hereby note the feature of this article
 which shows only a snapshot of the moving theoretical terrain.
 Keeping this caveat in mind,
 it is also true that a number of surprising GW features of core-collapse SNe
 have been reported recently both by the 
first-principle simulations (e.g., in Table 1), and also by idealized simulations 
 in which explosions are parametrically initiated mostly
 by the light-bulb scheme. As will be mentioned in 
  the next section, the latter approach
 is also useful to get a better physical understanding of the GW signatures
 obtained in the first-principle simulations\footnote{
 In this sense, these two approaches are complimentary in understanding
 the GW signatures.}. We are now
 ready to move on to focus on the GW signatures from the next section.

\section{GW signatures} \label{s3}

  We first start to present a short overview concerning possible emission cites of GWs
  in core-collapse SNe so far proposed by a number of extensive
  studies (section \ref{short}, see also \cite{kota06,ott_rev,fryer11} 
for recent reviews).
   The most up-to-date results regarding the GW signatures emitted 
  near bounce in the case of rapidly rotating core-collapse,
 have been already given in the recent review by Ott (2009)
  (e.g., section 4 in \cite{ott_rev}).
   So in this article, we mainly focus on the GWs
 in the postbounce phase in the context of neutrino-driven
  explosion models and also on the postbounce GWs expected in
 MHD explosions, which will be described separately
  in sections \ref{neutrino} and \ref{mhd_gw}.
\subsubsection{A short overview}\label{short}
 The paper in 1982 by E. M\"uller of the MPA entitled as
"{\it Gravitational Radiation from Collapsing Rotating Stellar Cores}" unquestionably
 opened our eyes to the importance of making the GW prediction based on realistic 
 SN numerical modeling (see e.g., section 2 in \cite{ott_rev} for a 
 summary of more earlier work which had mainly focused on the
  GW emission in very idealized systems such as in homogeneous spheroids and 
 ellipsoids)\footnote{Needless to say, this kind of approach is still very important to 
extract the physics of the GW emission mechanism.}.
 As one may expect from the title of his paper, rapid rotation, if it would exist in the precollapse iron core, leads to
 significant rotational flattening of the collapsing and bouncing core,
 which produces a time-dependent quadrupole (or higher) GW emission.
 Following the first study by M\"uller,  most studies of the past thirty 
have focused on the so-called bounce signals 
(e.g., \cite{ewald82,mm,yama95,zwer97,kotakegw,kota04a,shibaseki,ott,ott_prl,
ott_2007,dimm02,dimmelprl,dimm08,simon,simon1}) and references therein). 

 As summarized by Ott (2009) \cite{ott_rev}, a number of 
 important progresses have been recently made to understand features of 
 the bounce signals by extensive 2D GR studies using the conformal-flat-condition (CFC)
 approximation \cite{dimm02,dimmelprl,dimm08} and also by fully GR 3D
 simulations \cite{ott} both including realistic EOSs and
 a deleptonization effect based on 1D-Boltzmann
 simulations \cite{matthias_dep}.
 Due to the page limit of this article, we are only able to touch on them
 in section \ref{mhd_gw}. 

  For the bounce signals having a strong and characteristic 
signature, the iron core must rotate enough rapidly. 
 Although the role of rapid rotation in combination with magnetic fields
 is attracting great attention as an important key in understanding
  the dynamics of collapsars and magnetars (e.g., section \ref{mhd_gw}), recent
 stellar evolution calculations predict that such a extreme condition can be
  realized only in a special case \cite{woos06}\footnote{which experiences the
so-called chemically homogeneous evolution \cite{yoon}.} ($\sim$ 1\% of massive star population). 
In addition, the precollapse rotation periods are estimated to be larger than 
$\sim$ 100 sec \cite{ott_birth} to explain the observed rotation periods of radio pulsars. In such a slowly rotating case, the detection of the 
bounce signals becomes very hard even by next-generation detectors
for a supernova in our Galaxy ($\sim 10$ kpc, e.g., \cite{kota04a,ott}). 

Besides the rapid rotation, anisotropic matter motions associated with
convection and anisotropic
neutrino emission in the postbounce phase are expected to be primary GW
sources with comparable amplitudes to the bounce signals. 
Thus far, various physical ingredients have been studied
 for producing the post-bounce asphericities, including 
 postshock convection \cite{muyan97}, pre-collapse density inhomogeneities \cite{burohey,muyan97,fryersingle,fryer04a},
 moderate rotation of the iron core \cite{mueller04}, nonaxisymmetric rotational 
instabilities \cite{rampp,ott_3d}, g-modes \cite{ott_new} and r-modes pulsations 
\cite{nils} of PNSs, and more recently by the Standing-Accretion-Shock Instability
  (SASI, \cite{kotake07,kotake_ray,kotake09,marek_gw,murphy}).
 In the multi-D modeling of stellar evolution, significant progresses
 have been recently made (e.g., \cite{arnett} and references therein), 
however the degree of initial
 inhomogeneities
 in the iron core seems still uncertain. Studies of non-radial
 instabilities in PNSs and cold NSs with their emission processes of GWs
 have a long history (e.g., section 7 in \cite{ott_rev} for references
 therein). The numerical studies to this end generally treat 
 the microphysics in a very phenomenological manner (like by a
 polytrope EOS), so there may remain a further room for sophistication.
 Among the candidate GW emission mechanisms,
  we therefore choose to focus on relatively well understood parts at first 
in the next section (section 3.1),
 which is the GW produced by convection and
 anisotropic neutrino emission.

\subsection{Gravitational waveforms in neutrino-driven explosions}\label{neutrino}

As mentioned, the first-principle and experimental simulations have a complimentary 
role to understand the GWs in the context of neutrino-driven SN explosions.
  Since the latter approach is equivalently useful, 
 we firstly summarize recent findings obtained by the experimental simulations
 in the next sections \ref{exp1} and \ref{exp2}, then proceed to the first-principle results 
in section \ref{first}.
  For the matter GW\footnote{Note that "Matter GW" means the GW produced by 
 quadrupole matter deformation and that "neutrino GW" by 
 anisotropic neutrino emission.} signals, we review the results by Murphy et
 al. \cite{murphy} who conducted a systematic study by
 changing neutrino luminosity and progenitor models. For the neutrino GW 
 signals, a piece of our work \cite{kotake_ray} is summarized in which
  a ray-tracing analysis was performed to accurately estimate the neutrino GWs.

\subsubsection{Features of Matter GWs in experimental 2D simulations}\label{exp1}

\begin{figure}[hbtp]
\begin{center}
\includegraphics[width=1.2\linewidth]{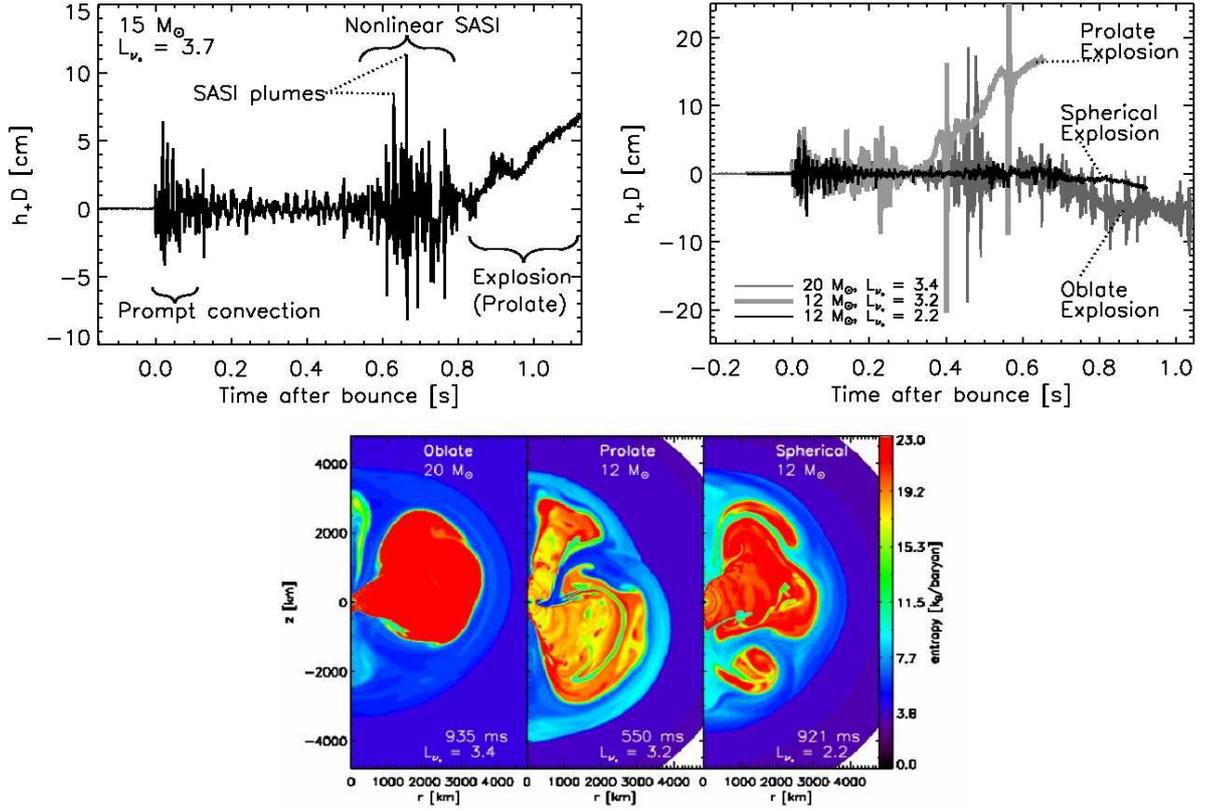}
\caption{Samples of gravitational waveforms in 2D parametric explosion
 models (by courtesy of Murphy and the coauthors \cite{murphy}).
 Top left panel shows matter 
GWs obtained for a 15 $M_{\odot}$ progenitor model \cite{woos02} with an input neutrino
  luminosity of $L_{\nu_e} = 3.7\times 10^{52}$erg/s. Three typical GW
  emission cites are clearly indicated, which are
 ''Prompt convection (after bounce up to $\sim$ 50 ms after bounce )'',
 ''Postshock convection'' (then till $550$ ms after bounce, unless the
  word is shown in the plot), 
 ''Non-linear SASI with SASI plumes''(then till $\sim$ 800 ms after bounce), and
 ''Explosion (then afterwards)''. The top right panel shows that the GW
  amplitudes monotonically increase or decease with time in the
  ''Explosion'' phase reflecting the geometry of the expanding shock (bottom panel),
 which is either prolate (increase), oblate (decrease), or
 spherical (in-between). In each panel, the chosen progenitor models of
 \cite{woos02} are given in each panel by 12, 20 $M_{\odot}$ and the
 input luminosity is indicated by the value of $L_{\nu_e}$ in unit of
 $10^{52}$ erg/s. }
\label{murphy}
\end{center}
\end{figure}
  
Figure \ref{murphy} shows samples of gravitational waveforms obtained in
  2D models by Murphy et al. \cite{murphy}.
 In their 2D models, the dynamics of a suite of progenitor models 
(12,15,20, and 40 $M_{\odot}$) is followed starting from gravitational collapse,
  through bounce, up to parametric explosions via the light-bulb scheme. 
In each panel of Figure
 \ref{murphy}, the taken progenitor and
  the input (electron-)neutrino luminosity in unit of $10^{52}$erg/s
   are indicated.
  For example, top left panel shows matter GWs obtained for a
 15 $M_{\odot}$ progenitor model \cite{woos02} with neutrino
  luminosity of $L_{\nu_e} = 3.7\times 10^{52}$erg/s. Three typical GW
  emission cites are clearly presented, which are
 ''Prompt convection'', ''Postshock convection''
 (unless it is shown in the plot), ''Non-linear SASI with SASI plumes'',
 and ''Explosion''.

 Shortly after the bounce shock is formed,
 negative entropy gradient behind the stalling shock
 predominantly gives rise to prompt convection.
 GWs indicated by ''Prompt convection'' (top left panel in Figure
 \ref{murphy}) comes from this. Later on, the
 PNS convection driven by the negative lepton gradient near its surface
 and the neutrino-driven convection in the postshock heating region develop.
 This corresponds to the GW emission by ''Postshock convection'' (unless
  shown in the plot) after the prompt convection till $\sim
 550$ ms after bounce. Subsequently, the GW amplitudes become much more 
 larger as the SASI enters to a non-linear phase 
with violent sloshing of the postshock material. Large spikes appearing
 in the "Non-linear SASI" phase come from the down-flowing "SASI plumes" 
 striking the PNS surface. Afterward when the sloshing shock turns to be explosion 
(indicated by ''Explosion''), the sign of the GW amplitudes change reflecting the
 geometry of the expanding shock, that is either prolate (increase),
 oblate (decrease), or spherical (in-between) as shown in the top right and
 bottom panels in Figure \ref{murphy}. On top of these illuminating findings, they argued that the
 characteristic GW frequency has a tight correlation with a deceleration
 timescale of the accreting material at the lower boundary of the
 postshock convective region. As will be mentioned in section \ref{stochastic},
 this trend is also observed in a recent 3D simulation by M\"uller et al.
  \cite{ewald11}.

\subsubsection{Features of Neutrino GWs in experimental 2D simulations}\label{exp2}
To understand the behavior of neutrino GWs\footnote{
As anisotropic matter motions generate GWs, anisotropic neutrino emission also gives 
rise to GWs, which has been originally pointed out in late 1970's by 
\cite{epstein,turner1979} (see recent progress in \cite{favata}). It is expected as a primary GW source also in 
 gamma-ray bursts \cite{hiramatsu,suwa09} and Pop III stars \cite{suwa07b}.} in 2D simulations, it is useful to look
 the stress formula. In 2D simulations, the nonvanishing component of the neutrino 
GWs can be expressed as \cite{muyan97} 
\begin{equation}
h^{\rm TT}_{\nu} = \frac{4G}{c^4 R} \int_{0}^{t} dt^{'}
\int_{0}^{\pi}~d\theta^{'}~\Phi(\theta^{'})~\frac{dl_{\nu}(\theta^{'},t^{'})}
{d\Omega^{'}},
\label{tt}
\end{equation}
where $G$ is the gravitational constant, $c$ is the speed of light, $R$
is the distance of the source to the observer, $dl_{\nu}/d\Omega$
represents the direction-dependent neutrino luminosity emitted per unit
of solid angle into direction of $\Omega$,  and $\Phi(\theta^{'})$ denotes
the quantity, which depends on the angle measured from the symmetry axis 
($\theta^{'}$),
\begin{eqnarray}
\Phi({\theta^{'}}) &=&  \pi \sin \theta^{'} ( - 1 + 2 | \cos  \theta^{'}| ).
\label{graph1}
\end{eqnarray} 
 
It is easy to show that the neutrino GW is zero if the neutrino emission is isotropic
 ($dl_{\nu}/(\theta^{'},t^{'})$=constant).
 This function has positive values in the north polar cap for 
$0 \leq \theta' \leq 60^{\circ}$ and in 
the south polar cap for $120^{\circ} \leq \theta' \leq 180^{\circ}$, 
but becomes negative values between $60^{\circ} < \theta' < 120^{\circ}$. 
 Therefore, the polar and equatorial excess in the neutrino emission tends to
 make a positive and negative change in the neutrino GWs, respectively.

\begin{figure}[hbtp]
  \begin{center}
    \begin{tabular}{cc}
\resizebox{70mm}{!}{\includegraphics{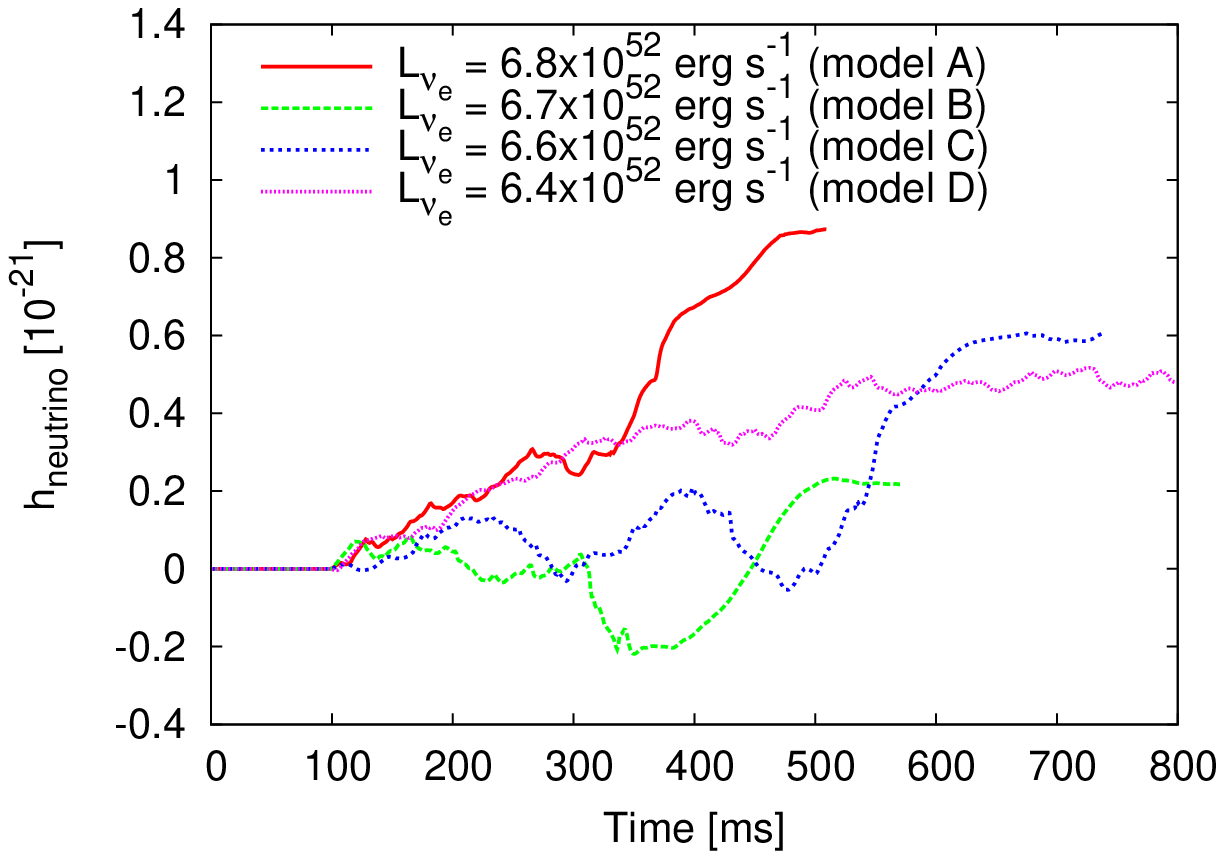}} &
\resizebox{70mm}{!}{\includegraphics{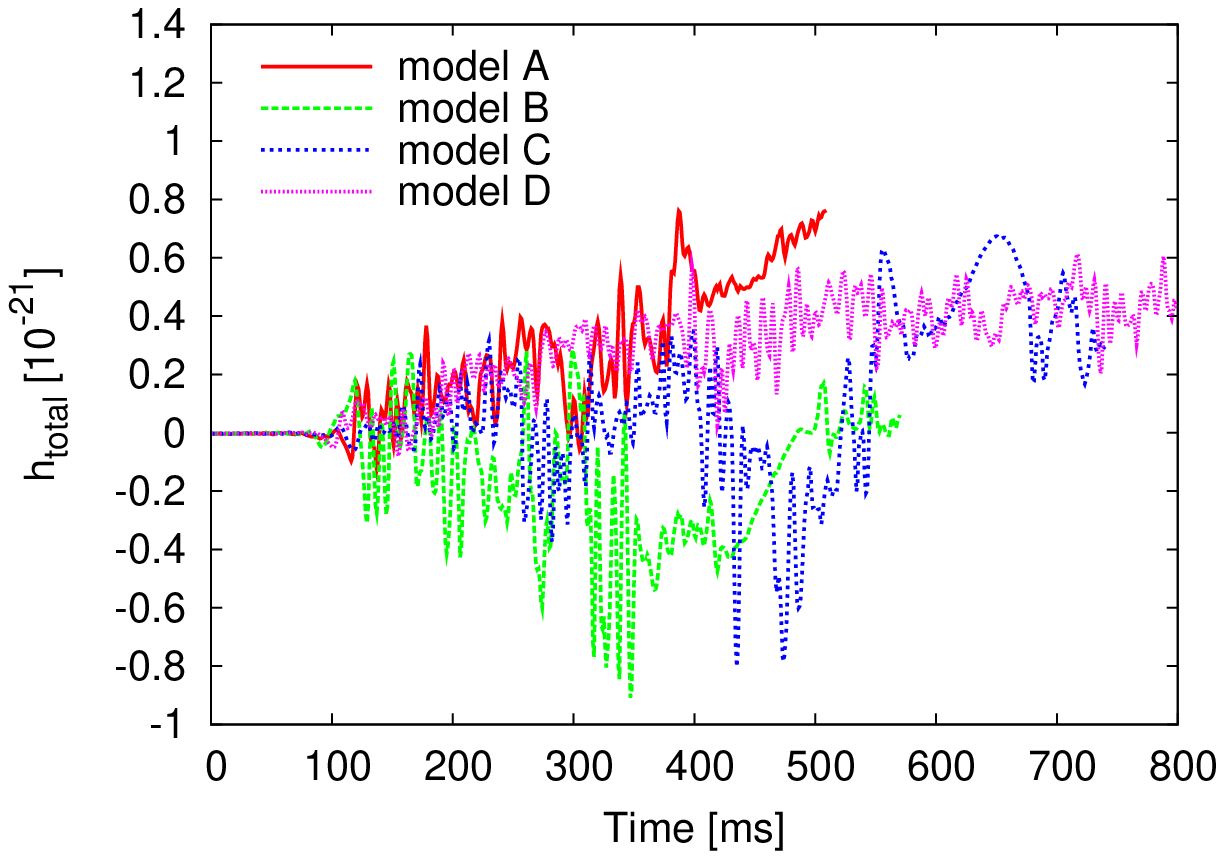}}  \\
\resizebox{70mm}{!}{\includegraphics{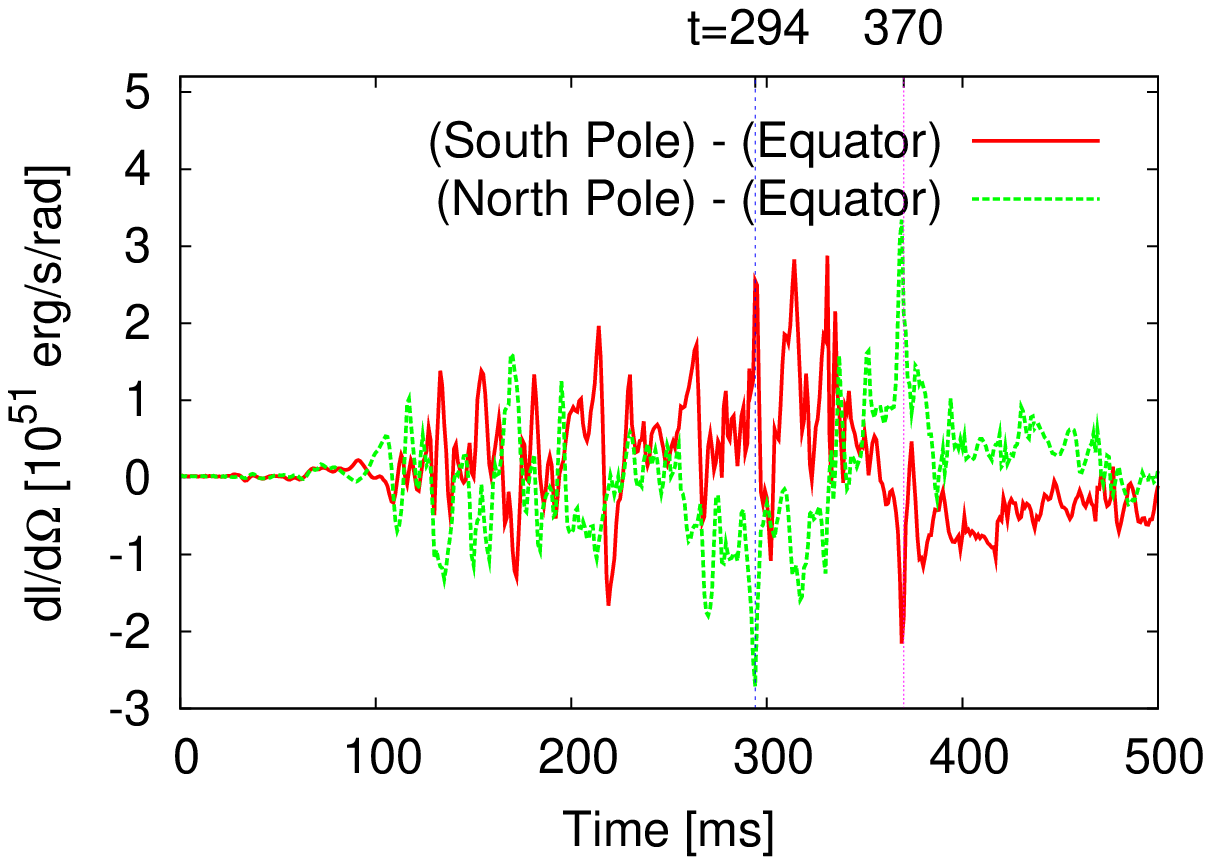}}  &
     \resizebox{90mm}{!}{\includegraphics{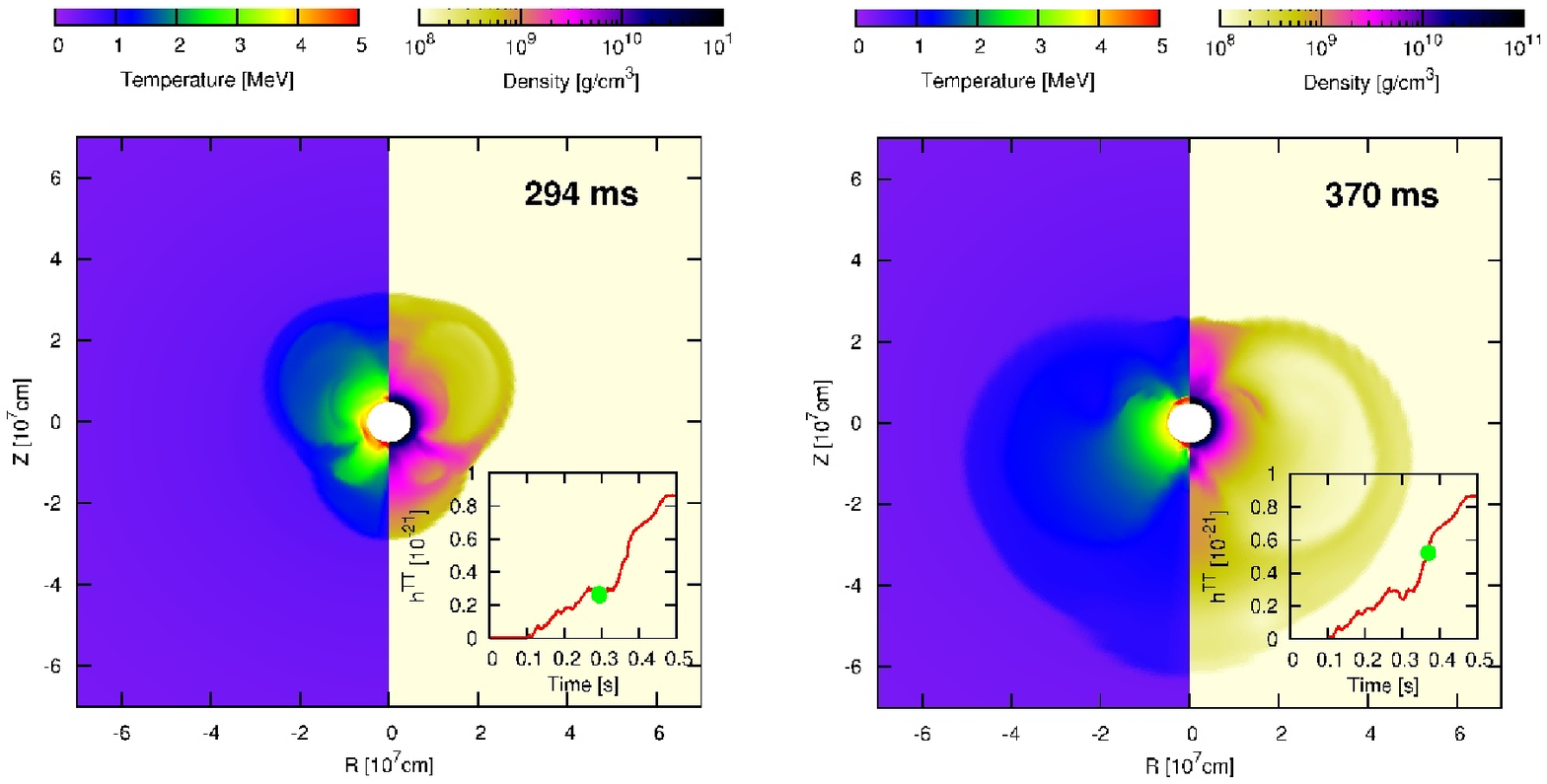}}  \\
    \end{tabular}
   \caption{Gravitational waveforms obtained in 2D parametric explosion models in 
 Kotake et al. (2009) \cite{kotake_ray} (for a distance of 10 kpc). For the four models (of A, B, C, and D) with different input 
luminosities (see text for more details), the top panels show the GWs only from neutrinos
 (top left) and from the sum of neutrinos and matter (top right).
 The time is measured from the epoch when the simulation starts from the 
 initial condition (see text). In all the models, the SASI enters to 
the non-linear regime at about $100$ ms, simultaneously making the amplitudes 
deviate from zero. The bottom left panel shows the directional dependent neutrino 
luminosity: $dl_{\nu}/d\Omega$ for 
model A, in which the polar excess from equator in the luminosity is indicated 
 by "(South pole) - (Equator)" or "(North pole) - (Equator)".
 Vertical lines represent the epochs of $t$ = 294 and 370 ms, respectively 
(see text for details). For the chosen two epochs, the bottom right panel shows 
temperature- (the left-half of 
each panel) and density- (the right-half) distributions in the meridian section.
 The insert of each panel shows the neutrino GW amplitudes, in which the green point 
indicates the time of the snapshot. 
The central region colored by white (50 km in radius) 
 represents the inner boundary taken in the 2D simulation.}
  \label{f3_GW}
  \end{center}
\end{figure}

 Figure \ref{f3_GW} shows properties of neutrino GWs obtained in the parametric
 2D simulations by Kotake et al. (2009) \cite{kotake_ray} who used the light-bulb scheme to obtain explosions.
 In their simulations, the initial conditions were derived from a steady-state 
approximation of the postshock structure (not from
 stellar evolution calculations) and the dynamics only outside an inner boundary
 at 50 km was solved. To estimate the neutrino anisotropy (e.g., $dl_{\nu}/(\theta^{'},t^{'})$ in Equation \ref{tt}), a ray-tracing analysis was performed. In the two top panels
  of Figure \ref{f3_GW},  the input electron neutrino luminosity ($L_{\nu_e}$) is 
indicated for four models A, B, C, and D, respectively corresponding to 
$L_{\nu_e} = 6.4, 6.6, 6.7,$ and $6.8\times 10^{52}$erg/s.

 The top left panel shows the GW amplitudes
contributed only from neutrinos. Comparing the right panel, which shows the 
  total amplitudes (neutrino + matter),
 it can be seen that the overall structures of the waveforms are 
 predominantly determined by the neutrino GWs with slower temporal 
 variations ($\gtrsim 50$ ms), to which the GWs from matter motions 
 with rapid temporal variations ($\lesssim 10$ ms) are superimposed.

The bottom left panel shows the angle dependent neutrino 
luminosity (e.g., $dl_{\nu}/d\Omega$ in Equation (\ref{tt})) for model A, 
in which the polar excess 
from equator in the luminosity is indicated by "(South pole) - (Equator)" or 
"(North pole) - (Equator)". As shown, the dominance of the 
neutrino emission in the north (green line) and south poles (red line) 
are closely anti-correlated. This is the consequence of the low-mode nature of the SASI, 
here of $\ell=1$ (in terms of an expansion in spherical harmonics with order $\ell$).
 In fact, the left-hand side in the bottom right panel 
shows that at 294 ms, 
the blob encompassing the regions inside the stalled shock is moving from the southern 
to the northern hemisphere, 
leading to the compression of the matter in the south hemisphere, 
which is vice versa at 370 ms (right-hand side).
 Recalling again that the function (Figure \ref{graph1}) is positive near poles,
 the polar excess in the neutrino GWs makes the positively growing feature depicted 
 in the neutrino GWs (e.g., top left panel in Figure \ref{f3_GW}).
In addition, the waveforms show
large negative growth for some epochs during the growth of SASI (e.g., 
 for models C and D in the top left panel). 
 Such a feature comes from a dominance of the neutrino emission in the equatorial
 direction\footnote{To capture this trend, the ray-tracing calculation would be   
 more accurate than a ray-by-ray approach.}. Reflecting the nature of the SASI
 and convection both of which grow chaotically with time, they pointed out that 
 there may be no systematic dependence of the input neutrino luminosities on the 
maximum GW amplitudes.

\subsubsection{GW signatures in 2D radiation-hydrodynamic
   simulations}\label{first}

 Having summarized the findings in parametric
 2D simulations, we are now ready to discuss GW
 signatures obtained in 2D radiation-hydrodynamic simulations in which
 a spectral neutrino transport is solved. Similar to Table 1, but Table 2 shows a 
summary for some selected GW predictions obtained in the first-principle 
2D models\footnote{Note that much more stronger GW emission than 
those in Table 2 was expected if the
 acoustic mechanism would work. See section 7.1 in \cite{ott_rev} for more details.}.
 Since the model predictions are not too many (as of 2011),
 the important GW features obtained mostly by the MPA simulation and also by the 
OakRidge+ simulation are all summarized in this section by reproducing their results 
 as much as possible.

\begin{table}
\begin{center}
\begin{tabular}{lcccccc}
\hline
Progenitor & Group & Mechanism & Dim. & $t_{\rm fin}$ & $|h_{\rm max}|$(cm) & $f_{\rm peak}$ \\ 
 (rotation)   & (Year) & (Domain)  & (Hydro) & (ms) & (from) & (Hz) \\ 

\hline
11.2 $M_{\odot}$\cite{woos02} & MPA\cite{buras06} & $\nu$-driven & 2D & $\sim$200 
    & $\sim$30 & $\sim$700 \\
 (No)    & (2004)  &(180$^\circ$wedge) & (PN)   & & (matter) &      \\    \cmidrule(l){1-7}
12 $M_{\odot}$\cite{woos02}    & Oak Ridge+ & $\nu$-driven & 2D & $\sim$550 & 80 & $\sim$900  \\
  (No)  &  \cite{bruenn}(2009) & (180$^\circ$wedge) & (PN)   & & (neutrino) &  \\    \cmidrule(l){1-7}
13 $M_{\odot}$\cite{nomo88}   & NAOJ+ & $\nu$-driven  & 2D & $\sim$700 & 18  & $\sim$600  \\
 (Both)  &  \cite{suwa}(2011) &(180$^\circ$wedge) & (N)   & & (neutrino) &     \\    \cmidrule(l){1-7}
 15 $M_{\odot}$ \cite{WW95} & MPA\cite{buras06} & $\nu$-driven & 2D & $\sim$270 
    & $\sim$150  & $\sim$600 \\
 (Yes)     & (2004)  &(90$^\circ$wedge) & (PN)   & & (neutrino) &      \\    \cmidrule(l){2-7}
 & MPA\cite{buras06} & $\nu$-driven & 2D & $\sim$400 
    & $\sim$130  & $\sim$600 \\
 (No)   & (2006)  &(180$^\circ$wedge) & (PN)   & & (neutrino) &      \\    \cmidrule(l){1-7}

15 $M_{\odot}$\cite{woos02}    & Oak Ridge+ & $\nu$-driven & 2D & $\sim$550 & $\gtrsim$200 & $\sim$900  \\
 (No)  &  \cite{bruenn}(2009) & (180$^\circ$wedge) & (PN)   & & (neutrino) &  \\    \cmidrule(l){1-7}
25 $M_{\odot}$\cite{woos02}    & Oak Ridge+ & $\nu$-driven & 2D & $\sim$550 & $\sim$200 & $\sim$900  \\
 (No)  &  \cite{bruenn}(2009) & (180$^\circ$wedge) & (PN)   & & (matter) &  \\    \cmidrule(l){1-7}
\end{tabular}
\end{center}
\caption{Similar to Table 1, but for selected GW predictions obtained in the 
 state-of-the-art 2D neutrino-radiation hydrodynamic simulations. See Table 1 
 for the details of the employed numerical techniques. The column of (Domain) indicates 
 that the computational volume in the lateral direction is solved fully $0 \leq \theta
 \leq 180^{\circ}$ ("180$^\circ$wedge") or with an assumption of certain imposed 
 symmetry ("90$^\circ$wedge"). $t_{\rm fin}$ represents the 
 final simulation time, and $|h_{\rm max}|$ indicates the absolute maximum GW amplitudes
 which come either from "neutrino" or "matter" GWs. $f_{\rm peak}$ represents
  the peak frequency in the GW spectra.
}
\end{table}

\begin{figure}[hbt]
\begin{center}
\includegraphics[width=.95\linewidth]{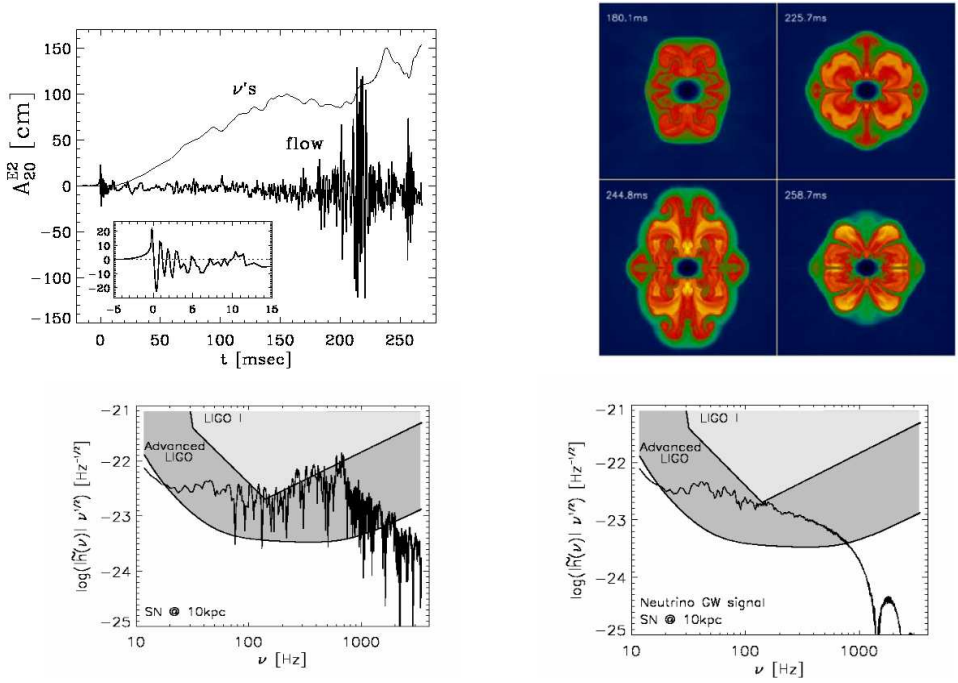}
\caption{Gravitational waveforms from neutrinos (indicated by
 ''$\nu$'s'' in the top left panel) or from matter motions (indicated
  by''flow'' in the panel) and their spectra
 (sum of matter and neutrino (bottom left) and contribution only
  from neutrino GWs (bottom right)) for a rotating model (s15r) of 15 $M_{\odot}$
  progenitor in M\"uller et al. (2004)\cite{mueller04}. The insert in the top left panel shows
  an enlargement of the signal near bounce, which comes from prompt convection (e.g., 
 section \ref{exp1}).
  Top right panel shows the growth of large scale, non-radial pulsations
 in the post-shock region (bright regions in the entropy plot (600x600 km)),
 which leads to a slower temporal evolution ($\gtrsim 10$ ms)
 of the neutrino GWs. All of these plots are by courtesy of M\"uller and 
 the coauthors \cite{mueller04}.}
\label{f1}
\end{center}
\end{figure}

The first example is for a rotating 15$M_{\odot}$ progenitor model by 
 the MPA simulation
 (e.g., \cite{buras06}, see the numerical details in Table 1), the detailed 
GW analysis of which was performed by M\"uller et al. (2004) (Figure \ref{f1}).
 The initial angular velocity for the model was 
 taken to be 0.5 rad/s, which is slightly faster (a factor 5) than the 
one predicted in a recent stellar evolution calculation \cite{hege05}.
 As shown in the top left panel, they found a quasi-monotonically increasing GW signal
   at the end of their simulation ($\sim 270$ ms after bounce), which predominantly 
 comes from neutrino GWs. In the case of rotational core-collapse, neutrino emission 
can be stronger along polar directions \cite{kota03a,jamu,walder}. 
As discussed in section \ref{exp2}, 
the polar excess in neutrino emission could be a possible explanation of the increasing
   trend. It is also worth mentioning that this feature can be seen in
   the waveform for a rapidly-rotating accretion-induced collapse model \cite{dessart_07}
 of the ''Princeton+'' simulation (e.g., Figure 10 in \cite{ott_rev}).
 
  The top right panel of Figure \ref{f1} 
shows the growth of large scale, non-radial pulsations
 in the post-shock region (bright regions in the entropy plot (600x600 km)),
 which leads to a slower temporal evolution ($\gtrsim 10$ ms)
 of the neutrino GWs (see the line labeled by "$\nu$'s" in the top left panel). 
As seen, an equatorial symmetry was imposed for the 2D
 simulation\footnote{Hence the dynamics is north-south symmetric.},
 which is indicated by "90$^{\circ}$wedge" in Table 2.
 Comparing the GW spectra from the sum of matter motions and neutrinos 
(bottom left panel in Figure \ref{f1}) to the one only 
from neutrinos (bottom right), one can clearly see that the low-frequency part of 
the spectrum (below $\sim$ 100 Hz) is dominated by the contribution of the 
neutrino GWs. 
As mentioned above, this is because their wave amplitude varies 
on much longer timescales ($\gtrsim 10$ ms) than that of the mass flow 
($\sim $ ms, e.g., the waveform in the top left panel). For this model,
  the maximum amplitude comes from the neutrino contribution (e.g., the top left panel 
 and $|h_{\rm max}|$ with the comment of (neutrino) in Table 2).

  It should be noted whether the 
 maximum amplitude comes from matter or neutrino contribution change from models to 
 models as shown in Table 2 (see the sixth column indicated by (from)), which may reflect
 the stochastic nature of the SASI and convection which is determined 
by the non-linear hydrodynamics. In contrast, the peak frequencies in the GW spectrum 
 (typically in the range of 600-1000 Hz) always come from the matter GWs (albeit
 not explicitly mentioned in the table) due to 
its short temporal variations ($f_{\rm  peak}$ in Table 2).
 The bottom panels of Figure \ref{f1} show that while the 
 signal-to-noise ratio is probably too small for this event to be detectable by
 LIGO I, it could be well detected by the Advanced LIGO for a Galactic source. This
 is a generic feature of the detectability of the GWs from neutrino-driven explosion
 models.

\begin{figure}[hbt]
\begin{center}
\includegraphics[width=1\linewidth]{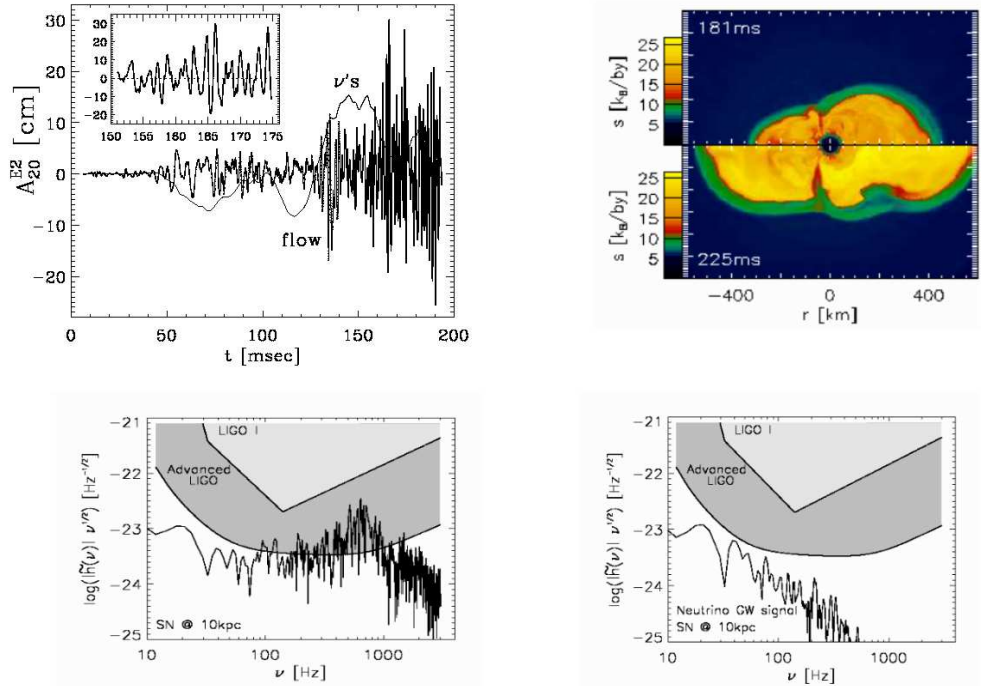}
\caption{Similar to Figure \ref{f1} but for a non-rotating explosion
  model in the MPA simulations of an 11.2 $M_{\odot}$ progenitor (taken 
from M\"uller et al. (2004)\cite{mueller04}).
 The top right panel indicates the success explosion of this model, in which 
 the entropy (bright) with different timescales 
(181 (top) or 225 ms (bottom) after bounce) are displayed. The polar 
 axis is directed horizontally.
  These figures are by courtesy of M\"uller and the coauthors \cite{mueller04}.}
\label{f2}
\end{center}
\end{figure}

\begin{figure}[hbt]
\begin{center}
\includegraphics[width=1.2\linewidth]{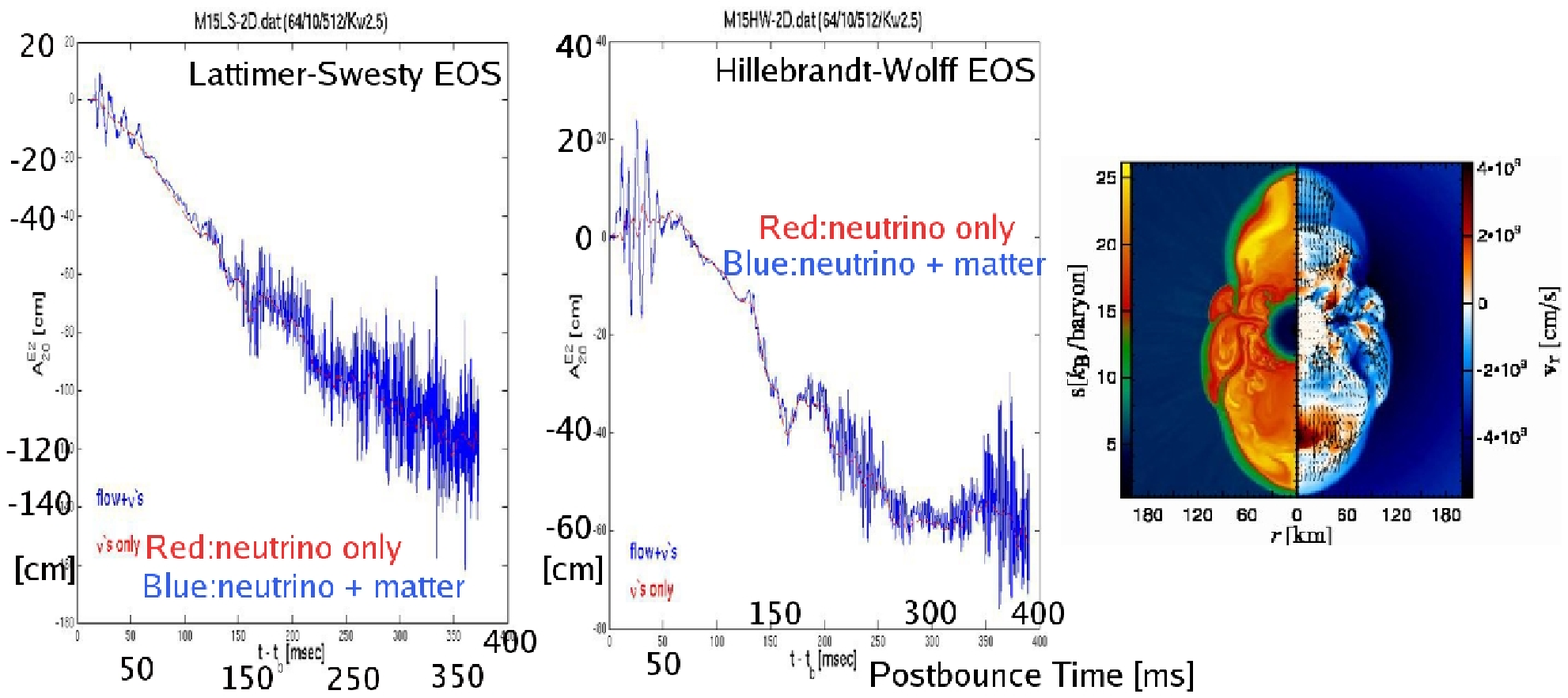}
\caption{Gravitational waveforms for a non-rotating
 exploding model in the MPA simulations of a 15 $M_{\odot}$ progenitor
  \cite{marek} in which the
  Lattimer-Swesty (left panel) or Hillebrant-Wolff EOS (middle panel) was employed
 (taken from Marek et al. (2009) \cite{marek_gw}).
  The right panel indicates the explosion of this model, in which the lefthand panel
 shows color-coded entropy distribution, the righthand panel the radial velocity 
components with white and whitish hues denoting matter at or near rest, while 
 black arrows indicate the direction of the velocity fields. These figures
  are by courtesy of Marek and the coauthors \cite{marek_gw}.
  }
\label{f3}
\end{center}
\end{figure}

Figure \ref{f2} is the same as Figure \ref{f1} but for a non-rotating {\it exploding} 
model of an 11.2 $M_{\odot}$ progenitor of Buras et al.\cite{buras06}. In contrast 
 to the rotating model mentioned above, no signal is emitted near at bounce for 
 this non-rotating model (compare
 the insert in the top left of Figure \ref{f1} and the waveform near $t=0$ in
 the top left panel of Figure \ref{f2}). 
For this light progenitor model, the wave amplitude from 
 prompt convection (till $\sim 50$ ms after bounce) is very small.
 Only after the neutrino-driven convection becomes 
 stronger ($\sim 80$ ms after bounce), the maximum (absolute) amplitudes rise to 
the values of several 10 cm. Although the slower temporal evolution of the neutrino GWs
 is similar to the one in Figure \ref{f1}, the monotonically growing 
 trend is not seen for this model. The GW spectrum (bottom panels in Figure \ref{f2})
is qualitatively quite similar to that of the rotating model (Figure \ref{f1}),
 but the GW amplitudes become much smaller. This is probably because 
 of a smaller core mass of the 11.2 $M_{\odot}$ progenitor, which could 
potentially make the mass quadrupole and the emergent neutrino luminosity smaller
 (compare the spectrum of neutrino GW in the bottom right panel
of Figure \ref{f1} and \ref{f2}). 

The left and middle panels in Figure \ref{f3} show the wave amplitude 
(blue line (total), red line (only from neutrinos)) for a non-rotating {\it 
 exploding} 15 $M_{\odot}$ progenitor in Marek \& Janka (2006) \cite{marek} in which 
 a soft variant of the Lattimer-Swesty EOS (left) and a rather stiff Hillebrant-Wolff
 EOS (middle) was employed.\footnote{An incompressibility at nuclear
densities ($K$) is 180 MeV and 281 MeV for the Lattimer-Swesty and Hillebrant-Wolff
 EOS, respectively.} Although an overall trend is similar with each other, they found 
 several important EOS effects on the GW signatures. The first one is about the GWs from
 prompt convection (e.g., the signals before 50 ms after bounce). As seen, the wave 
amplitude for the stiffer EOS ($\sim$ 20 cm) becomes up to a factor of 2 larger maximum 
 amplitudes than its softer EOS counterpart  ($\sim$ 10 cm). They pointed out that 
 the prompt convection exhibits larger amplitude, slightly higher frequencies, and 
 more power for the stiffer EOS. They found that this is because the region of prompt 
convection involves more mass and extends to larger radii. This may agree with one's 
 intuition that the inner-core mass near bounce becomes larger in the case
 of stiffer EOS (due to the suppressed electron capture), potentially providing a larger quadrupole GW emission.

 The maximum (absolute) amplitudes come from the neutrino 
 contribution irrespective of the employed EOSs (see the column of (from) for 
 the non-rotating $15 M_{\odot}$ progenitor in Table 2). The maximum (absolute)
 amplitude is about 130 and 60 cm for the softer and stiffer EOS, 
respectively.\footnote{Note that the result only for the softer EOS
 is given in Table 2.}
  The higher neutrino luminosities that are radiated from a more compact
 PNS in the softer EOS predominantly give a favourable condition for a more efficient
 emission of the neutrino GWs. However, their peak frequencies are typically 
 below $\sim$ 100 Hz (e.g., the bottom right panel in Figure \ref{f1}), making the component of the neutrino GWs very difficult to 
 be detectable for ground-based detectors whose sensitivity is limited mainly by 
 seismic noises at low frequencies. Therefore the matter GW signals are more 
 important in discussing the detectability. For a Galactic source, the matter signal
 are of the orders of $10^{-22}$ regardless of the different EOSs. 
They discussed that they should be marginally observable
 with LIGO I but it is not easy to tell the EOS difference. This
 is because although the spectral peak become (slightly) higher if the EOS is soft, 
 this simultaneously sticks out of the
 highest sensitivity frequency domain of the LIGO instrument.
 
In contrast to an anticipation in section \ref{exp2}, 
 their neutrino GWs decrease with time.
 They argued that this comes from the excess of neutrino emission in the equatorial 
 direction, which was argued to be
 determined by the anisotropic transport of heavy-lepton neutrinos
 in the vicinity of the PNS \footnote{On the other hand, they did see the 
stronger emission in the polar regions regarding $\nu_e$ and $\bar{\nu}_e$, which 
 is consistent with the discussion in section \ref{exp2}.}.
 To pin down the GW signal from neutrinos, one apparently needs to precisely 
 determine the neutrino anisotropy which should require multi-angle neutrino 
 transport calculation. Currently it would be practicable in 2D simulations 
 \cite{ott_multi} if enough 
 number of momentum-space angles could be cast (probably) 
 by using the next-generation supercomputers. But before that, we may need to be 
 a little bit careful in interpreting the neutrino GWs.

\begin{figure}[hbt]
\begin{center}
\includegraphics[width=.9\linewidth]{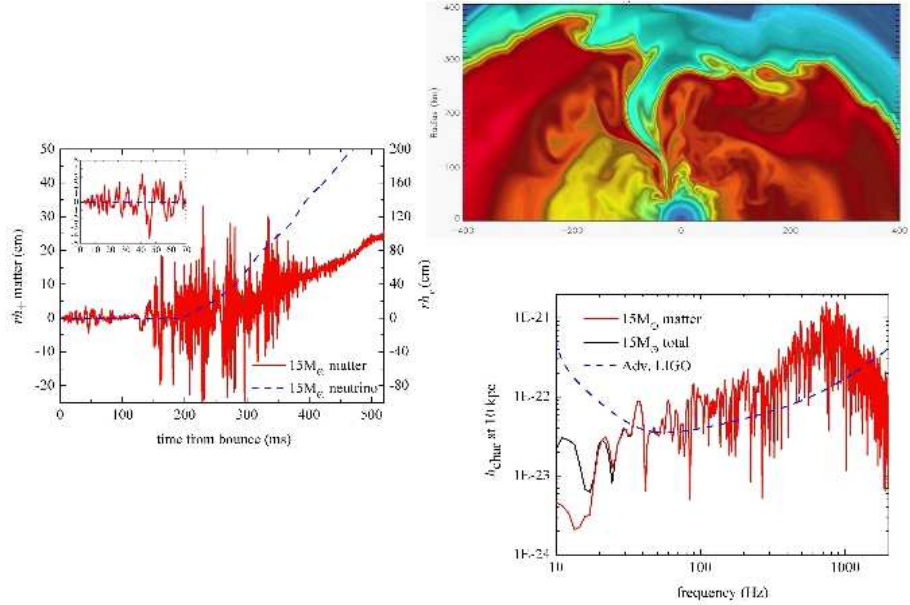}
\caption{Similar to Figure \ref{f1} but for a non-rotating exploding
  model in the OakRidge+ simulations of a non-rotating 15 $M_{\odot}$ progenitor
 (from Yakunin et al. (2009) \cite{yakunin}). The left panel shows the gravitational waveforms
  contributed from matter motions (indicated by matter) and from
 neutrinos (indicated by neutrino). The insert shows an enlargement of
 the signal near bounce, which comes from prompt-convection. The top right 
 panel shows entropy distribution at 244 ms after bounce for the 
progenitor. A bipolar explosion with high-entropy outflows (yellow-orange-red) 
is shown along the symmetry axis. The bottom
  right panel shows the spectra with the sensitivity curve of the
  advanced LIGO. These figures are by courtesy of Yakunin and the coauthors \cite{yakunin}.}
\label{f4}
\end{center}
\end{figure}
 
 Keeping this in mind, the neutrino GWs (blue line in the left panel of Figure \ref{f4}) 
for a non-rotating 15$M_{\odot}$ exploding model in the 
 OakRidge+ simulation (e.g., Table 1 for the numerical details)
 show again a monotonically increasing trend. They obtained a similar 
 trend in the waveforms of their exploding models of 12 and 25 $M_{\odot}$ progenitors
\cite{yakunin}. In consistent with the analysis 
in section \ref{exp1},  the matter GWs show 
a monotonic increase after 
 the SASI-driven sloshing shock turns into 2D prolate explosions
 ($\gtrsim 300$ ms postbounce in the left panel of Figure \ref{f4}). 
For the first time this model succeeded in capturing 
 the feature, which was not seen in other 
radiation-hydrodynamic simulations. To produce the increasing trend, the anisotropic 
 kinetic energy of the expanding shock needs to be enough large. Relatively earlier 
shock revival with larger explosion energies (e.g., Table 1) obtained in the OakRidge+
 simulations could satisfy the condition. The maximum amplitudes for a Galactic supernova
 reach to the orders of $10^{-21}$ regardless of the employed progenitors, which could 
be a promising target for the advanced LIGO (bottom right panel in Figure \ref{f4}).

\begin{figure}[hbt]
  \begin{center}
    \begin{tabular}{ccc}
\resizebox{55mm}{!}{\includegraphics{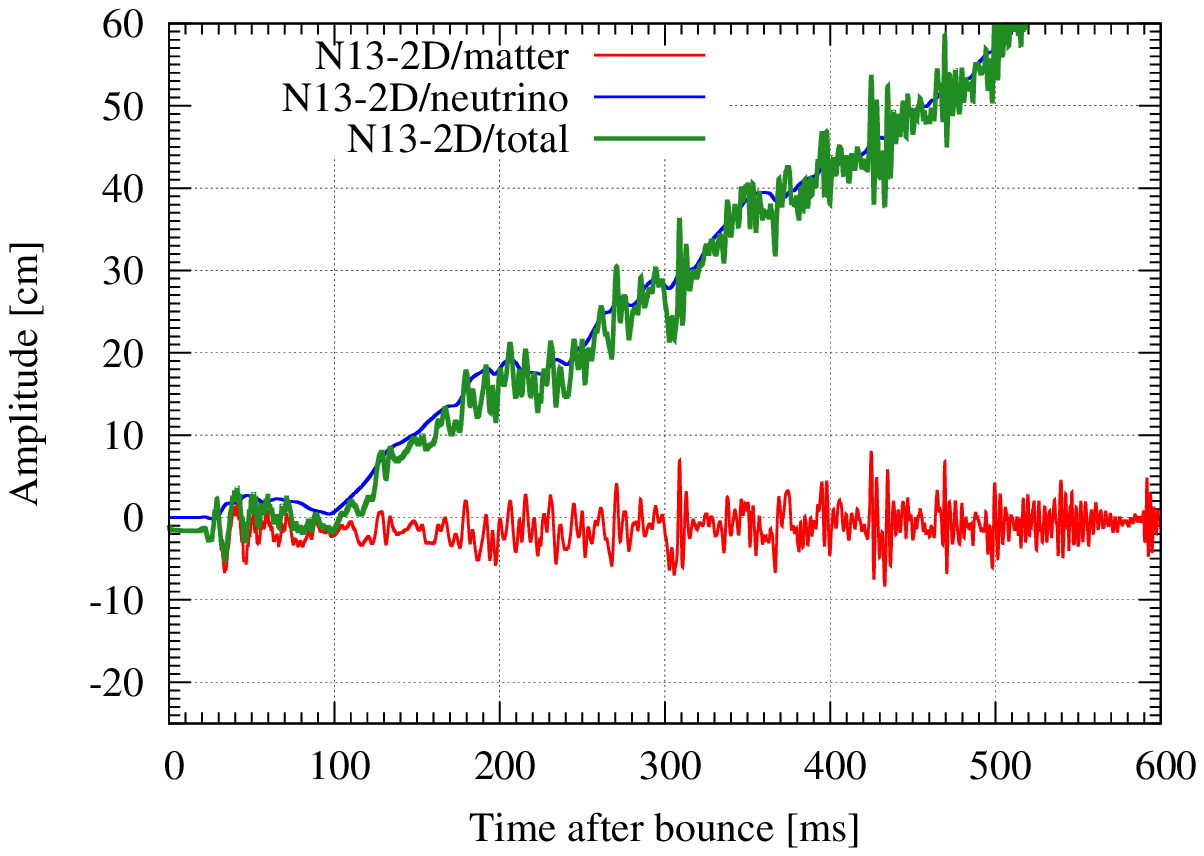}} &
\resizebox{55mm}{!}{\includegraphics{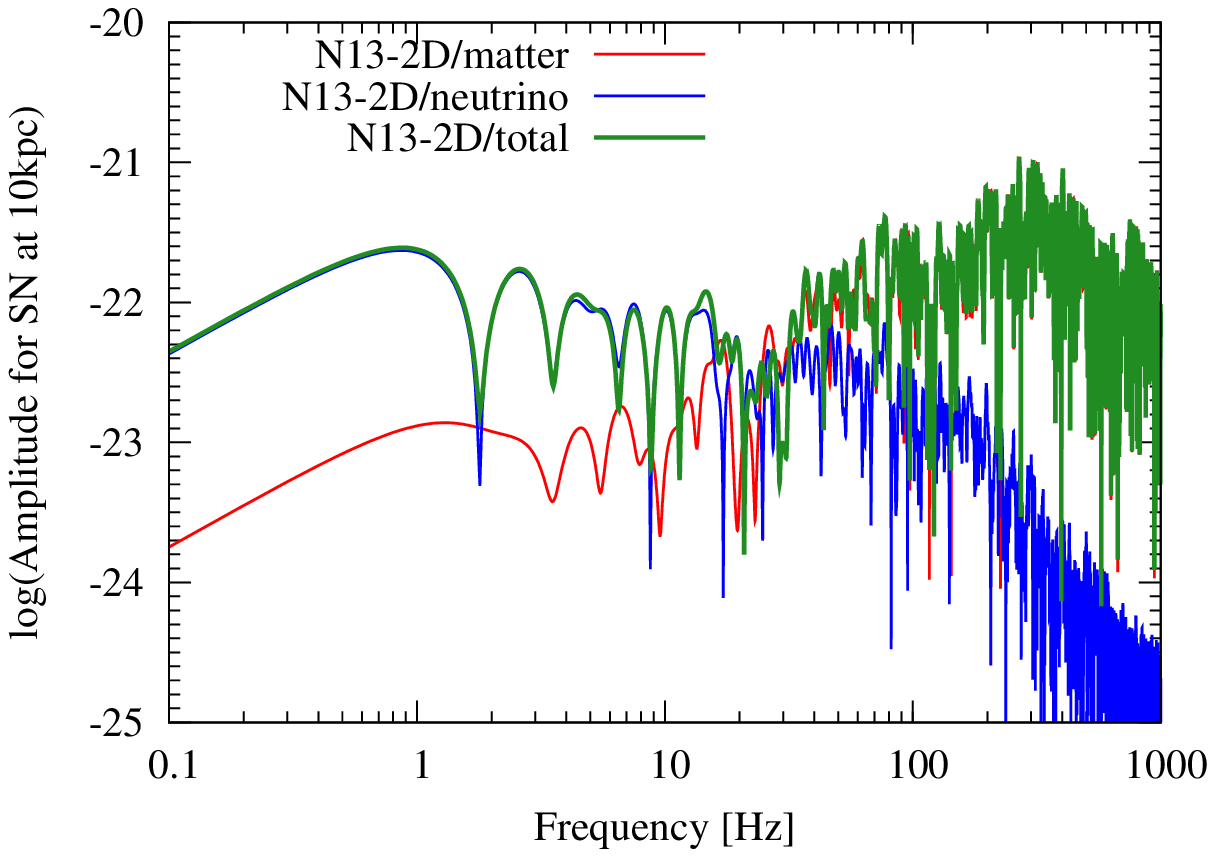}} &  
\resizebox{40mm}{!}{\includegraphics{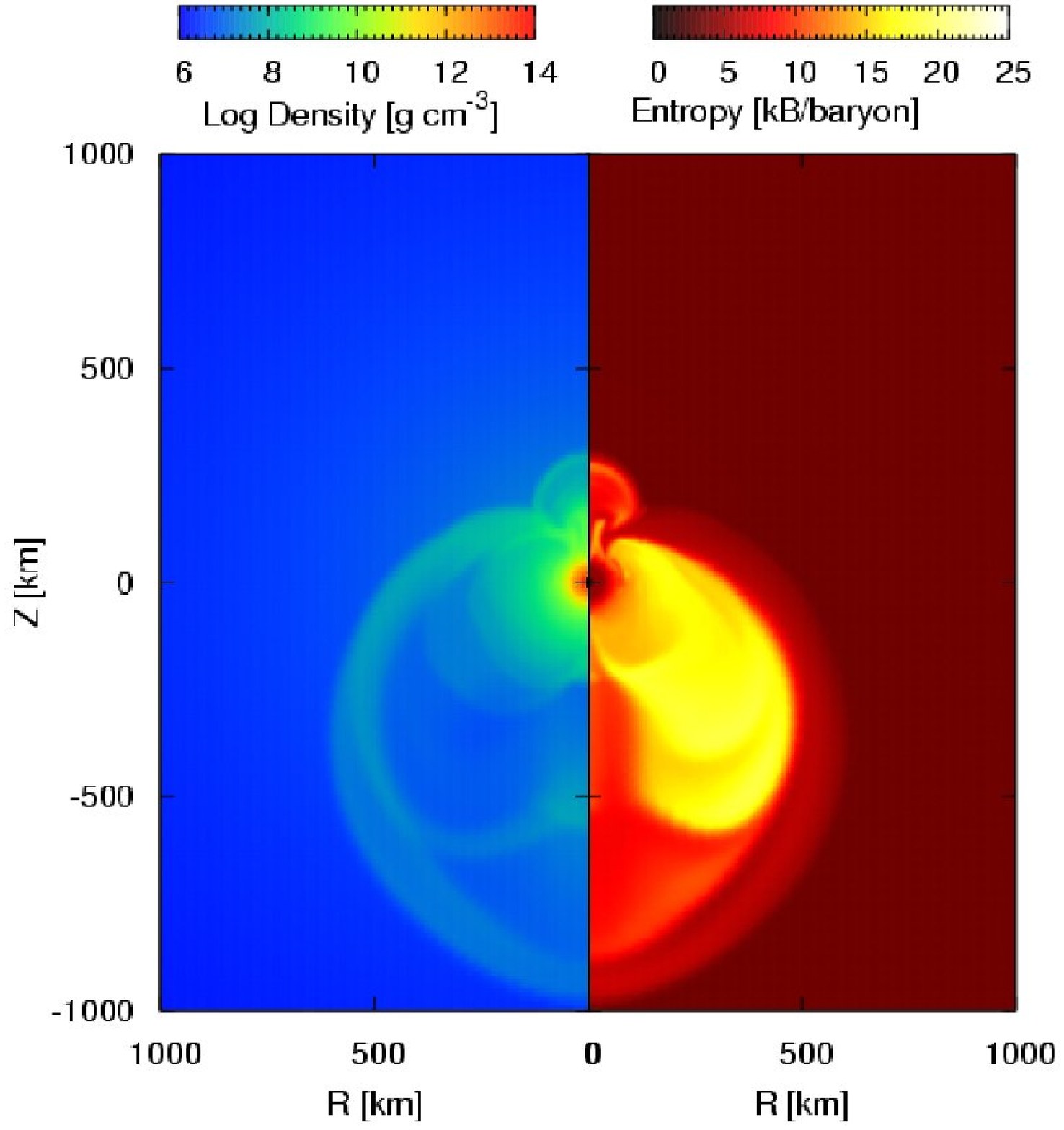}} \\
\resizebox{55mm}{!}{\includegraphics{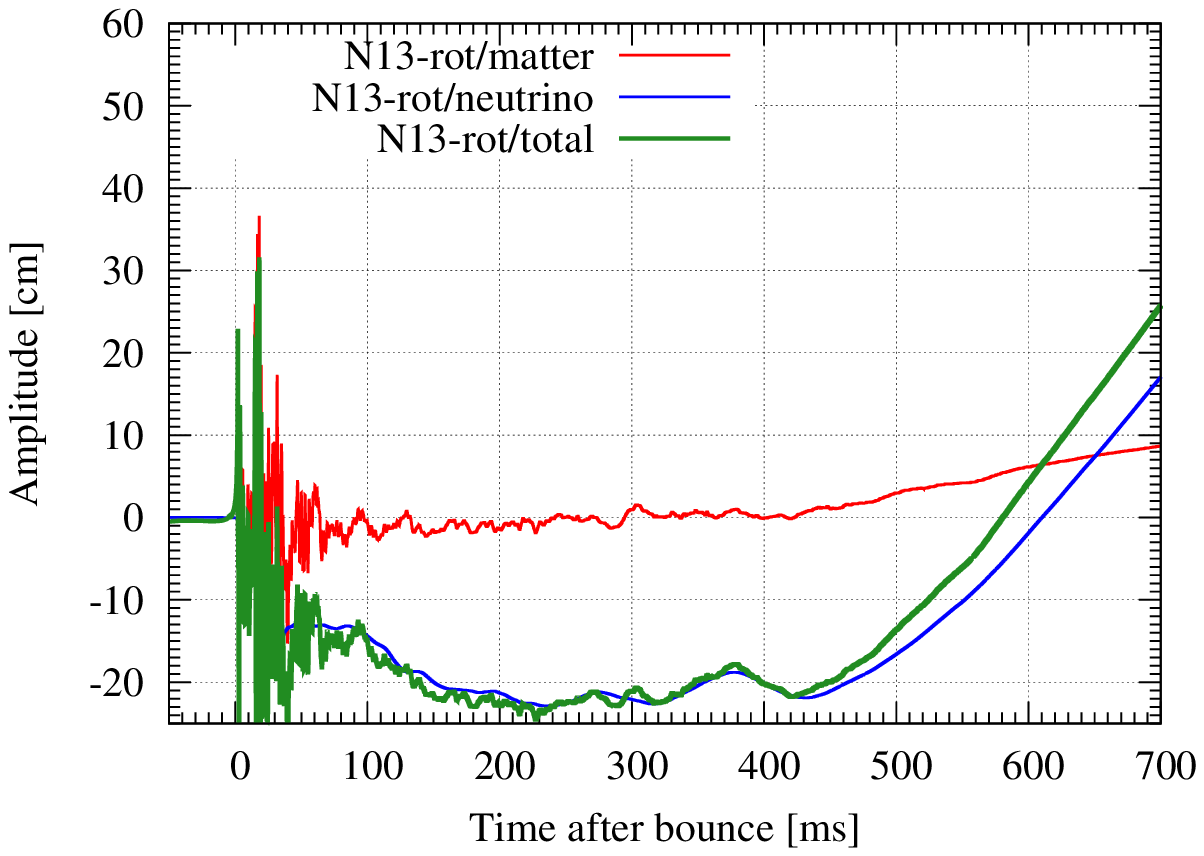}}  &
     \resizebox{55mm}{!}{\includegraphics{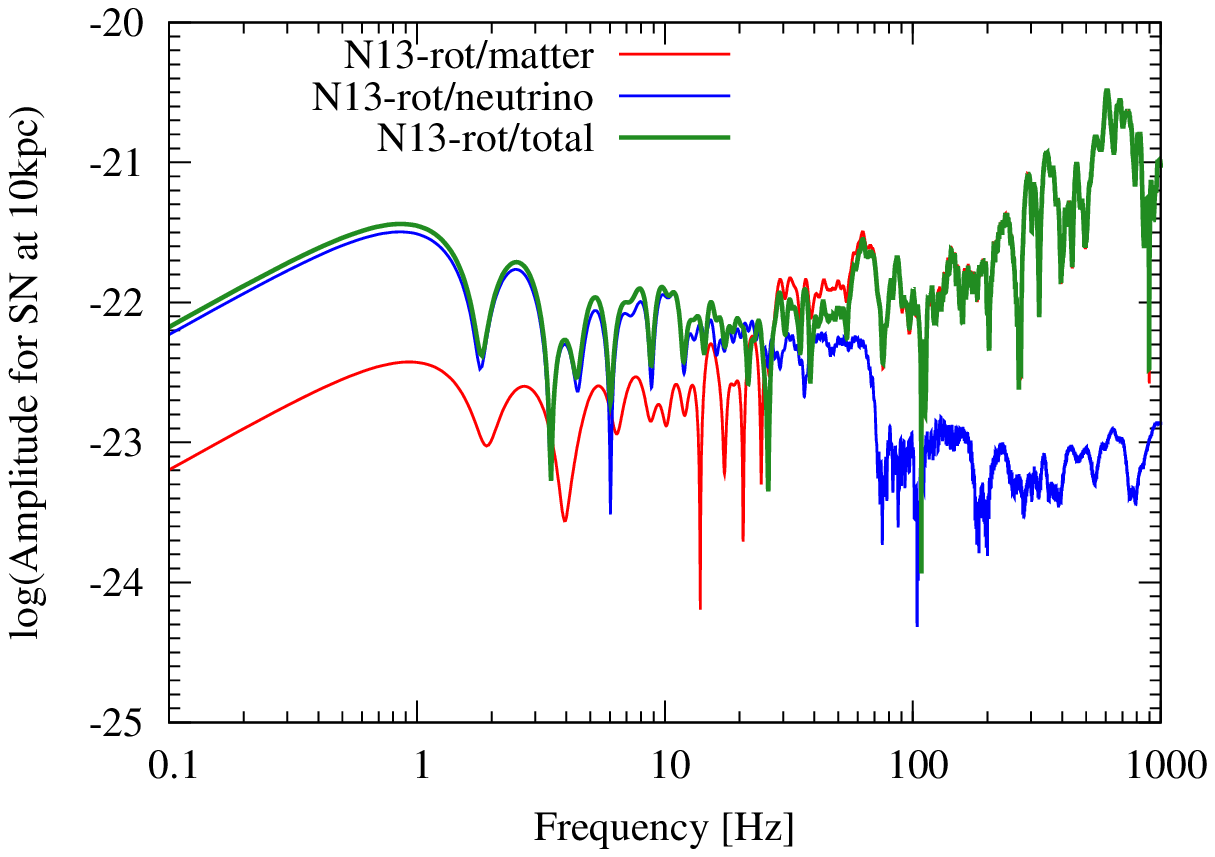}}  &
\resizebox{40mm}{!}{\includegraphics{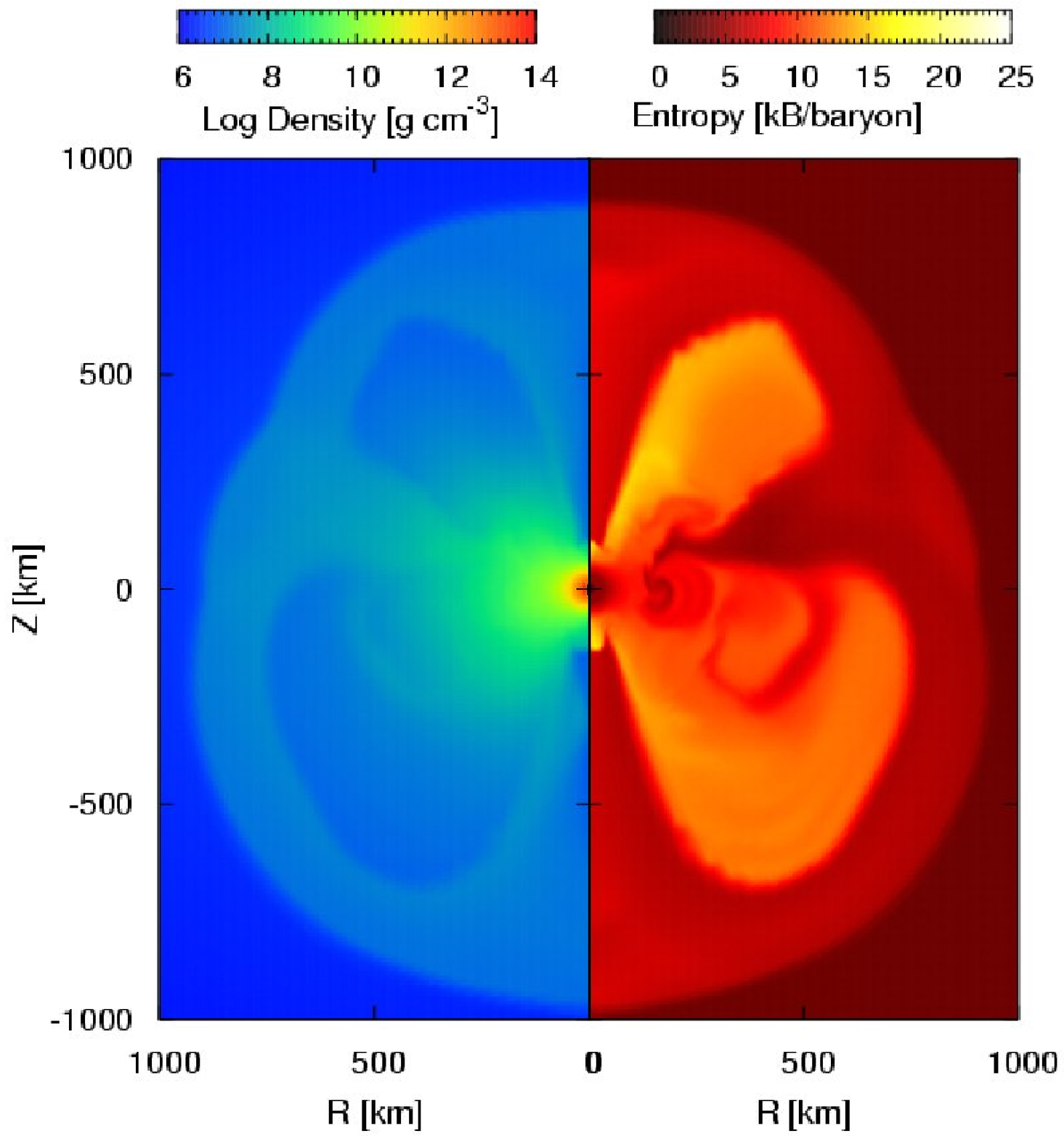}} \\
    \end{tabular}
   \caption{Gravitational waveforms (left panels) with their spectra 
(middle panels) for a non-rotating (top panels) and rapidly rotating model 
(bottom panels) for a 13 $M_{\odot}$ progenitor \cite{nomo88} obtained in the 
NAOJ+ simulation (e.g., Table 1). For the rotating model, the precollapse angular 
 velocity is taken to be 2 rad/s. Right panels show snapshots of the density 
(left half) and the entropy (right half) for models without (top) or with rotation
 (bottom) at the epoch when
  the shock reaches 1000 km, corresponding to $\sim$ 470 ms after
  bounce in both cases. The middle panels are for a supernova 
 at a distance of 10 kpc. These figures are by courtesy of Suwa and the 
 coauthors \cite{suwa_in_prep}.
}
  \label{f5}
  \end{center}
\end{figure}

Figure \ref{f5} shows the wave amplitudes (left panels) with their spectra 
(middle panels)  for a non-rotating (top panels) and rapidly rotating model 
(bottom panels) for a 13 $M_{\odot}$ progenitor \cite{nomo88} obtained in the 
NAOJ+ simulation (e.g., Table 1 for the numerical details).
 For the non-rotating model, the neutrino GWs 
 show the increasing trend (green line in the top left panel). This model
 explodes in a unipolar manner (top right panel), expelling material
 only in the northern part of the core. Due to this one-sided explosion with 
smaller mass ejection, this model explodes only weakly with decreasing explosion 
energies less than $\sim 4 \times 10^{49}$ erg at the final simulation time 
(see Figure 4 in \cite{suwa}). This should be the reason why the matter GW signals
 (red line in the top left panel) are much smaller than those obtained in the 
 Oak Ridge+ simulation (e.g., red line in Figure \ref{f4}). 

On the other hand, a bipolar explosion is seen to be obtained for their 2D model
 with rapid rotation\footnote{The precollapse angular velocity was  taken to be 2 rad/s.}
(bottom right panel in Figure \ref{f5}). 
Since the north-south symmetric ($\ell =2$) explosion can expel more
material than for the unipolar explosion, the explosion energies becomes larger
 (e.g., Figure 4 in \cite{suwa}), simultaneously making the matter GWs larger. In fact, the increasing trend in the matter 
 GWs is seen after $\sim 400$ ms after bounce (red line in the bottom left panel 
 in Figure \ref{f5}). The neutrino GWs firstly decrease after bounce (blue line
 in the same panel), but later shift to exhibit the increasing trend ($\gtrsim 500$ ms 
 after bounce). The middle panels (for a supernova 
 at a distance of 10 kpc) show that the spectrum rises to a broad peak between
 $\sim 600$ and 900 Hz, which could be visible to the 
 advanced-class detectors for the galactic source 
(see further discussions in \cite{suwa_in_prep}).

\subsubsection{Stochastic GW nature in experiment 3D simulations}\label{stochastic}

 (By definition) 2D simulations so far mentioned in previous sections have an 
 assumption of axisymmetry. Due to the symmetry axis, the growth of the SASI and the 
large-scale postshock convection could develop along the axis preferentially, thus 
 suppressing anisotropies in explosions. Since the GW emission is 
 very sensitive to the degree of the explosion anisotropies, 3D simulations are
 apparently needed for a more accurate GW prediction.

 Since it is still computationally very expensive to solve the spectral
 neutrino transport,
a light-bulb scheme (e.g., \cite{jankamueller96}) has been often employed 
 in 3D simulations to trigger explosions, 
in which the heating and cooling by neutrinos 
  are treated by a parametric manner (e.g., section \ref{s2} and 
 references in \cite{iwakami1,nordhaus,annop2}).
Using the recipe, a handful of GW predictions based on the 3D simulations
 have been reported so far, which we are going to review in this section.
As a preface, we just like to mention that 
 although these studies use experimental approaches and thus their findings 
should be tested by the first-principle 3D simulations in the future 
(e.g., \cite{taki11}), they have indeed shed a new light towards 
 a better understanding of the GW signatures as they have done so in 2D
 simulations (see discussions in sections \ref{exp1} and \ref{exp2}).


M\"uller and Janka (1997) \cite{muyan97} coined the first study to analyze 
the GW signature of 3D non-radial
 matter motion and anisotropic neutrino emission from prompt convection in the outer 
 layers of a PNS during the first 30 ms after bounce. 
 Their first 3D calculations using the light-bulb recipe 
were forced to be performed in a wedge of opening 
angle of $60^{\circ}$.
 Albeit with this limitation (probably coming from
 the computer power at that time), they obtained important findings that because of smaller convective activities inside the cone with slower 
overturn velocities, the GW amplitudes of their 3D models are more than a factor of
 10 smaller
 than those of the corresponding 2D models, and the wave amplitudes from neutrinos
 are a factor of 10 larger than those due to non-radial matter motions.
  With another pioneering (2D) study by Burrows \& Hayes (1996) \cite{burohey},
 it is worth mentioning that
 those early studies had brought new blood into the conventional GW predictions, 
 which illuminated the importance of the theoretical prediction of the 
neutrino GWs.

A series of findings obtained by Fryer et al. in early 2000s \cite{frye02,chris04} 
have illuminated also the importance of the 3D modeling. By running
 their 3D Newtonian Smoothed Particle Hydrodynamic (SPH) code coupled to 
 a gray flux-limited neutrino transport scheme, they studied the GW emission
 due to the inhomogeneous core-collapse, core rotation, low-modes convection, 
and anisotropic neutrino emission.
 Although the early shock-revival and the subsequent
 powerful explosions obtained in these 
 SPH simulations have yet to be confirmed by other groups, their approach
 paying particular attention to the multiple interplay between the explosion dynamics, 
the GW signatures, the kick and spins of pulsars, and also the non-spherical explosive 
nucleosynthesis, blazed a new path on which current supernova studies are
 progressing.



\begin{figure}[hbtp]
\begin{center}
 \begin{tabular}{cc}
\resizebox{80mm}{!}{\includegraphics{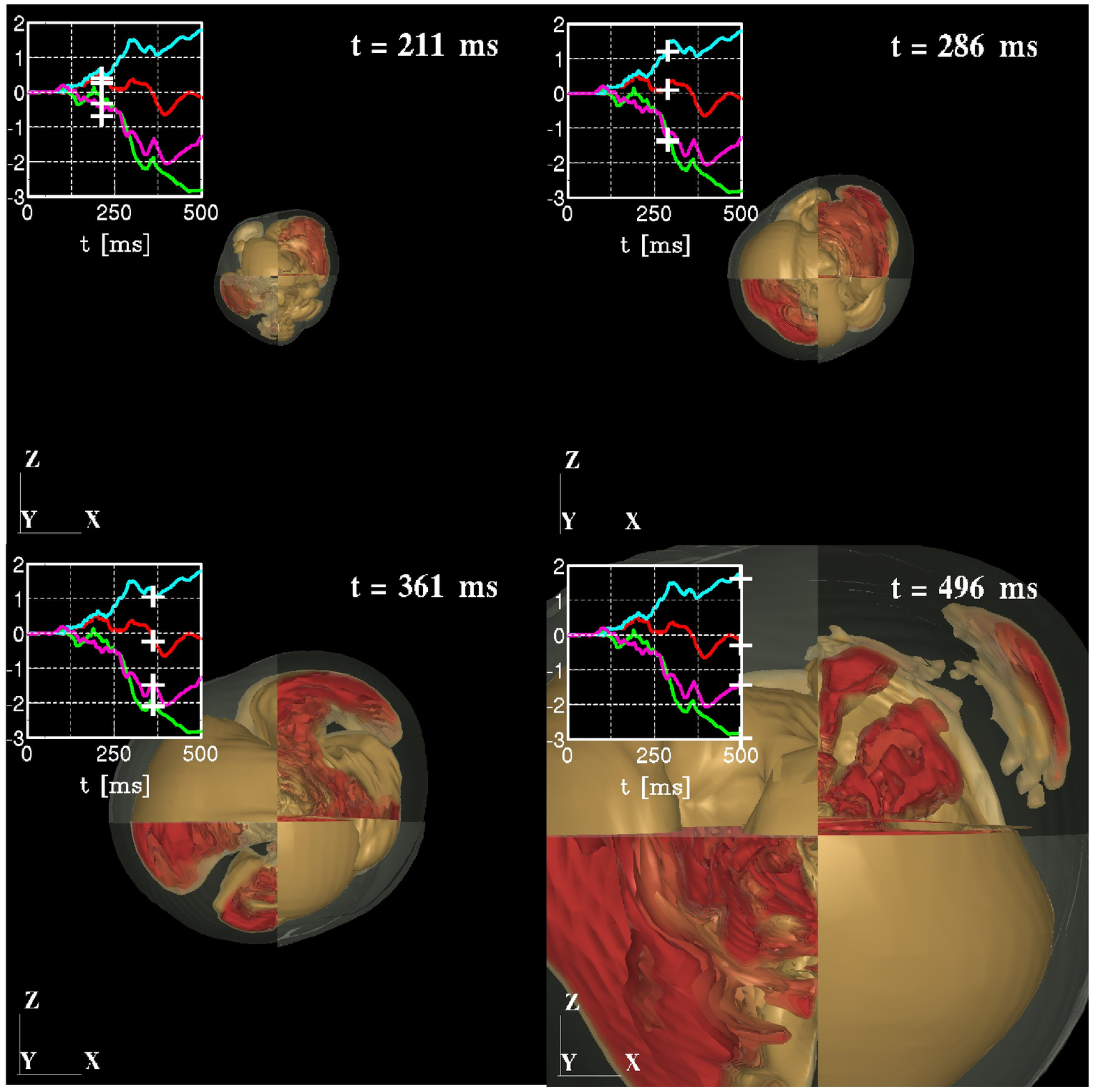}} &
\resizebox{59mm}{!}{\includegraphics{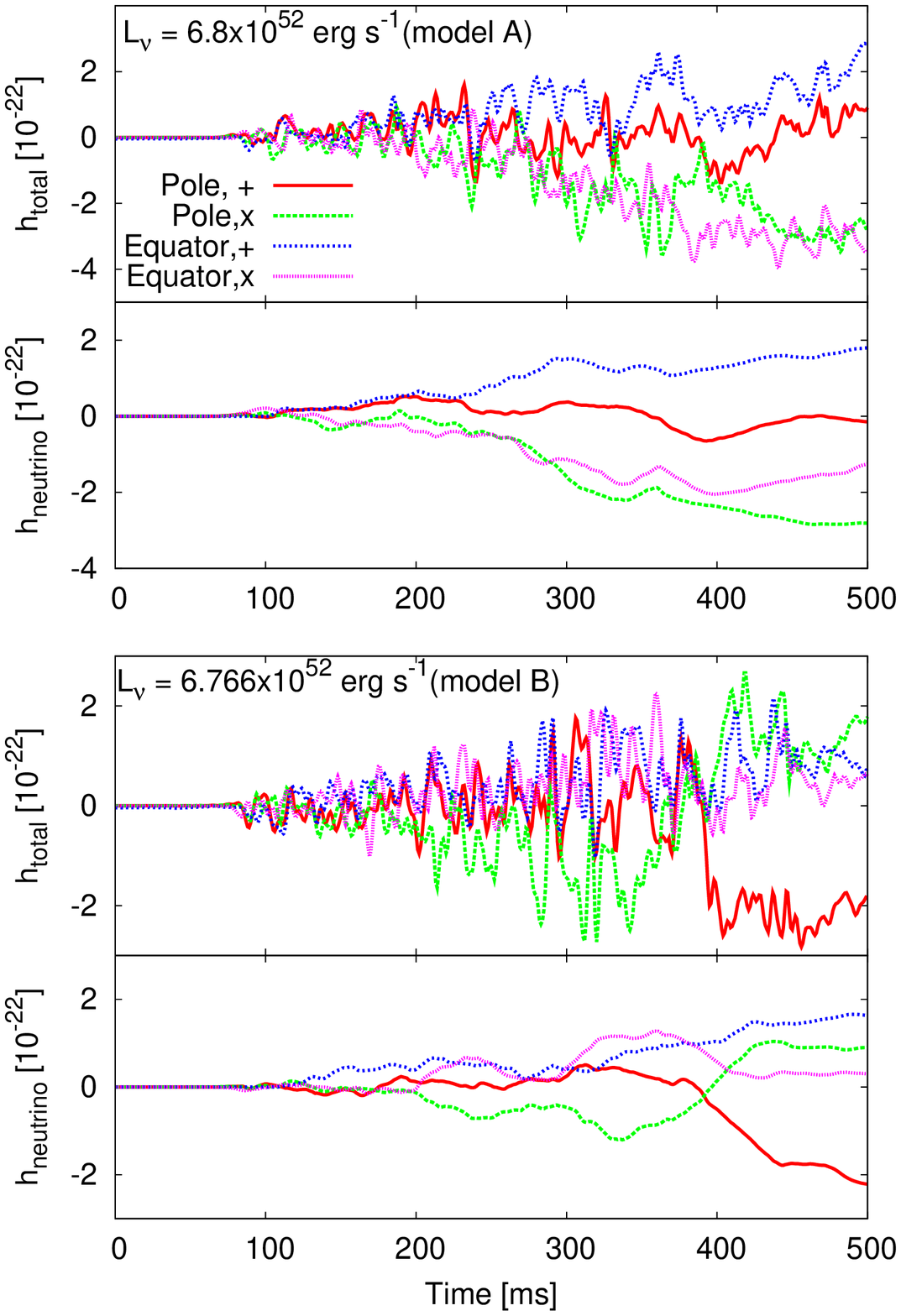}} \\  
\end{tabular}
\caption{Left panel shows four snapshots of the entropy distributions of 
a representative 3D supernova explosion model (corresponding to model A in 
 Kotake et al. (2009) \cite{kotake09}).
The second and fourth quadrant of each panel shows 
the surface of the standing shock wave.
In the first and third
quadrant, the profiles of the 
high entropy bubbles (colored by red) inside the section cut by the 
$ZX$ plane are shown. The side
length of each plot is 1000km.
The insets show the gravitational waveforms from anisotropic neutrino emissions, 
with '$+$' on each curves 
representing the time of the snapshot. Note that the colors of the curves are
taken to be the same as the top panel. 
After about $100$ ms, the deformation of the standing 
shock becomes remarkable marking the epoch when the SASI enters the non-linear regime.
At the same time, the gravitational amplitudes 
begin to deviate from zero.
Right panel shows gravitational waveforms 
from neutrinos (bottom) and from the sum of 
 neutrinos and matter motions (top), seen from the polar
axis and along the equator (indicated by 'Pole' and 'Equator')
with polarization ($+$ or $\times$ modes) for two representative 3D models 
 of A and B (see \cite{kotake09} for details), in which 
the input luminosity for the two pair 
panels differs only $0.5\%$. 
From the right panels,
 it can be seen that the overall structures of the waveforms are 
 predominantly determined by the neutrino GWs which was also the case in 2D simulations
 (e.g., section \ref{exp2}).
The distance to the source is assumed to be 10 kpc.}
\label{f8}
\end{center}
\end{figure}

We also studied the GW signals from 3D models that mimic neutrino-driven explosions
 aided by the SASI \cite{kotake_ray,kotake09,kotake11}.
 These studies were an extension of the 2D simulations already mentioned in
 section \ref{exp2}, in which the light-bulb scheme was used to obtain explosions and 
  the initial conditions were derived from a steady-state 
approximation of the postshock structure and the dynamics only outside an inner boundary
 at 50 km was solved.
The left panel of Figure \ref{f8} shows the evolution of 3D hydrodynamic features
 from the onset of the non-linear regime of SASI (top left) 
until the shock break-out (bottom right) with the gravitational waveform from neutrinos 
inserted in each panel. As seen, the major axis of the growth of SASI 
is shown to be not aligned with the symmetric axis (:$Z$ axis in the figure)
and the flow inside the standing shock wave is not symmetric with
respect to this major axis (see the first and third quadrant).
 This is a generic feature obtained in their 3D
models, which is in contrast to the axisymmetric case.
As discussed in section \ref{exp2}, the GW amplitudes from the SASI 
in 2D showed an increasing trend with time due to the symmetry axis, 
along which SASI can develop preferentially.
Free from such a restriction, 
 a variety of the waveforms is shown to appear (see waveforms inserted in 
Figure \ref{f8}).
Furthermore, the 3D standing shock can also oscillate in all
 directions, which leads to the smaller explosion anisotropy than 2D.
 With these two factors, the maximum amplitudes seen either from the
equator or the pole becomes smaller than 2D.
On the other hand, their sum in terms of the total radiated
energy are found to be 
almost comparable between 2D and 3D models, which is likely to imply the
energy equipartition with respect to the spatial dimensions.

 The right panel of Figure \ref{f8} shows the gravitational waveforms 
for models with different neutrino luminosity. 
 The input luminosity for the bottom panel is smaller than that for 
 the top panel only by $0.5\%$. Despite the slight difference,
the waveforms of each polarization 
are shown to exhibit no systematic similarity when seen either from the pole or equator.

 The stochastic nature of GWs produced by the interplay 
of the SASI and neutrino-driven explosions have been also confirmed 
more recently by M\"uller et al. (2011) \cite{ewald11}. They analyzed the GW signatures 
based on their 3D models of parametric neutrino-driven explosions \cite{annop}
in which the gray spectral neutrino transport is solved in a ray-by-ray manner 
\cite{scheck06}\footnote{It is worth mentioning that a complete derivation 
 of formulae for extracting the neutrino GWs is given
 for arbitrary positioning between the source and observer 
coordinate systems (e.g., section 4 in \cite{ewald11}).}. At the sacrifice of cutting out the high-density core 
(at around 80 - 120 km in radius),
  their 3D models succeeded in following the dynamics
 from the post-bounce accretion phase through 
 the onset of the explosion more than one second for the first time. Among a number of
 interesting findings obtained in their work, we here focus on the progenitor dependence
 that could not be studied by Kotake et al. (2009,2011)
 \cite{kotake09,kotake11} who used the idealized initial conditions.

\begin{figure}[hbtp]
  \begin{center}
    \begin{tabular}{cc}
\resizebox{55mm}{!}{\includegraphics{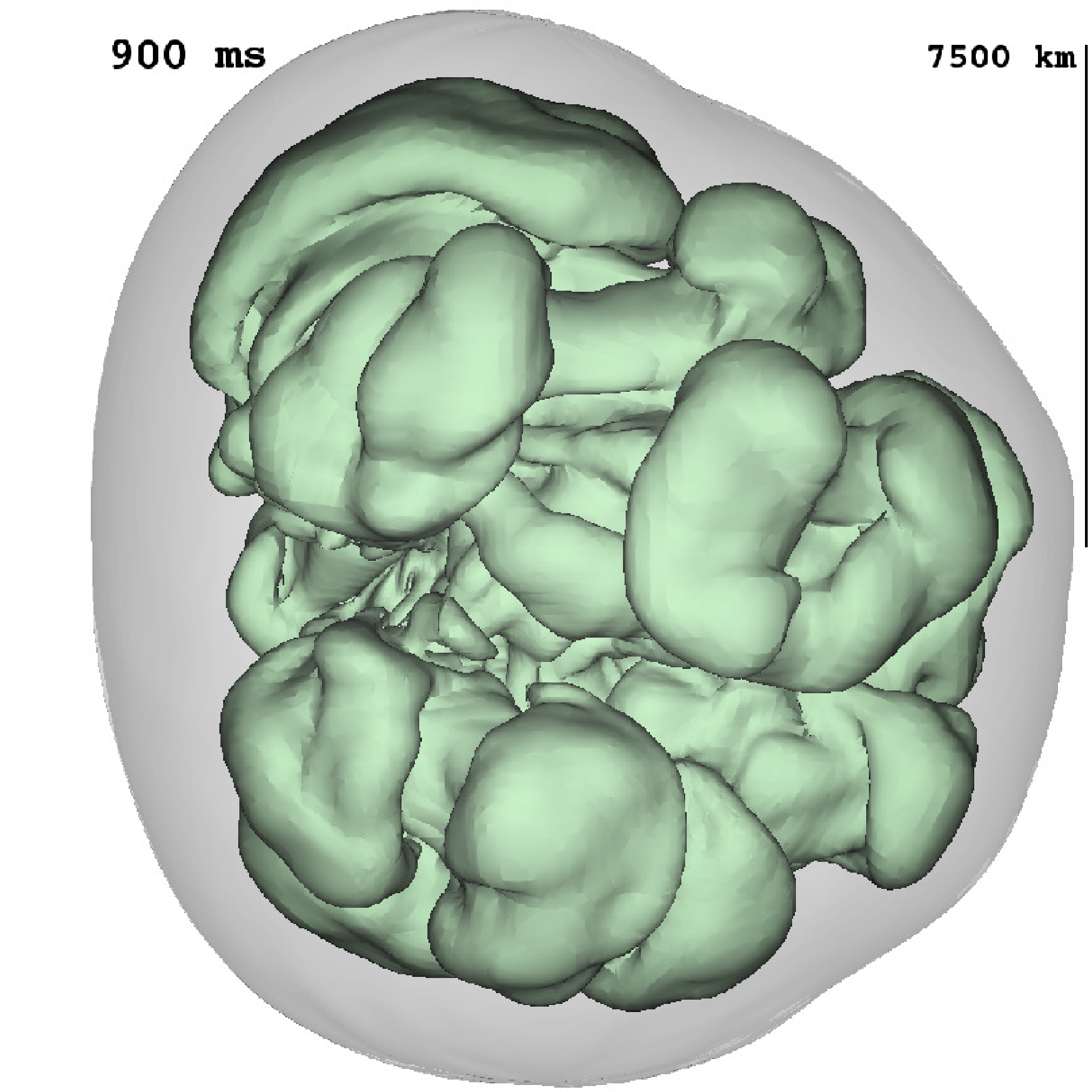}} &
\resizebox{55mm}{!}{\includegraphics{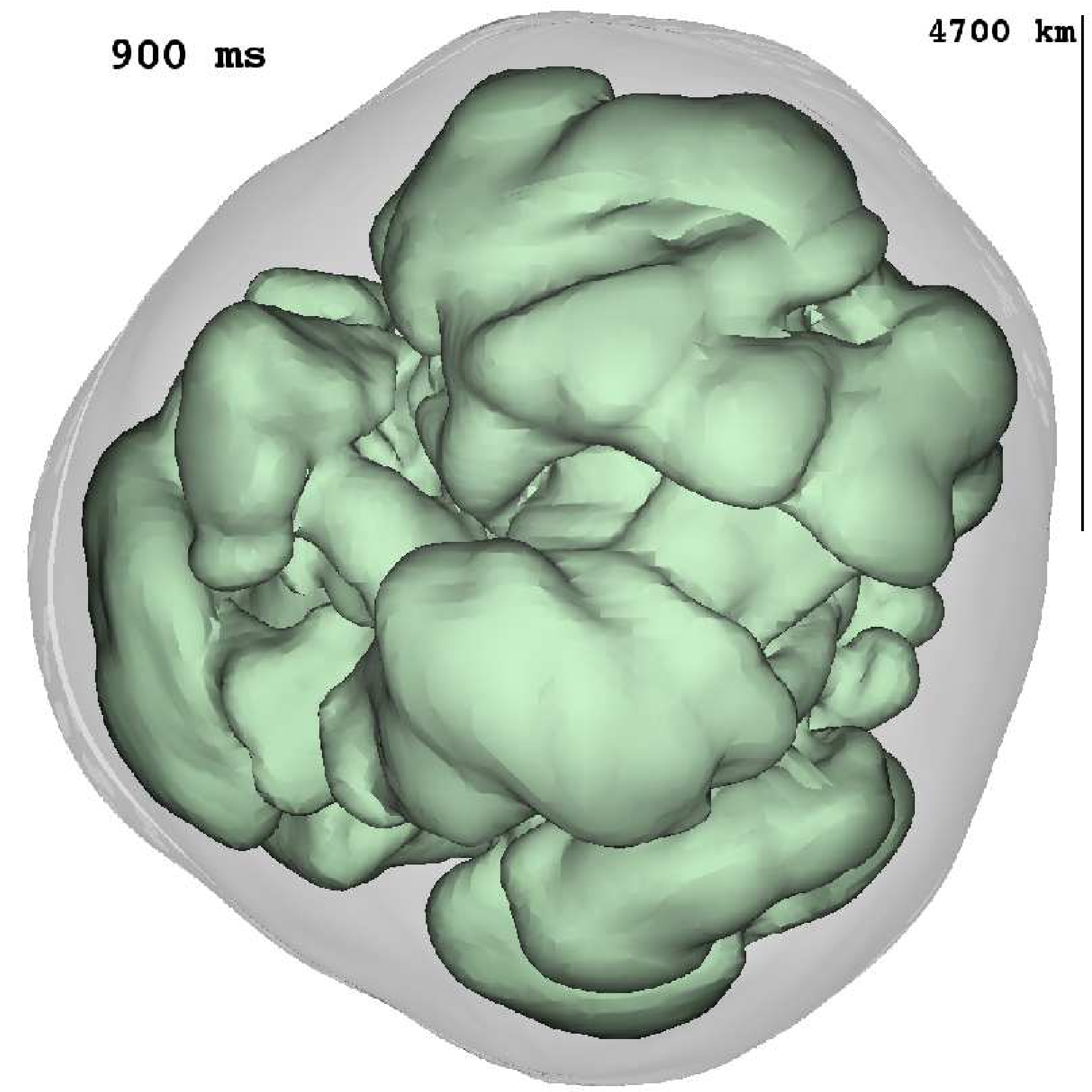}} \\
  \resizebox{48mm}{!}{\includegraphics{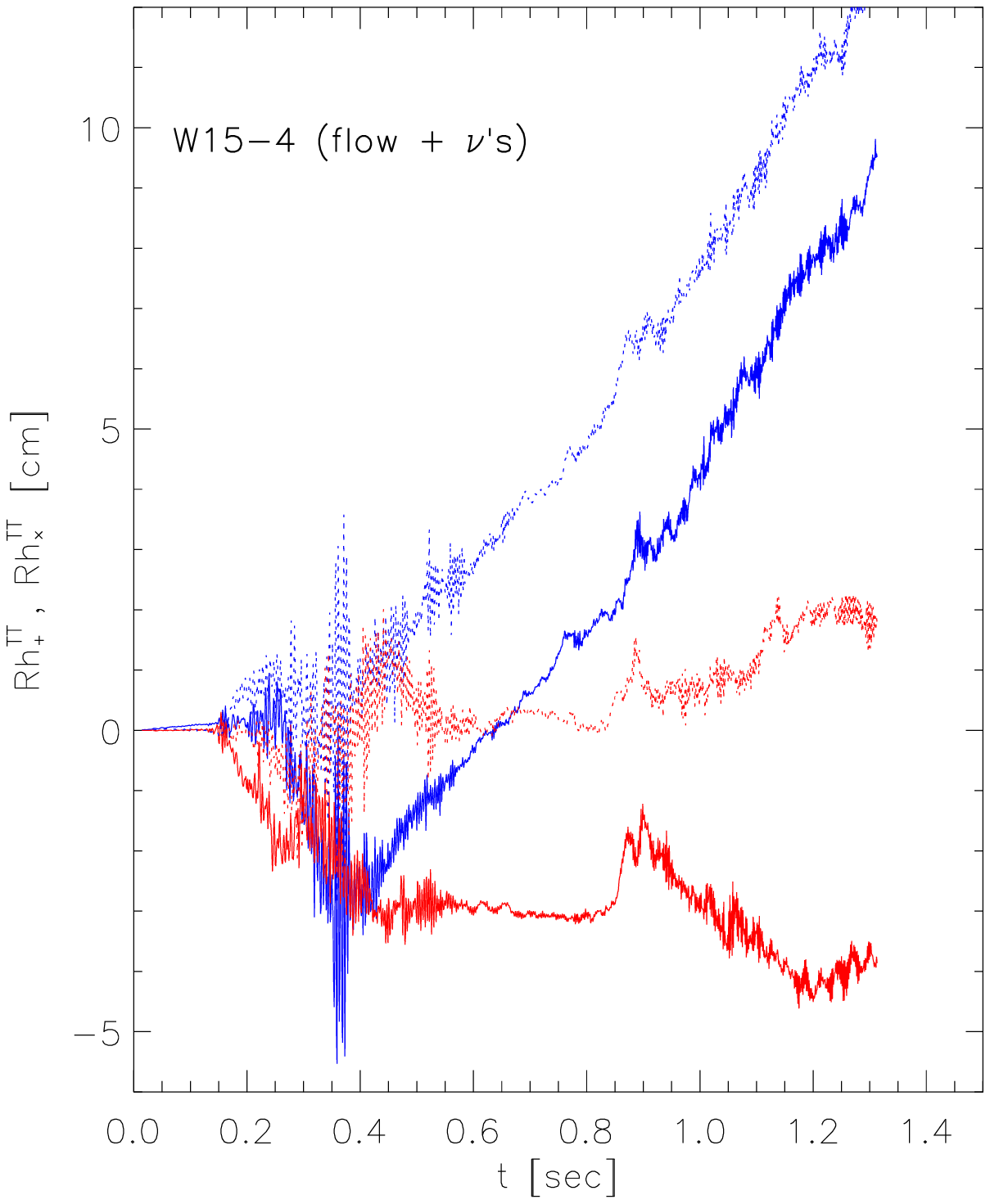}} &
      \resizebox{48mm}{!}{\includegraphics{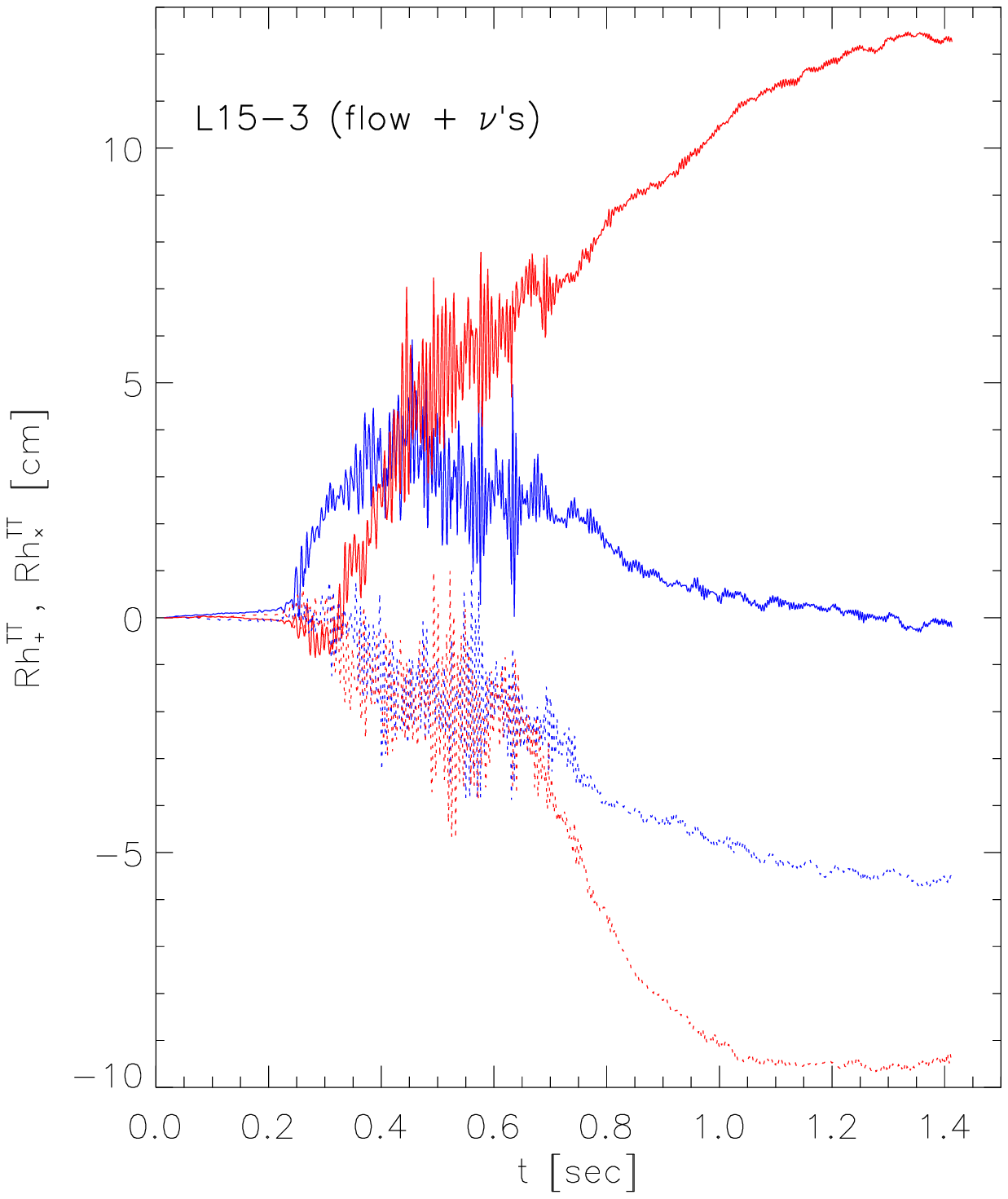}} \\
\resizebox{50mm}{!}{\includegraphics{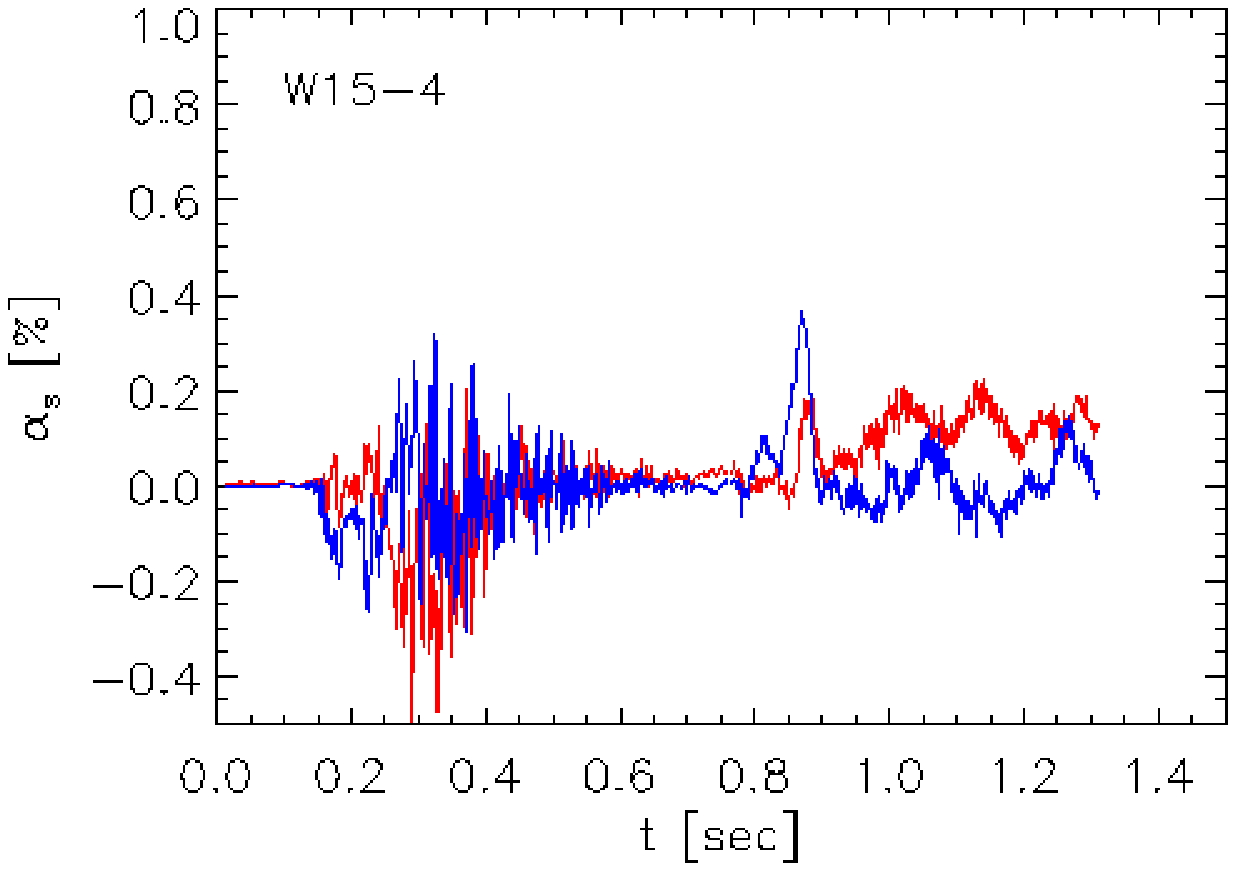}} &
\resizebox{50mm}{!}{\includegraphics{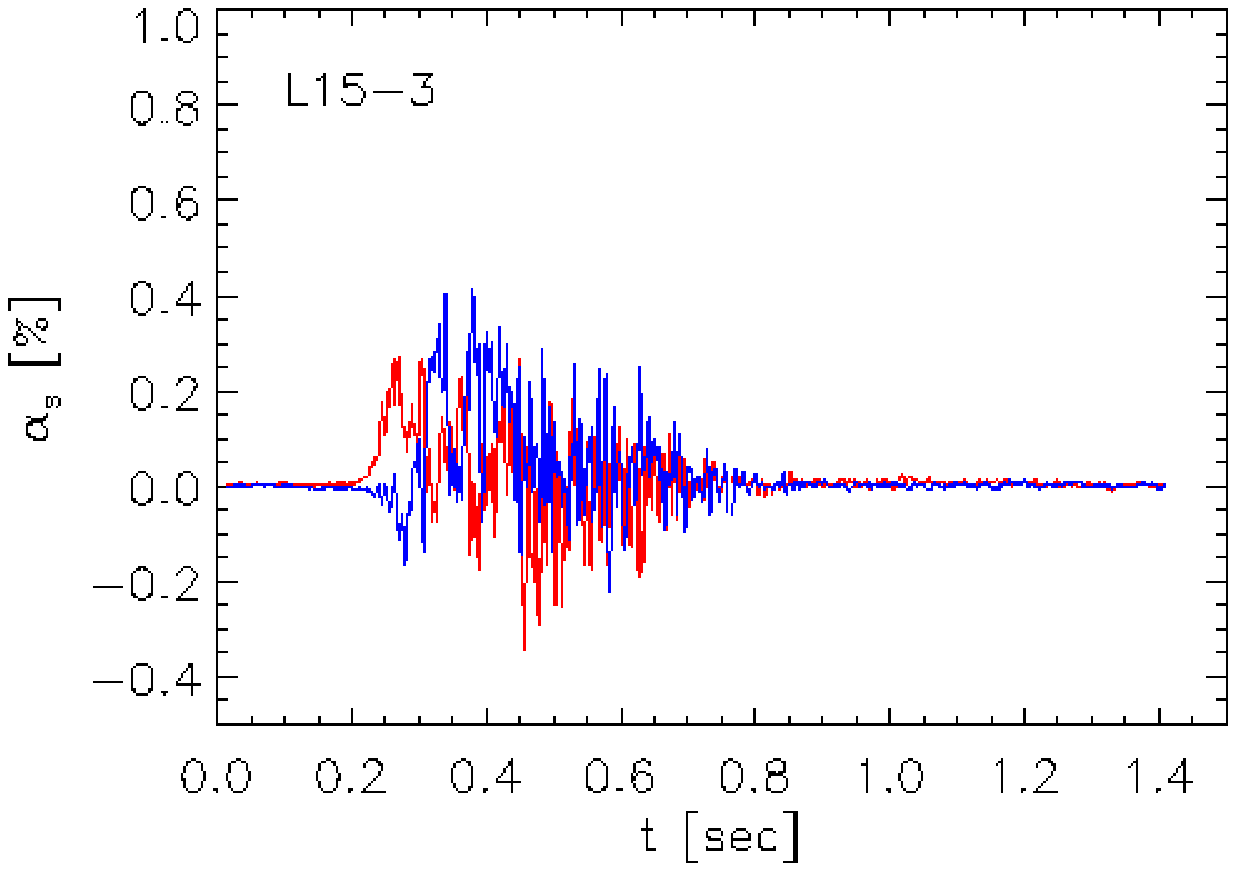}} \\
    \end{tabular}
   \caption{Top two panels show snapshots illustrating a globally asymmetric 3D 
explosion in the post-accretion phase of two different progenitor models 
 using the Woosley-Weaver (left) or Limongi progenitor (right) 
of a 15 $M_{\odot}$ progenitor
 (taken from M\"uller et al. (2011) \cite{ewald11}).
 Note that "-4" or "-3" differ only by the initial seed perturbations 
(see \cite{ewald11} for more details). The time and linear scale are indicated in 
each plot. Middle panels show the plus (blue) or cross (red) mode of the wave 
amplitude due to anisotropic mass flow and neutrino emission for models 
 W15-4 (left) and L15-3 (right), respectively. The solid curves show the amplitudes 
 for an observer above the north pole of the source, while the other curves give 
 the amplitudes at the equator. Similar to the middle panels, the bottom panels 
shows the anisotropy parameter of neutrino emission (only the plus mode is shown).
 These figures are by courtesy of M\"uller and the coauthors \cite{ewald11}.
} 
  \label{f8added}
  \end{center}
\end{figure}

The top two panels in Figure \ref{f8added} 
show snapshots of blast morphologies between two models,
 in which a non-rotating 15 $M_{\odot}$ 
progenitors either evolved by Woosley and Weaver (1995) (left)
 or by Limongi et al. (1995) (right) was employed,
 respectively. Note that they succeeded in following an unprecedentedly 
long-term 3D evolution (up to 1.4 s after bounce) by utilizing 
an axis-free coordinate system; the so-called Yin-Yang grids (see
 \cite{annop} for more details). The middle panels in Figure \ref{f8added} 
indicate that the total 
 GW amplitude is significantly different from that of the flow-only
 GW amplitudes (not shown) for models which do or do not
 exhibit PNS convection below
 the neutrino sphere. Particularly at late times 
($t \gtrsim 0.8$ s in the middle panels), anisotropic neutrino emission causes a 
 continuing growth of the GW amplitudes for the model using the Woosley-Weaver
 (indicated by W15) progenitor
 (left) instead of saturation seen for the model using the Limongi progenitor (
 indicated by L15, right panel).
 By a detailed spectrogram analysis of the GW energy
 distributions (see their Figure 12),
 they revealed that the primary agent to make the model discrepancy
 is the PNS convection, which develops much more 
 vigorously for the W15 progenitor model than for the L15 progenitor model. 
This can be also depicted in the bottom panels of Figure \ref{f8added} which shows 
 bigger neutrino anisotropy for the W15 progenitor model (left) in the late 
 postbounce phase ($t \gtrsim 0.8$ s) than for the L15 progenitor model (right). 
 They furthermore pointed out that 
 in their 3D models, very prominent, quasi-periodic sloshing motions due to the 
 SASI are absent and the 
 emission from different surface areas facing an observer adds up incoherently,
 so that the measurable modulation amplitudes of the GW signals 
(as well as neutrinos) are significantly smaller than predicted by 2D simulations
 (\cite{murphy,yakunin}).

\begin{figure}[hbtp]
  \begin{center}
    \begin{tabular}{cc}
\resizebox{70mm}{!}{\includegraphics{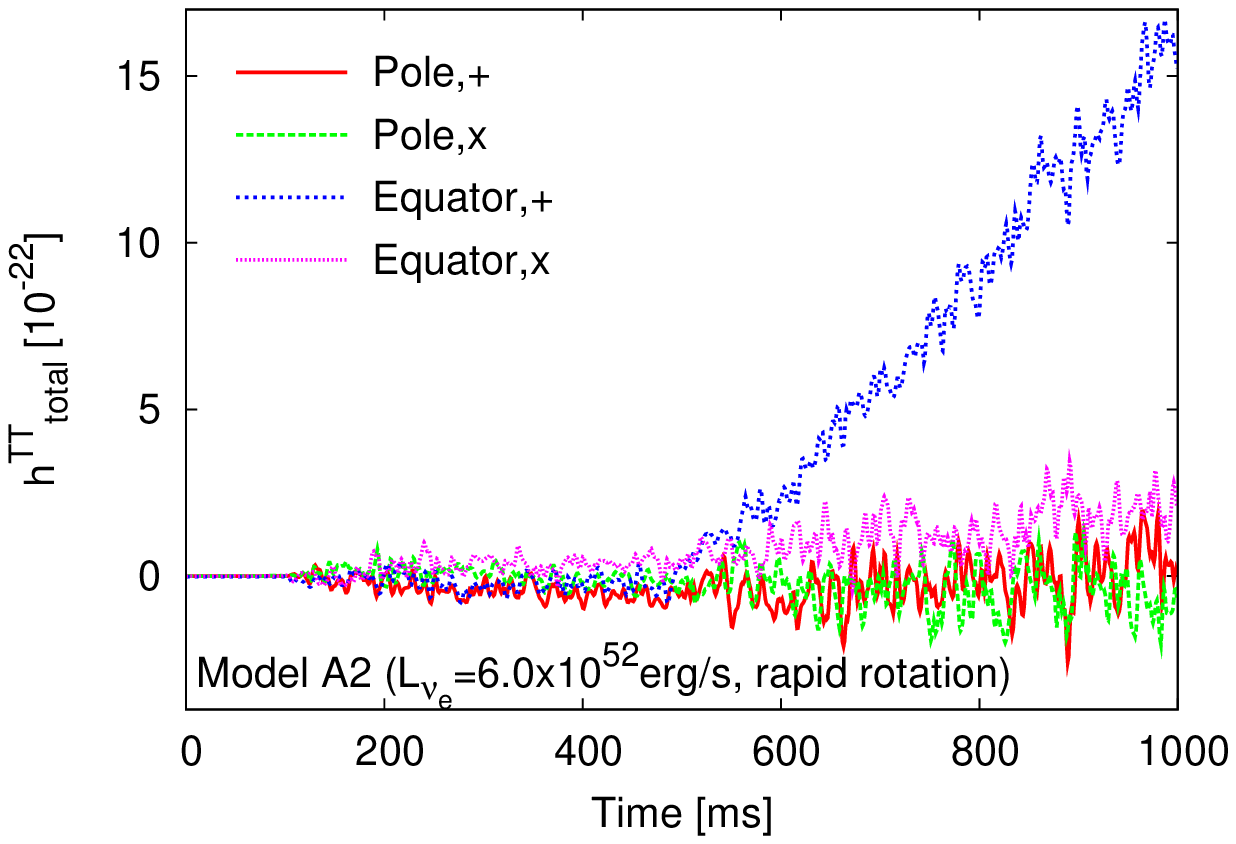}} &
\resizebox{70mm}{!}{\includegraphics{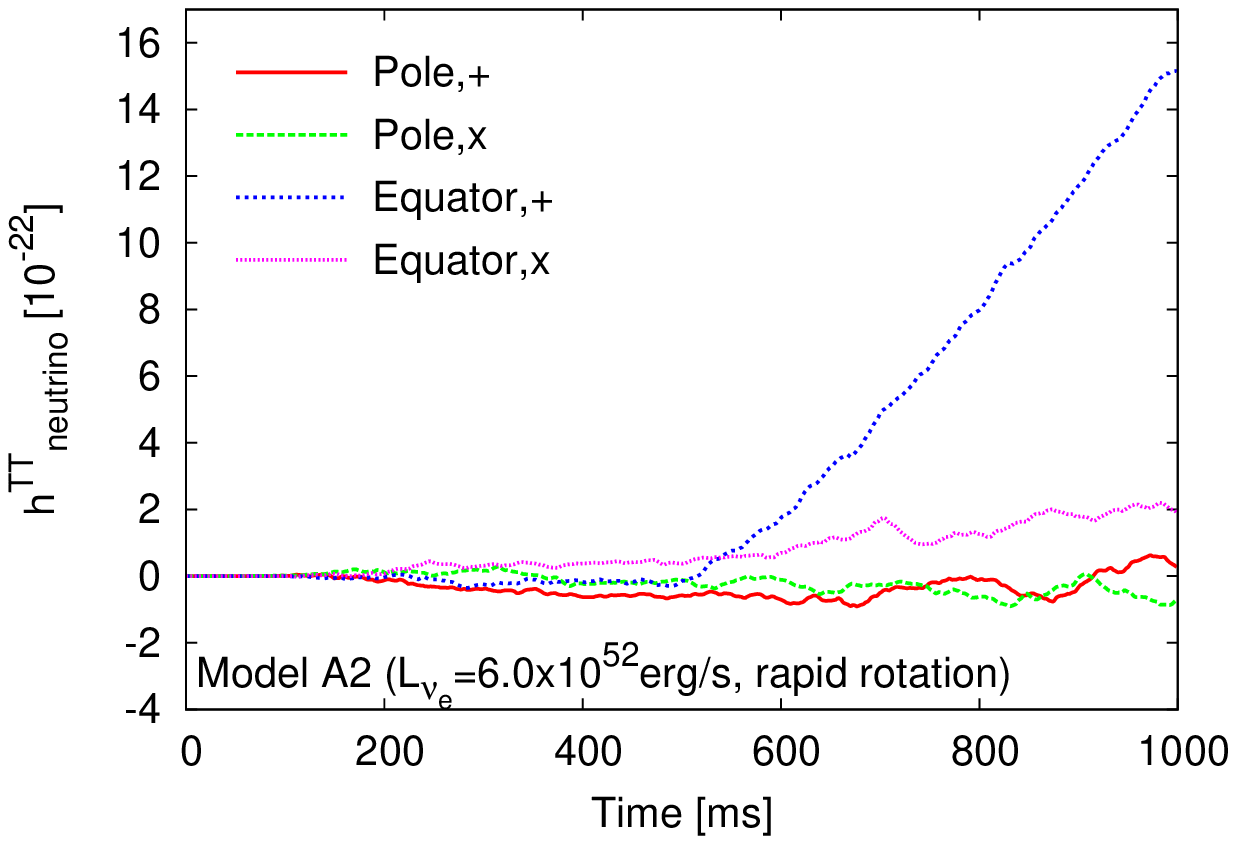}} \\
  \resizebox{50mm}{!}{\includegraphics{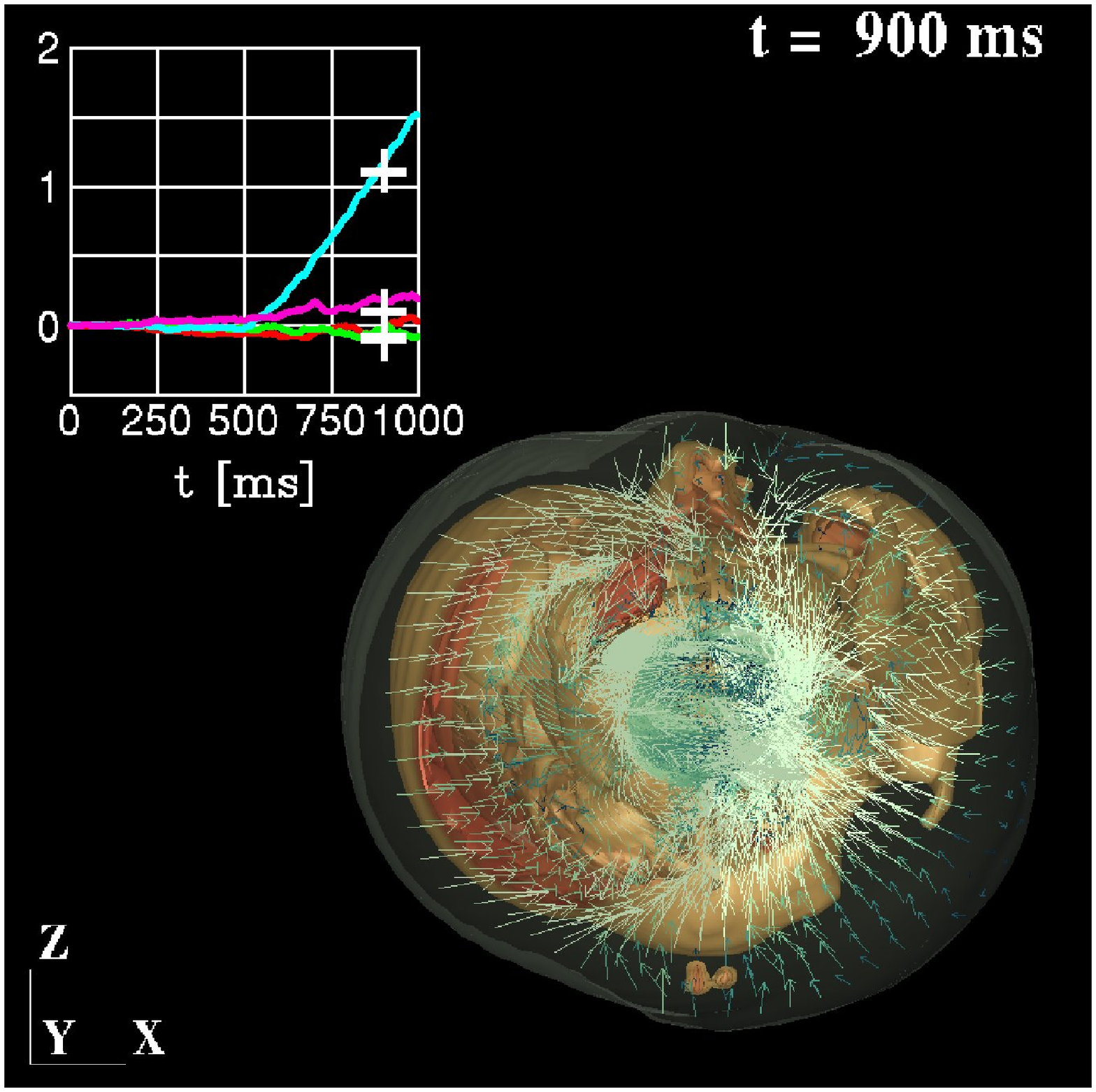}} &
      \resizebox{50mm}{!}{\includegraphics{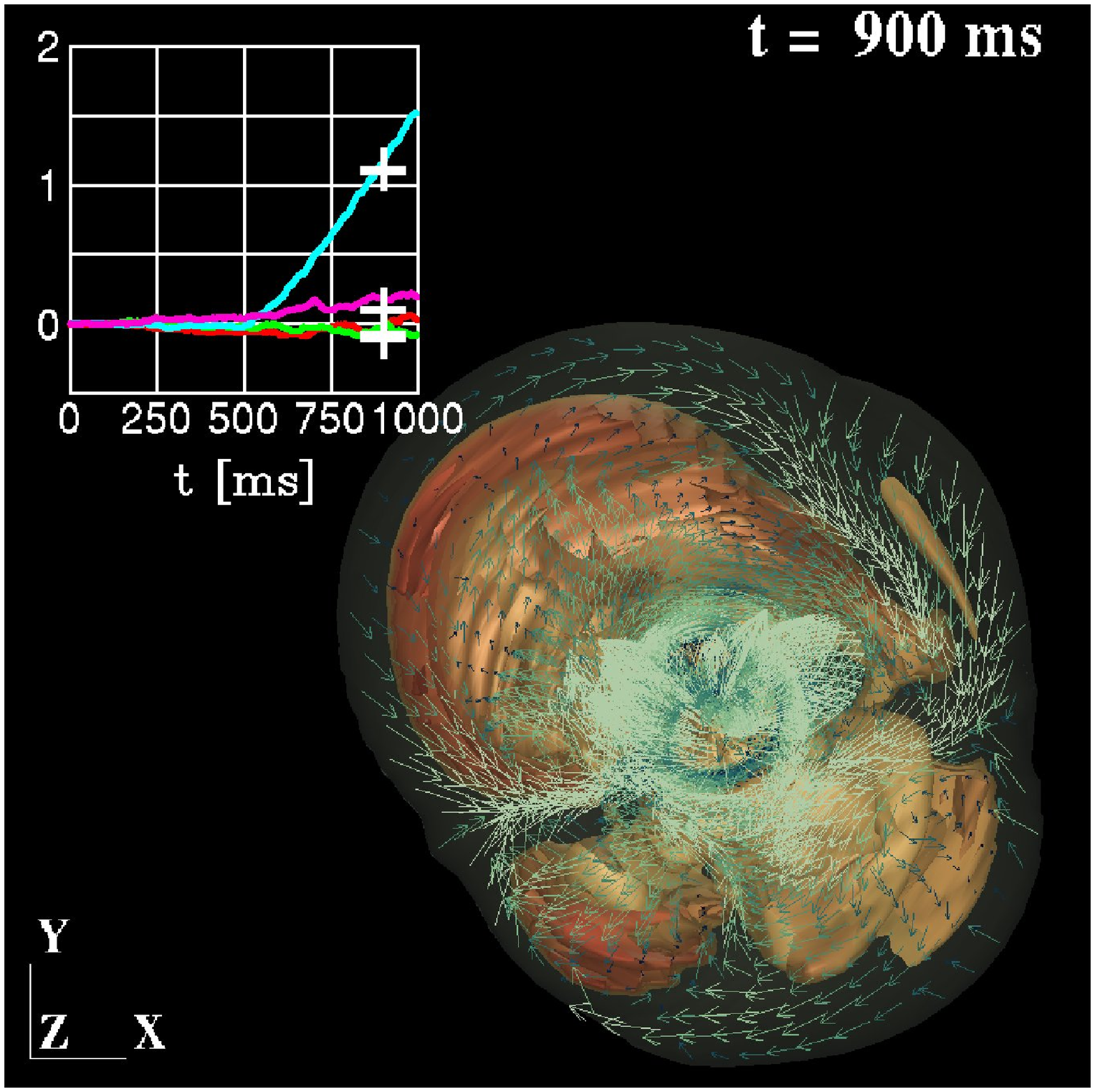}} \\
    \end{tabular}
\includegraphics[width=.5\linewidth]{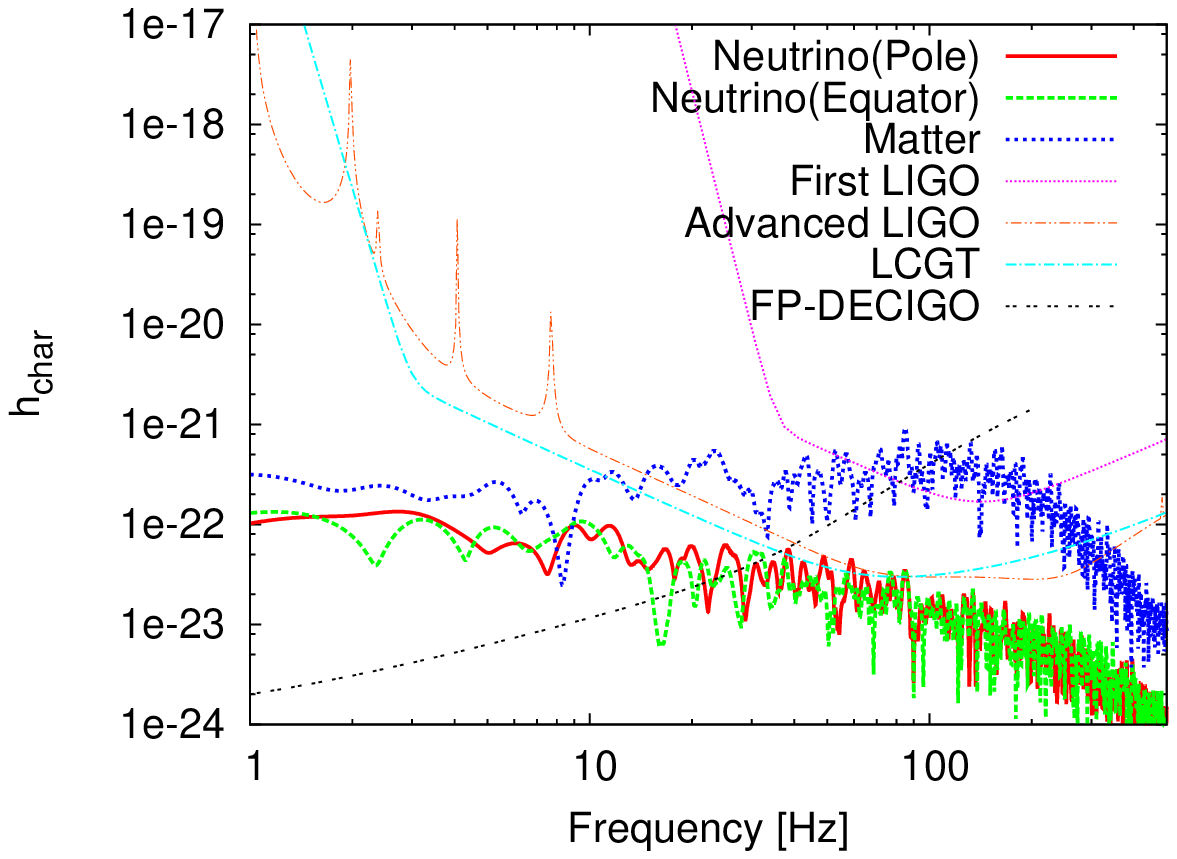}
   \caption{Top panels show gravitational waveforms from the sum of neutrinos and matter motions 
(left) and only from neutrinos (right) obtained in a parametric explosion 3D model 
with rotation (taken from Kotake et al. (2011) \cite{kotake11}). 
 Similar to the left panel in Figure \ref{f8}, 
 the middle panels show a partial cutaway of the
 entropy isosurfaces and the velocity vectors on the 
 cutting plane for the equatorial (left) and polar observer (right), 
respectively. The bottom panel shows spectral distributions for the model
  from matter motions (``Matter'') and neutrino emission (``Neutrino'') 
seen from the pole or the equator (e.g., \cite{kotake11}) with 
the expected detection limits of TAMA300 \cite{tamanew}, first LIGO and 
 advanced LIGO \cite{firstligo}, LCGT \cite{lcgt}, and Fabry-Perot type 
 DECIGO \cite{fpdecigo}.
 The distance to the supernova is assumed to be 10 kpc.
 Note that for the matter signal, the $+$ mode seen from the polar direction is 
plotted.} 
  \label{f9}
  \end{center}
\end{figure}

The effects of stellar rotation on the stochastic 
nature of the GWs have been recently studied by Kotake et al. (2011) \cite{kotake11} who
used the same numerical techniques that was already mentioned 
in explaining Figure \ref{f8}. 
The top two panels in
 Figure \ref{f9} show the gravitational waveform for a typical 3D with rotation 
(left:total amplitudes,
 right:neutrino only). To construct a model with rotation,
 a uniform rotation was manually given on the flow advecting from the outer boundary 
of the iron core as in \cite{iwakami2}, whose specific angular momentum is 
assumed to agree with recent stellar evolution models \cite{hege05}.
 Comparing to the right panel in Figure \ref{f8} (i.e., in the absence of rotation),
  one can clearly see a sudden rise in the GW amplitude after around 500 ms
 for the rotating model (blue line in the top panels in Figure \ref{f9}), which is 
 plus mode of the neutrino GWs seen from the equator.
 These features were found to be common in the fifteen 
 3D models computed in their study.

  Seen from the equatorial direction (the middle left panel of Figure \ref{f9}), 
one may guess the presence of
 the sloshing modes, but it just happens
 to develop along the rotational axis ($z$-axis) at this 
 epoch. It should be emphasized that the dominance of $h^{\rm equ}_{\nu,+}$ 
 observed in the current 3D simulations have nothing to do with the one
 discussed in section \ref{exp2}. 
Free from the 2D axis effects, the major axis of the SASI changes stochastically
 with time, and the flow patters behind the standing shock also change 
in every direction.
 As a result, the sloshing modes can make only a small contribution to the GW emission.
  The remaining 
possibility is that the spiral flows seen in the middle right panel should be a 
key importance to understand the GW feature mentioned above. In fact, by analyzing 
 the matter distribution on the equatorial plane, it was found
 that the compression of matter 
is more enhanced in the vicinity of the equatorial plane due to the growth of the 
spiral SASI modes, leading to the formation of the spiral flows circulating around the 
spin axis with higher temperatures. As a result, the neutrino emission seen parallel 
 to the spin axis becomes higher than the ones seen from the other direction.
 Remembering that the lateral-angle ($\theta$) dependent function 
 of the GW formulae (e.g., in Equation (\ref{tt})) is positive near 
the north and south polar caps, the dominance of the polar neutrino luminosities 
leads to make the positively growing feature of $h^{\rm equ}_{\nu,+}$ in the top
 panels of Figure 
\ref{f9} (blue line).
 From the spectral analysis of the gravitational waveform 
(the bottom panel of Figure \ref{f9}), it can be readily seen that 
 it is not easy to detect these neutrino-originated GW signatures
 with slower temporal evolution ($\gtrsim O(10)$ms) by ground-based detectors whose 
sensitivity is 
limited mainly by the seismic noises at such lower frequencies. 
However these signals may be detectable by the recently proposed future space 
interferometers like Fabry-Perot type DECIGO (\cite{fpdecigo}, black line in the bottom
 panel). Contributed by the neutrino GWs in the lower frequency domains,
 the total GW spectrum tends to become rather flat over a broad frequency range
 below $\sim 100$ Hz. These GW features obtained in the context of the 
SASI-aided neutrino-driven mechanism are different from the ones expected in the 
other candidate supernova mechanism, such as the MHD mechanism (e.g., 
 section \ref{mhd_gw}) and
 the acoustic mechanism (\cite{ott_new}, e.g., section 7.1 in \cite{ott_rev}).
 Therefore the detection of such signals 
 could be expected to provide an important probe into the explosion mechanism.

Finally, most of the 3D models summarized in this section
  cut out the PNS and the neutrino transport is approximated by a simple light-bulb
  scheme \cite{kotake11} or by the gray transport scheme \cite{ewald11}. Needless
 to say, these exploratory approaches are but 
the very first step to model the neutrino-heating 
 explosion and to study the resulting GWs.
 As already mentioned, the excision of the central regions inside PNSs 
 truncates the feedback between the mass accretion to the PNS and the resulting 
 neutrino luminosity, which should affect the features of the neutrino GWs.  
 By the cut-out, efficient GW emission of the oscillating neutron star
\cite{ott_new} and non-axisymmetric instabilities 
 \cite{ott_3d,simon,simon1} of the PNSs, and the enhanced neutrino emissions inside 
the PNSs \cite{marek_gw} cannot be treated in principle.
 To elucidate the GW signatures in a much more 
quantitative manner,
 full 3D simulations with a spectral neutrino transport 
 are apparently needed. This is unquestionably a vast virgin territory awaited to be 
 explored for the future.



\subsection{GWs from MHD Explosions}\label{mhd_gw}

As already mentioned in the beginning of section \ref{s3}, state-of-the-art results 
 concerning the GW signatures emitted 
  {\it near bounce} in the case of rapidly rotating core-collapse
 have been given in the recent review by Ott (2009) \cite{ott_rev} (e.g., section 4).
 Conventionally the waveforms of the bounce signals are categorized into the three types, 
namely types I, II, and III. As explained in Ott (2009) \cite{ott_rev}, type II and III 
waveforms are 
 shown less likely to appear than type I, because a combination of general 
relativity and electron capture near core bounce suppresses 
 multiple bounce in the type II waveforms
 \cite{dimmelprl,ott_prl,ott_2007}. In general, a realistic nuclear 
 equation of state (EOS) is stiff enough to forbid the type III waveforms.
 So the generic type of the bounce signals is now known to take the type I 
waveform.

 For avoiding overlap, we mainly focus on the {\it postbounce} GW emission
 in the context of rapidly rotational core-collapse 
 in this section.
  The major GW emission cites have been proposed to be 
 magnetohydrodynamic (MHD) outflows 
 and nonaxisymmetric instabilities, which will be separately described in 
 sections \ref{sec_mhd} and \ref{nonaxi}, respectively.
 
\subsubsection{MHD outflows}\label{sec_mhd}

Numerical simulations of MHD stellar explosions 
have a long history\footnote{almost as old as the neutrino-driven model in Colagate
 and White (1966).} started already in early 1970's by LeBlanc and Wilson (1970) 
\cite{leblanc} shortly after the discovery of 
pulsars  \cite{bisno76,muller79,symb84}. 
 However, it is rather only recently that the MHD studies come back to
 the front-end topics in the supernova research followed by a number of 
 extensive MHD simulations
 (e.g., \cite{arde00,yama04,kota04a,kota04b,kotake05,sawa05,ober06a,ober06b,
burr07,cerd07,suwa07a,suwa07b,taki04,taki09,scheid,martin11} for references therein). 
Main reasons for this activity are observations indicating very asymmetric 
explosions \cite{wang01,wang02}, and the 
interpretation of magnetars \cite{dunc92,latt07} and gamma-ray bursts 
(e.g., \cite{woos06,yoon}) as a possible outcome of the magnetorotational core-collapse
 of massive stars.

The MHD mechanism of stellar explosions 
relies on the extraction of rotational free energy of 
collapsing progenitor core via magnetic fields. Hence a high angular momentum of 
the core is preconditioned for facilitating the mechanism \citep{meier}. 
Given (a rapid) rotation of the 
 precollapse core, there are at least two ways to amplify the initial 
magnetic fields to a dynamically important strength, namely by the field wrapping 
by means of differential rotation that naturally develops in the collapsing core, 
and by the magnetorotational instability (MRI, see \cite{balb98}).

 In the case of canonical initial magnetic fields ($\sim 10^{9}$G)\cite{hege05}, the fastest growing modes of the MRI 
 are estimated to be at most several meters in the collapsing iron core \cite{MRI}. 
 At present, it is generally computationally too expensive to resolve those small 
scales in the global MHD simulations, typically more than two or three 
orders-of-magnitudes smaller than their typical finest grid size.
 Let us remind first that the MHD explosions presented in the rest of 
 this section are predominantly generated by the field wrapping mechanism
assuming a very strong precollapse magnetic field ($B_0 \gtrsim 10^{11-12}$ G).
 Recent stellar evolution calculations show that 
 this extreme condition could be really the case, albeit minor
 ($\sim$ 1\% of massive star population \cite{woos_blom}), for
 progenitors of rapidly rotating metal-poor stars,
 which experience the so-called chemically homogeneous evolution \cite{woos06,yoon}.  

\begin{figure}[hbtp]
  \begin{center}
    \begin{tabular}{cc}
\resizebox{60mm}{!}{\includegraphics{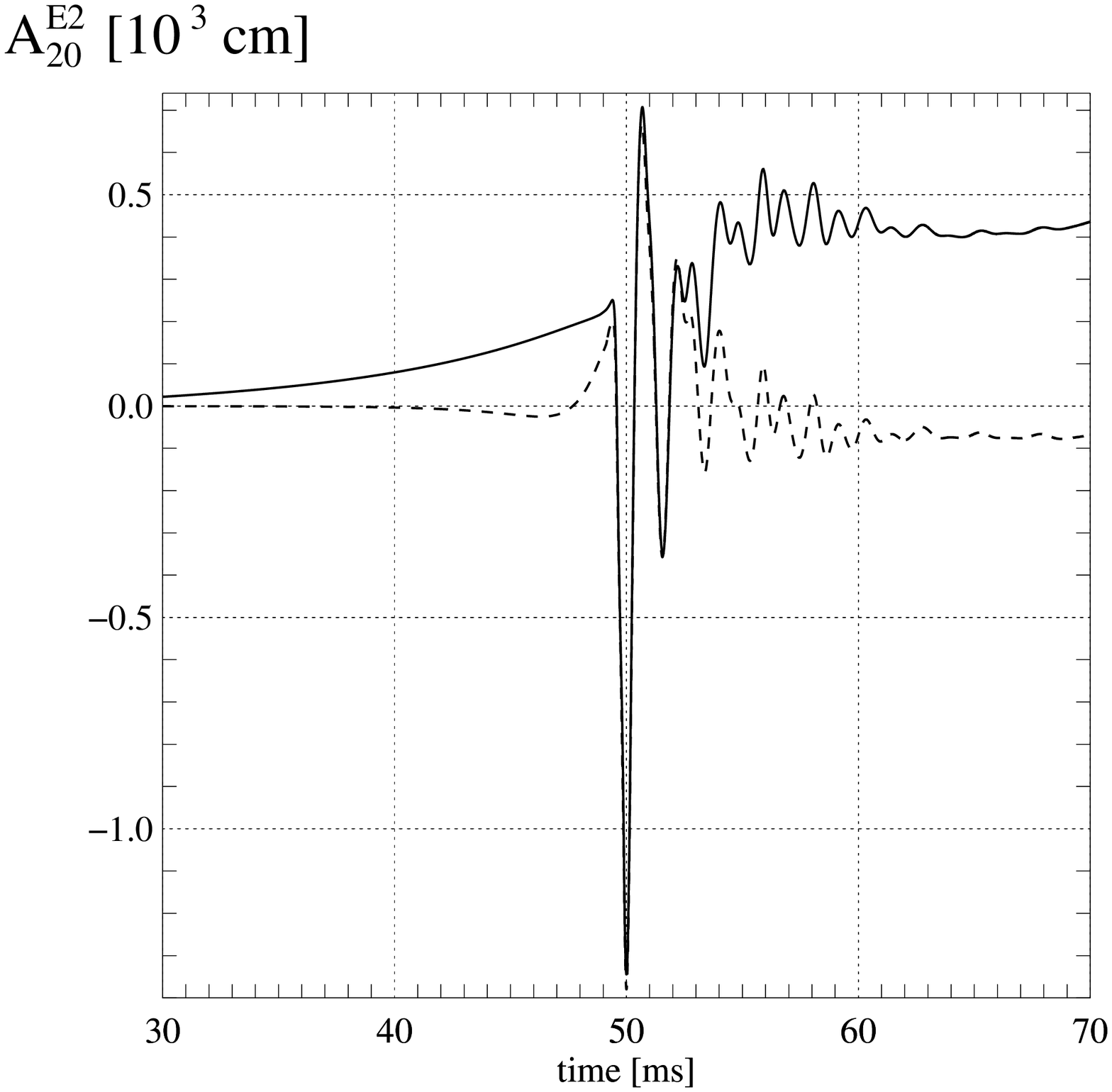}} &
\resizebox{60mm}{!}{\includegraphics{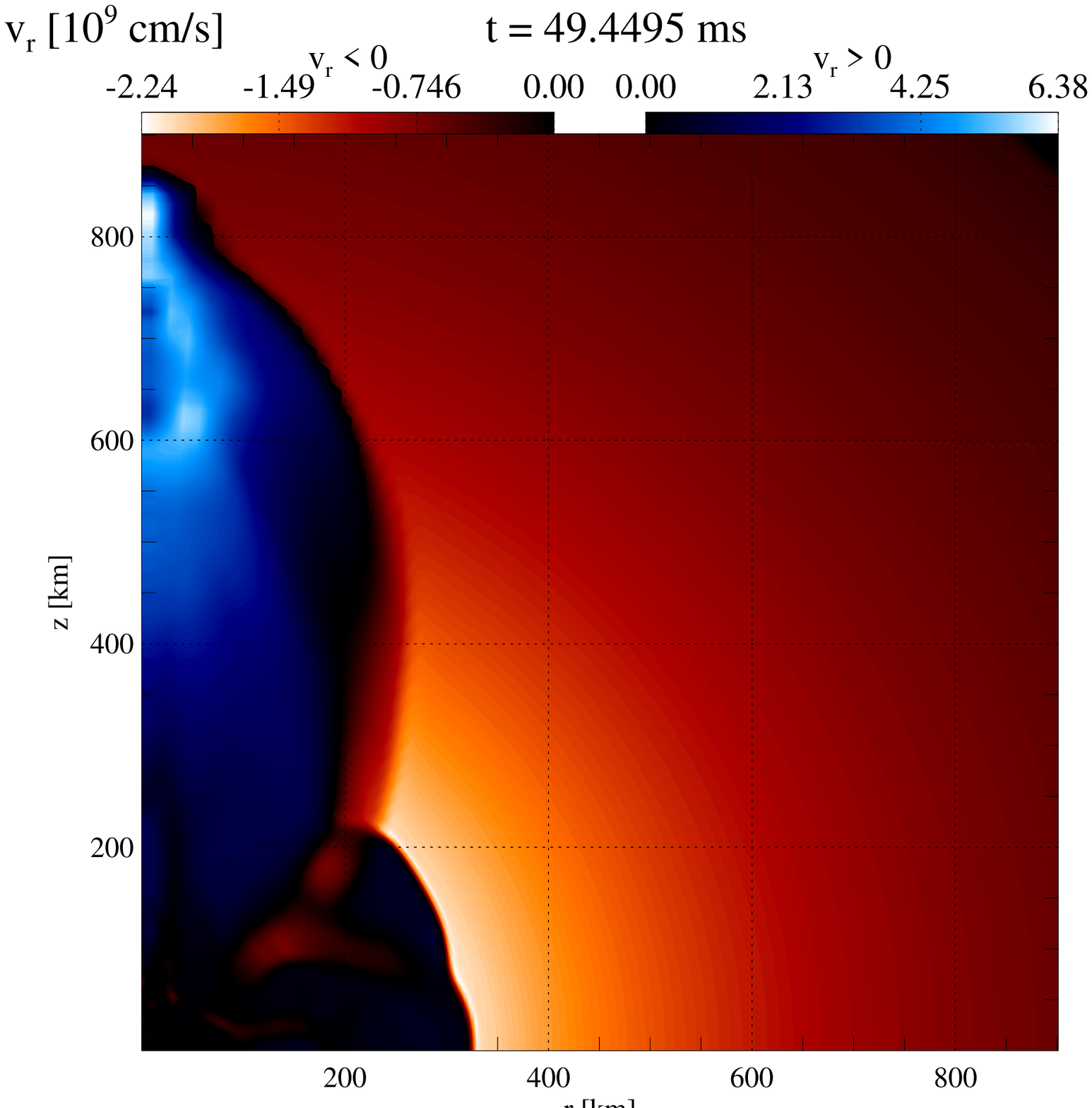}} \\
    \end{tabular}
   \caption{Left panel shows a gravitational waveform in the 2D post-Newtonian 
 simulations
 by Obergaulinger et al.
 (2006) \cite{ober06b} for their model A1B3G3-D3M13 in terms of the quadrupole  
 amplitude $A^{E2}_{20}$. The dashed line comes from the contribution of regions 
 inside 60 km in radius while the solid line shows the total signal. Comparing 
 to the right panel (velocity contours at $\sim$ 50 postbounce),
 the postbounce offset in the waveform is shown to come from the outer regions 
 with bipolar outflows. These figures are by courtesy of Obergaulinger and 
 the coauthors.}
  \label{f10}
  \end{center}
\end{figure}

\begin{figure}[hbtp]
  \begin{center}
    \begin{tabular}{cc}
\resizebox{60mm}{!}{\includegraphics{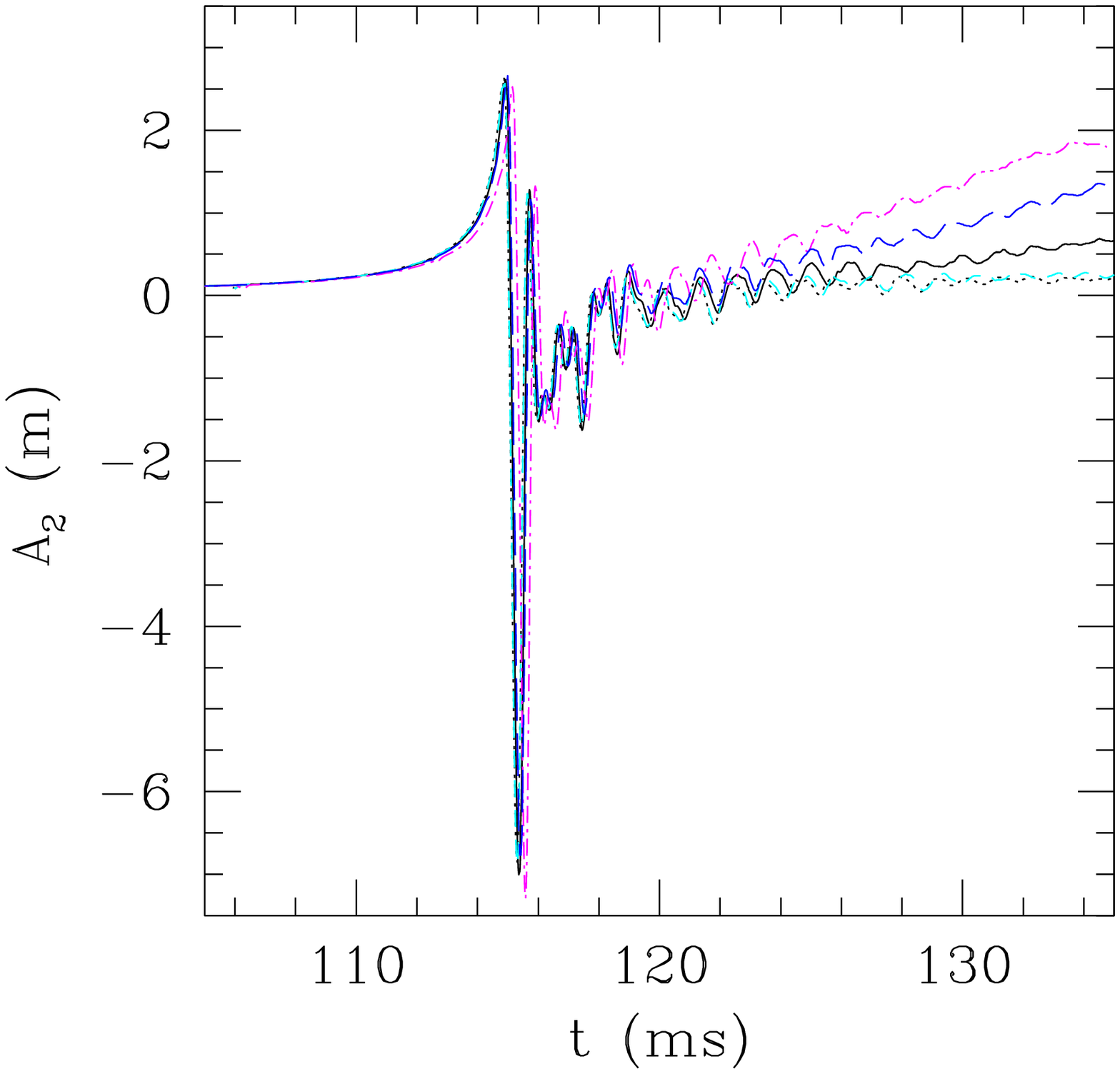}} &
\resizebox{60mm}{!}{\includegraphics{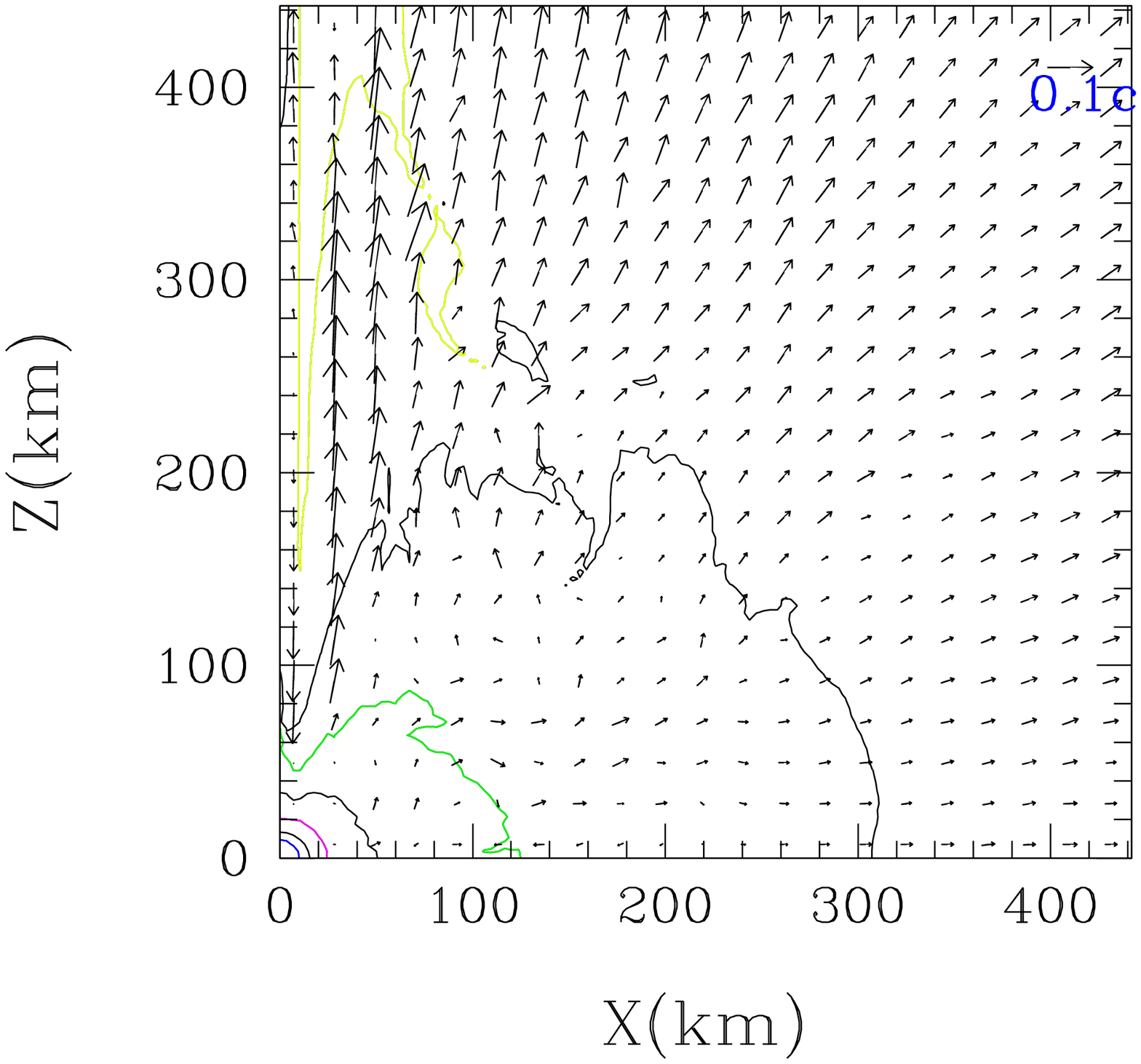}} \\
    \end{tabular}
   \caption{Left panel shows a gravitational waveform in the 2D GRMHD 
simulations by Shibata et al. (2006) 
\cite{shib06} for their models A1 to A4. For models with stronger precollapse 
magnetic fields (pink and blue lines), the quasi-increasing trend can be 
 clearly seen.  Right panel 
shows density contours with velocity vectors for a snapshot ($t=152$ ms for model A4) illustrating MHD-driven jet-like outflows. These figures are by courtesy of Shibata and
 the coauthors.}
  \label{f11}
  \end{center}
\end{figure}

\begin{figure}[hbtp]
  \begin{center}
    \begin{tabular}{cc}
\resizebox{85mm}{!}{\includegraphics{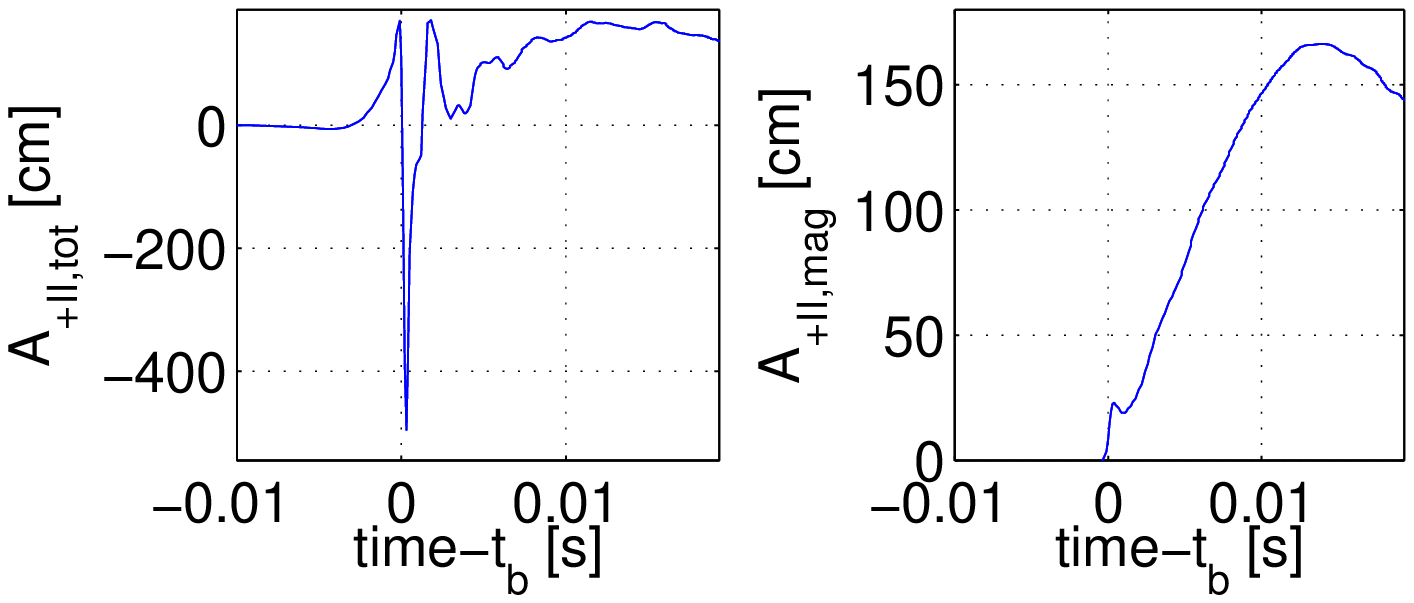}} &
\resizebox{50mm}{!}{\includegraphics{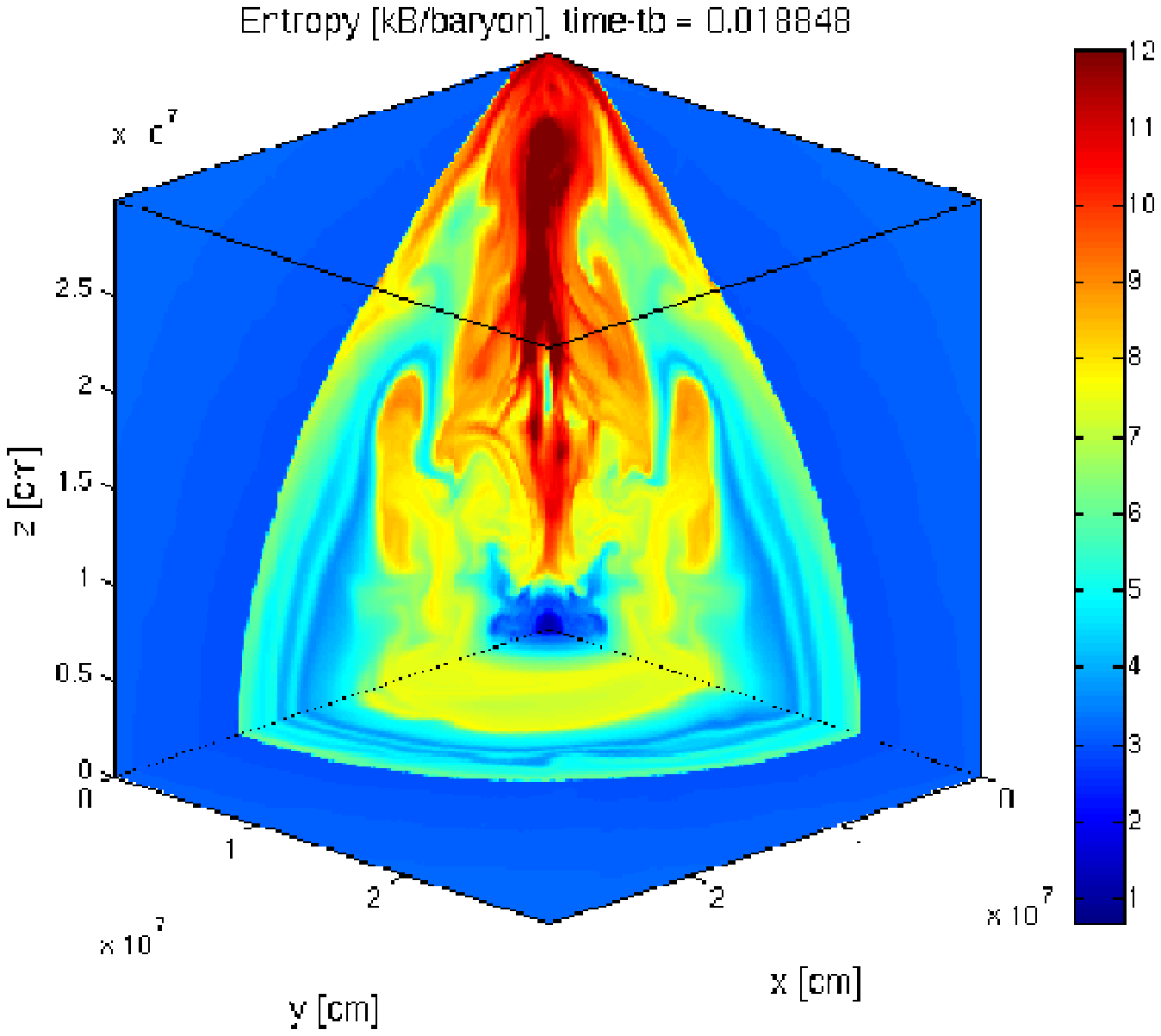}} \\
    \end{tabular}
   \caption{Left pair panels show gravitational waveform in the 3D post-Newtonian 
simulations by Scheidegger et al. 2010) \cite{simon1} for their model R4E1CF in 
 which the left- and right- handside corresponds to the total amplitude and contribution
 only from the magnetic field, respectively. 
The right panel shows the 3D blast morphology of 
 their MHD explosion (entropy distribution at $\sim$18 ms postbounce for a cubic of 
 $300^3$ km$^3$). These figures are by courtesy of Scheidegger and the coauthors.}
  \label{f12}
  \end{center}
\end{figure}

Figures \ref{f10} to \ref{f12} show several examples of gravitational waveforms 
obtained in representative MHD explosion models,
 in which the left and right panel shows
 the waveform and the blast morphology, respectively. 
After the bounce and ring-down signals with their typically duration of $\sim$ 10 ms
 after bounce, a quasi-monotonically growing 
trend can be seen in the every waveform in the left panels of 
Figure \ref{f10} to \ref{f12}.

 Figures \ref{f10} and \ref{f11} are from Obergaulinger et al. (2006) and 
 Shibata et al. (2006) who obtained 2D MHD explosions either 
in Newtonian and approximate GR or full GRMHD simulations, respectively. 
Both of them employed polytropic precollapse models, a phenomenological EOS that 
 mimics deleptonization and neutrino cooling. They pointed out that the increasing 
 trend comes from bipolar flows driven by MHD explosions,
 which can be visible only for cores with precollapse magnetic fields over 
$B_{0} \gtrsim 10^{12}$ G. 3D MHD simulations by Scheidegger et al. (2010) \cite{simon1}
 included realistic EOSs and a deleptonization effect based on 
1D-Boltzmann simulations \cite{matthias_dep}. Compared to the corresponding 
2D models, they pointed out that the jet-like explosions in their 3D models
 are much more difficult to obtain because the wind-up of the poloidal into 
the toroidal field does not proceed efficiently enough due to the growth of spiral 
SASI modes. But when a very rapid
 precollapse angular velocity of  $\Omega_0 = 3 \pi$ rad/s and (strong) magnetic field of $B_0 = 10^{12}$ G were assumed,
 they observed the secularly growing trend associated with the MHD explosions also 
 in their 3D simulations
 (Figure \ref{f12}).

In most of the mentioned MHD simulations, the bounce shock generally does not stall and 
a prompt explosion occurs within a few ten milliseconds after bounce. 
 In fact, the typical timescales shown in Figures \ref{f10} to \ref{f12} are at most 
 20 ms after bounce.
This is because some of them followed adiabatic core-collapse \cite{ober06b,shib06}, 
in which a polytropic EOS is employed to mimic supernova microphysics, and the 
deleptonization scheme in \cite{simon1} cannot capture a drastic energy-loss
 behind the bounce shock due to neutrino cooling.
 However, for models with weaker precollapse magnetic fields akin to the current GRB
 progenitors, the prompt shocks stall firstly in the core like a conventional 
 supernova model with more sophisticated neutrino treatment (e.g., the MHD version of the
 Princeton+ simulations presented in \cite{burr07}). 
In such a case, the onset of MHD explosions, depending on the initial 
 rotation rates, can be delayed till $\sim 100$ ms after bounce \cite{burr07,taki09}.

\begin{figure}[htbp]
    \centering
  \begin{tabular}{cc}
\includegraphics[width=.4\linewidth]{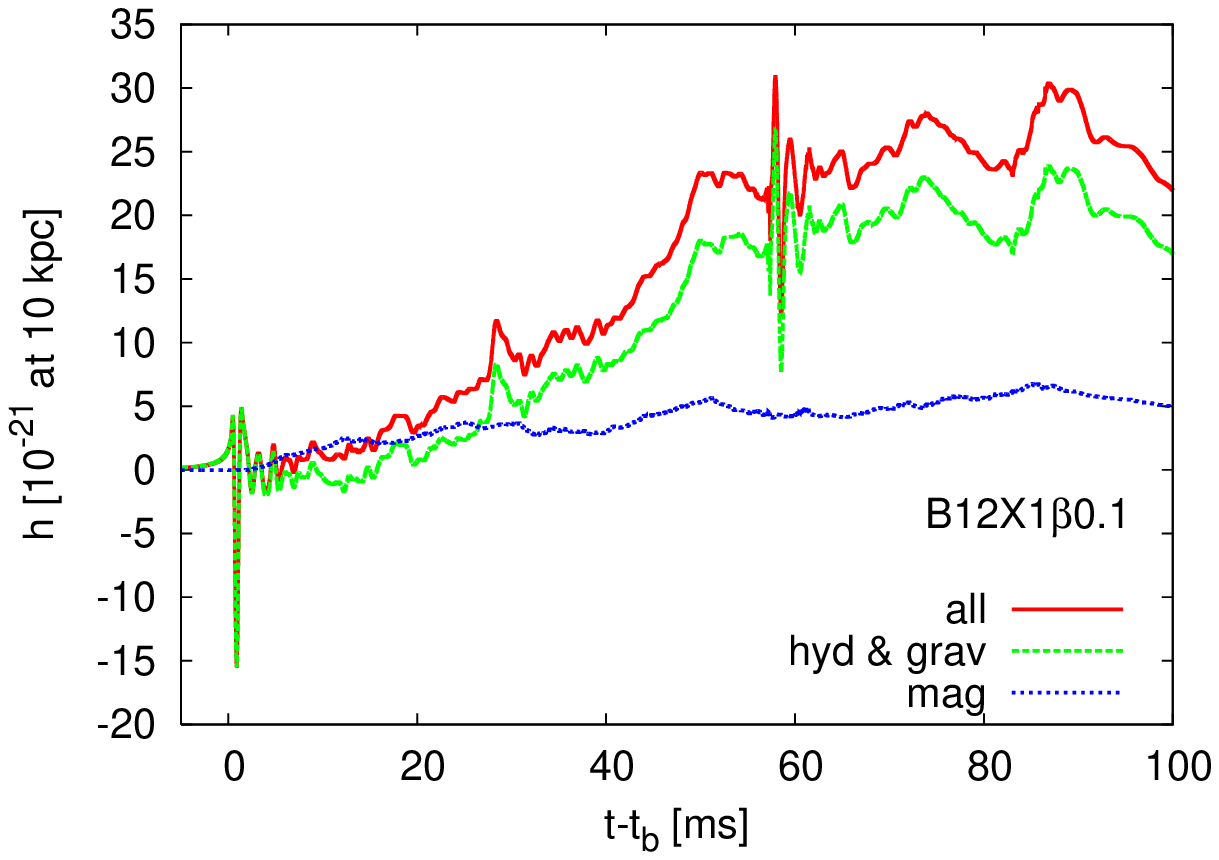} &
\includegraphics[width=.4\linewidth]{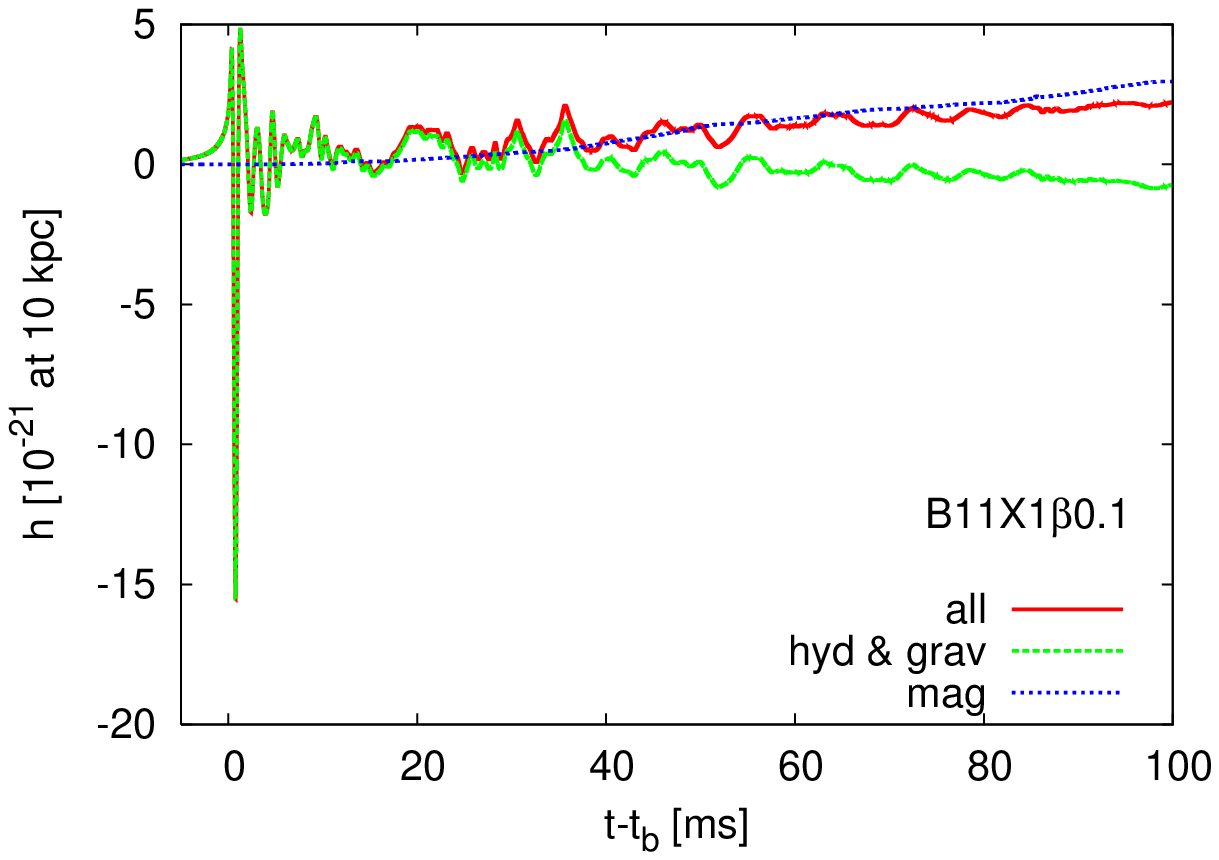}\\
\includegraphics[width=.4\linewidth]{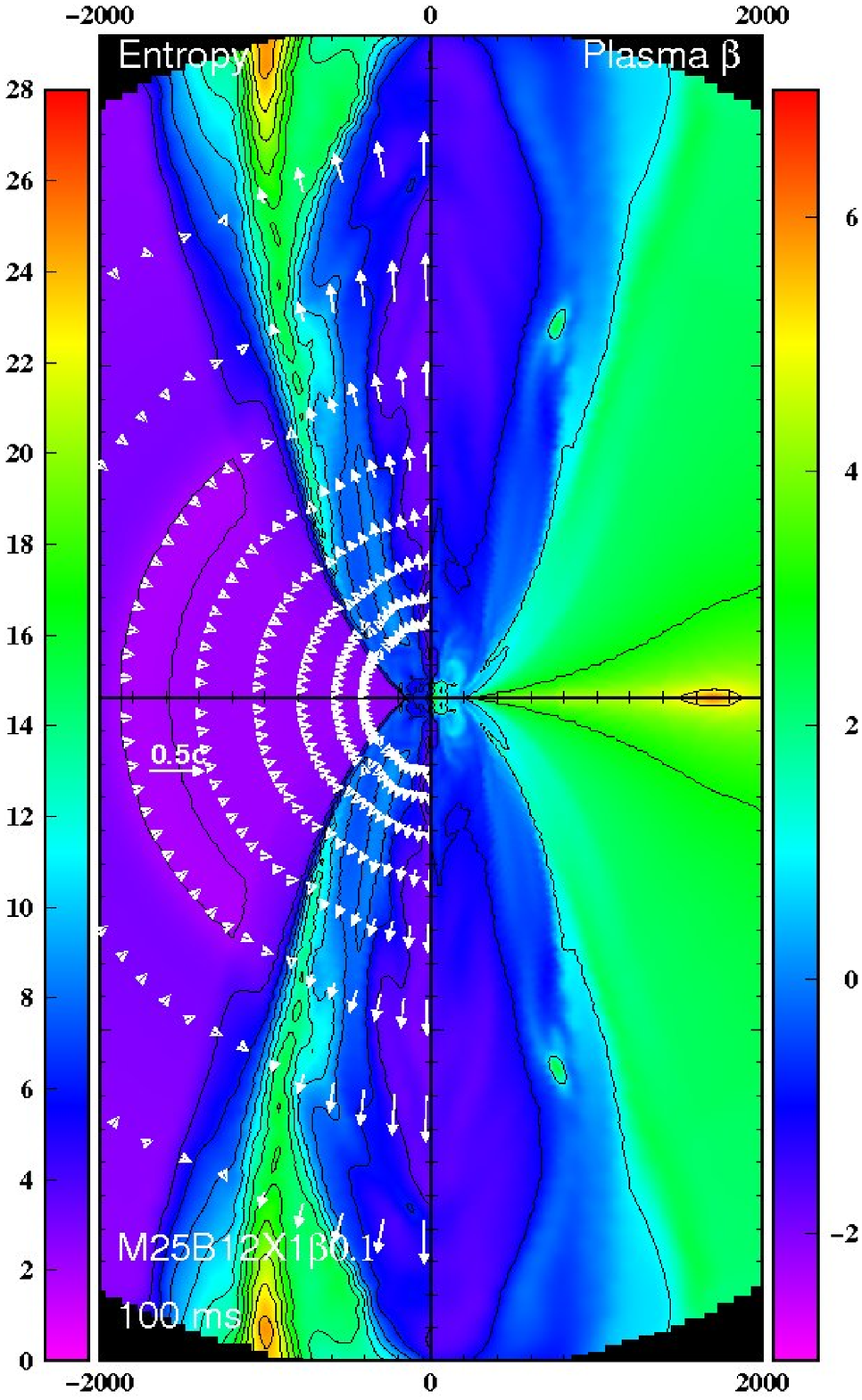} &
\includegraphics[width=.4\linewidth]{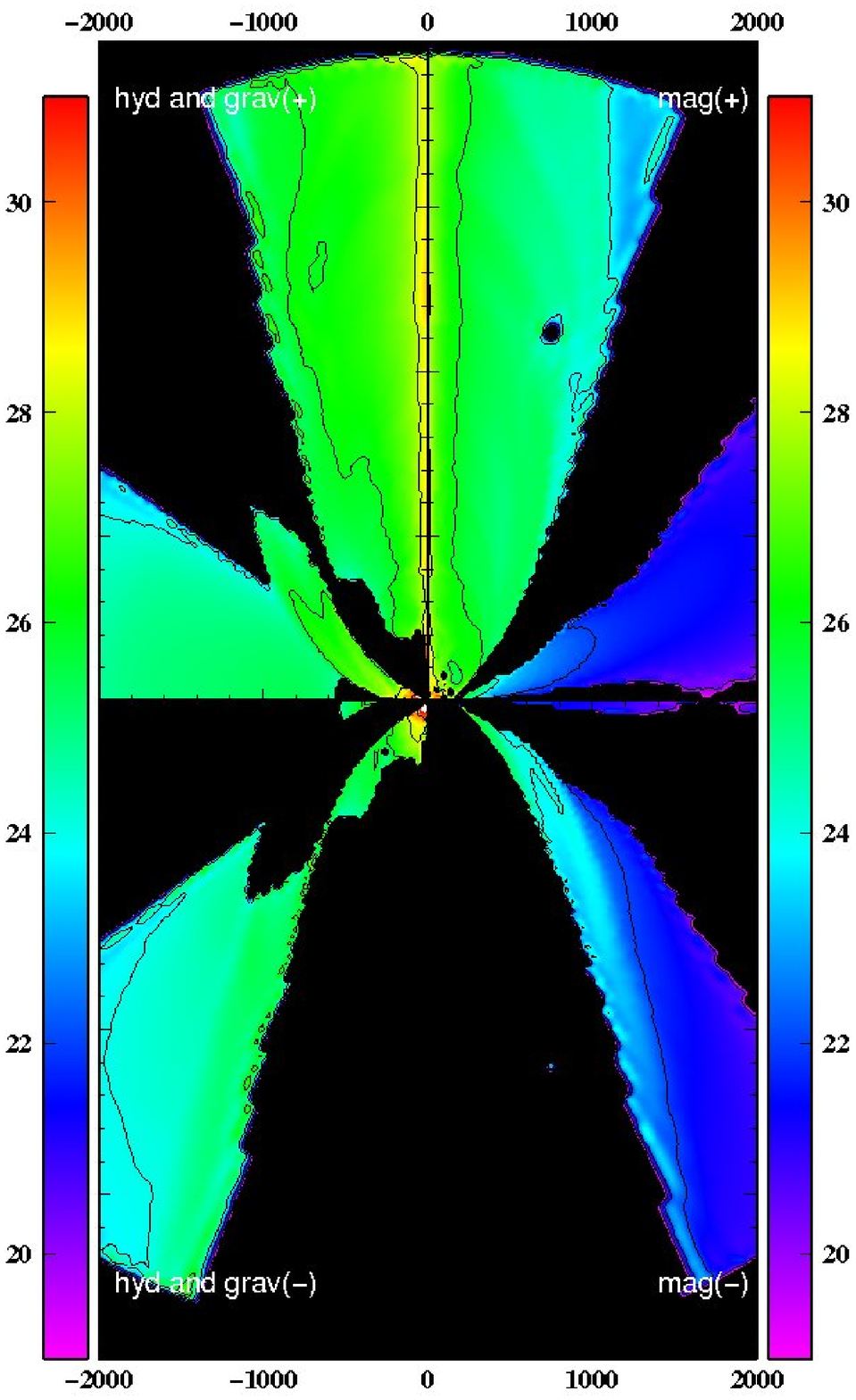} \\
\includegraphics[width=.4\linewidth]{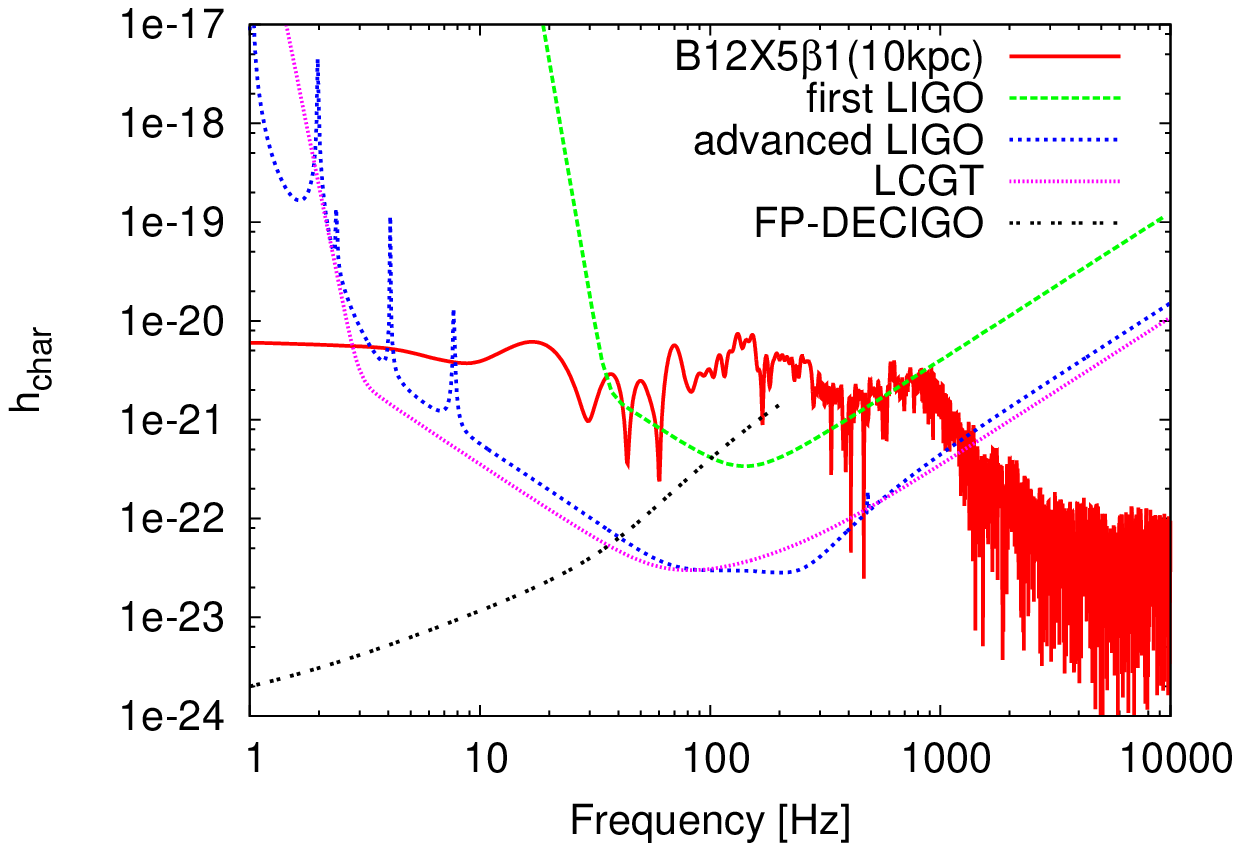}&
\includegraphics[width=.4\linewidth]{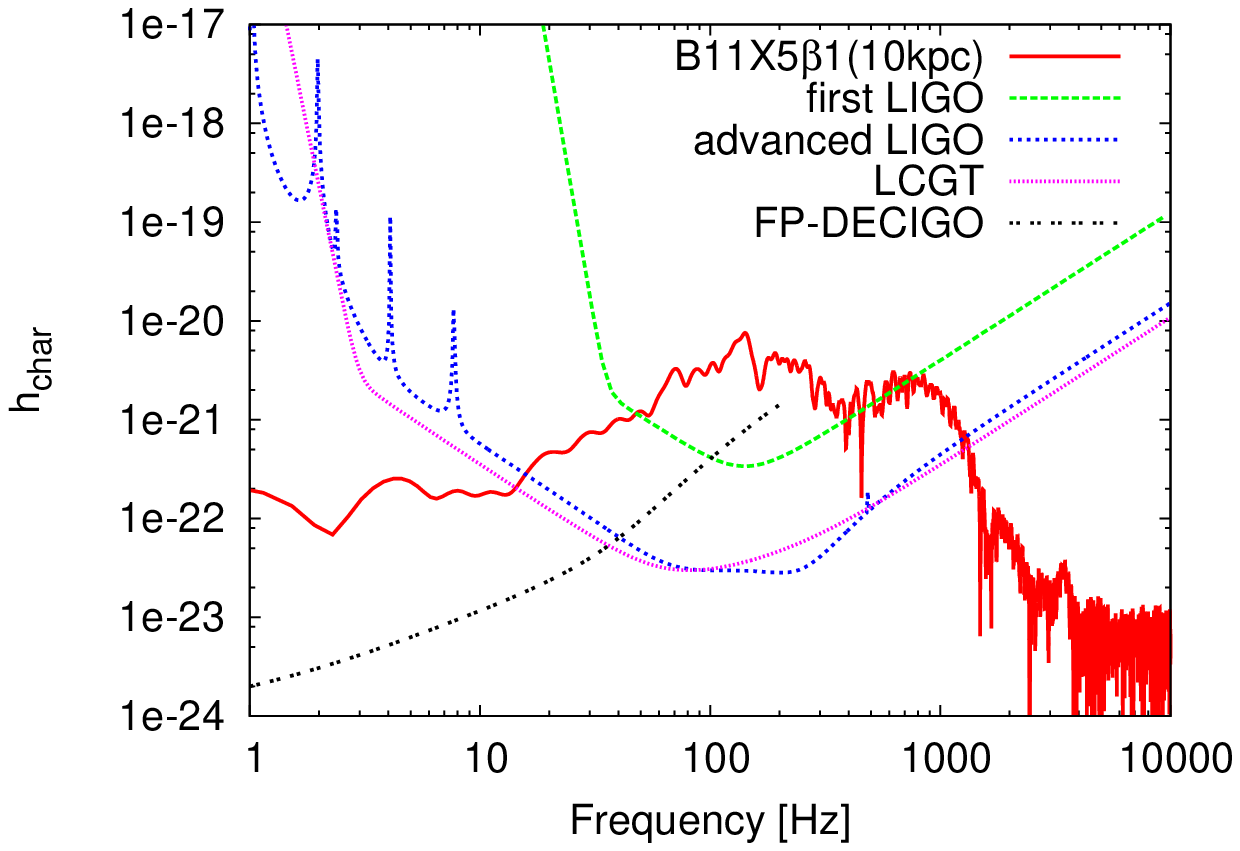}\\
    \end{tabular}
    
    \caption{Top panels show examples of gravitational waveforms with the
 quasi-monotonically increasing trend (left) or not (right panel) obtained in the 
2D SRMHD simulations in Takiwaki and Kotake (2011) \cite{taki_kota}.
 Initial rotation parameter is both set to be  $\beta = 0.1$ \%,
 while the precollapse magnetic field is taken as $10^{12}$ G (left) and 
 $10^{11}$ G (right), respectively.
 The total wave amplitudes are shown by the red line, 
while the contribution from the magnetic fields and from the sum of 
hydrodynamic and gravitational parts
 are shown by blue and green lines, respectively.
 The middle panels show various quantities
 for analyzing the increasing trend taking the top-left-panel model as a reference.
 Left panel shows the distributions of entropy [$k_B$/baryon] (left-half) and 
logarithm of plasma $\beta$ (right-half) at 100ms after bounce. 
The white arrows (left-half) show the velocity fields, 
which are normalized by the scale in the middle left edge ($0.5 c$).
The middle right panel shows the sum of the hydrodynamic and gravitational parts 
(indicated by ``hyd and grav'' in the left-hand side) and the magnetic part 
 (indicated by "mag" in the right-hand side), respectively. The top and bottom 
panels represent the positive and negative contribution (indicated by (+) or (-))
 to ${A_{20}^{\rm{E} 2}}$,
 respectively (see text for more details).  The side length of the middle panels is 
4000(km)x8000(km). The bottom panels indicate the detectability of the 
 GW spectra for a model with (left)
 or without (right) the increasing trend, respectively.
The supernova is assumed to be located at the distance of 10 kpc.}
    \label{f13}
\end{figure}

 Takiwaki and Kotake (2011) \cite{taki_kota} addressed this issue by performing 
 2D special relativistic MHD (SRMHD) simulations with the use of an approximate GR potential
 \cite{ober06a} in which a neutrino leakage scheme was employed to mimic neutrino 
cooling.  The top left panel of Figure \ref{f13} shows the quasi-monotonically 
increasing trend, which 
 is obtained for a model with strong precollapse 
magnetic field ($B_{0} = 10^{12}$G) also with rapid rotation initially imposed
 ($\beta$ parameter = 0.1 \% with $\beta$ representing ratio of 
the rotational energy to the absolute value of the gravitational energy prior 
 to core-collapse).
Such a feature cannot be 
 observed for a weakly magnetized model ($B_{0} = 10^{11}$G. top right panel).

To understand the origin of the increasing trend, it is most straightforward to 
 look into the quadrupole GW formula, which can be expressed in 2D as \cite{thorne},
\begin{equation}
h =  \frac{1}{R}A^{E2}_{20} \frac{1}{8}\sqrt{\frac{15}{\pi}}\sin^2 \theta,
\label{htt}
\end{equation}
where $R$ is the distance to the source, $\theta$ is the viewing angle of the 
 source, and ${A_{20}^{\rm{E} 2}}$ is the quadrupole harmonic amplitude consisting 
 of the following three parts \cite{kota04a,yama04,ober06a,taki_kota}, 
\begin{equation}
 {A_{20}^{\rm{E} 2}} =  {A_{20}^{\rm{E} 2}}_{\rm (hyd)} + 
 {A_{20}^{\rm{E} 2}}_{\rm (mag)} + {A_{20}^{\rm{E} 2}}_{\rm (grav)}.
\label{A20}
\end{equation}
On the right hand side, the first term is related to anisotropic 
 kinetic energies which we refer to as hydrodynamic part, 
\begin{eqnarray}
 {A_{20}^{\rm{E} 2}}_{\rm(hyd)} &=& \frac{G}{c^4} \frac{32 \pi^{3/2}}{\sqrt{15 }} 
\int_{0}^{1}d\mu 
\int_{0}^{\infty}  r^2  \,dr  {f_{20}^{\rm{E} 2}}_{\rm(hyd)}, \\
{f_{20}^{\rm{E} 2}}_{\rm(hyd)} 
&=& \rho_{*}W^2( {v_r}^2 ( 3 \mu^2 -1) + {v_{\theta}}^2 ( 2 - 3 \mu^2)
 - {v_{\phi}}^{2} - 6 v_{r} v_{\theta} \,\mu \sqrt{1-\mu^2}),
\label{quad}
\end{eqnarray}
 where $\rho_{*}$ is effective density $\rho_{*}= \rho + (e + p + |b|^2)/c^2$
 with $\rho, e, p$ and $b^2 = b_{i} b^{i}$ 
representing the baryon density, internal energy,  pressure, and the magnetic 
 energy, respectively. $\mu = \cos \theta$ is the direction cosine and 
 $W= 1/\sqrt{1-v^kv_k}$ is the Lorentz boost factor 
 with $v_{i}$ representing the spacial velocities in the spherical coordinates 
 (see \cite{taki_kota} for
 more details).
 The second term is related to anisotropy in gravitational potentials 
which we call as the gravitational part,
\begin{eqnarray}
 {A_{20}^{\rm{E} 2}}_{\rm(grav)} &=& \frac{G}{c^4} \frac{32 \pi^{3/2}}{\sqrt{15 }} 
\int_{0}^{1}d\mu  \int_{0}^{\infty}  r^2  \,dr
 {f_{20}^{\rm{E} 2}}_{\rm(grav)} ,\\
{f_{20}^{\rm{E} 2}}_{\rm(grav)}
&=&
 \left[\rho h (W^2+(v_k/c)^2) + 
 \frac{2}{c^2}\left( p+\frac{\left|b\right|^2}{2}\right)
-\frac{1}{c^2}\left((b^{0})^2+(b_{k})^2 \right)\right]\nonumber \\
& & \times \left[- r \partial_{r} \Phi (3 \mu^2 -1) + 3 \partial_{\theta} \Phi \,\mu
\sqrt{1-\mu^2}\right],
\label{grav}
\end{eqnarray}
 where $\Phi$ is the gravitational potential of self-gravity, 
and finally the third term is related to anisotropy in magnetic energies that 
 we refer to as magnetic part,
\begin{eqnarray}
 {A_{20}^{\rm{E} 2}}_{\rm(mag)} &=& - \frac{G}{c^4} \frac{32 \pi^{3/2}}{\sqrt{15 }} 
\int_{0}^{1}d\mu  \int_{0}^{\infty}  r^2  \,dr {f_{20}^{\rm{E} 2}}_{\rm(mag)}, \\ 
 {f_{20}^{\rm{E} 2}}_{\rm(mag)}
&=&
 [{b_r}^2 ( 3 \mu^2 -1) + {b_{\theta}}^2 ( 2 - 3 \mu^2)
 - {b_{\phi}}^{2} - 6  b_{r} b_{\theta} \mu \sqrt{1-\mu^2}].
\label{mag}
\end{eqnarray} 
 

The middle right panel of Figure \ref{f13} shows contributions
 to the total GW amplitudes (equation (6)) for the strongly magnetized model
 (corresponding to the top left panel in Figure \ref{f13}),
 in which the left-hand-side panels are for the sum of the hydrodynamic and 
gravitational part,
 namely $\log\left(\pm\left[{f_{20}^{\rm{E} 2}}_{\rm (hyd)} + {f_{20}^{\rm{E} 2}}_{\rm (grav)}\right]\right)$ 
(left top($+$)/bottom($-$)(equations (\ref{quad},\ref{grav})), and the right-hand-side 
panels are for the magnetic part, namely 
$\log\left(\pm{f_{20}^{\rm{E} 2}}_{\rm (mag)}\right)$ 
(right top($+$)/bottom($-$)) (e.g., equation (\ref{mag})).
 By comparing to the middle left panels,
 it can be seen that the positive contribution is  
 overlapped with the regions where the MHD outflows exist.
The major positive contribution is from the kinetic term of the MHD outflows 
 with large radial velocities (e.g., $ + \rho_{*}W^2 {v_r}^2$  in equation (\ref{quad})).
 The magnetic part also contributes to the positive trend (see top right-half
  in the middle right panel (labeled by mag(+))). This comes 
from the toroidal magnetic fields (e.g., $ + {b_{\phi}}^{2}$ in equation
 (\ref{mag})), which dominantly contributes to drive MHD explosions. 

The bottom panels in Figure \ref{f13} show
 the GW spectra for a pair of models that does or does not
 have the increasing trend (left and right) as in the top panels.
 Regardless of the increasing trend,
  the peak amplitudes in the spectra are rather broad-band around $100-1000$ Hz. 
 On the other hand, the spectra for lower frequency domains (below $\sim 100$ Hz) 
are much larger for the model with the increasing 
 trend (left) than without (right).
 This reflects a slower temporal variation of the secular drift 
inherent to the increase-type waveforms (e.g., top panels in Figure \ref{f13}).
Similar to the neutrino GWs mentioned in section \ref{neutrino}, 
it is true that the GWs in the low frequency domains are
  difficult to detect due to seismic noises, but
 a recently proposed future space interferometers like 
Fabry-Perot type DECIGO is designed to be sensitive in the frequency regimes
\cite{fpdecigo,kudoh} (e.g., the black line in the bottom panels). 
 These low-frequency signals, if observed, could be one 
 important messenger of the increase-type waveforms that are likely to be 
associated with MHD explosions exceeding $10^{51}$ erg. 

Concerning the microphysical aspects, the MHD simulations presented in this section
 are far behind the most up-to-date simulations in which the spectral 
 neutrino transport is solved (e.g., section \ref{s2} and Table 1). This means that the GW predictions 
in the MHD explosions are also still in their infancy. One may easily guess that 
 it is numerically challenging to solve the neutrino transport in the
 highly non-spherical environments that are inherent to MHD explosions.
 The PNS is deformed to be like a dumbbell and the regions outside the PNS
 become highly anisotropic such as by the formation of the polar funnel region 
after the bipolar flows pass by\footnote{In such a case, the ray-by-ray approximation
 would be not good. Instead, the MGFLD scheme or a recently proposed M1 closure
 technique \cite{martin11} would be much better.}. It is an urgent task 
to sophisticate the GW predictions in the MHD 
explosions in the same level as those in the neutrino-driven explosion models. 


\subsubsection{Non-axisymmetric instabilities}\label{nonaxi}

As the textbook by Chandrasekhar (1969) \cite{chandra69} says
 (see also \cite{tass78,shap83}), 
rapidly rotating compact objects can be subject to 
 non-axisymmetric rotational instabilities when the ratio of rotational to 
 gravitational potential energy ($T/|W| = \beta$) exceeds a 
 certain critical value. Since the growing instabilities carry the object's
 spheroidal into a triaxial configuration with a time-dependent quadrupole(or higher)
moment, strong GW emission can be expected (see Fryer and New (2011) \cite{fryer11}
 for a detailed
 review of the uncovered topics in this article).

Probably the best understood type of instability is the classical dynamical bar-mode 
instability with a threshold of $\beta_{\rm bar} \gtrsim 0.27$. Rampp et al. (1998) \cite{ramp98}
 reported the first 3D hydrodynamic core-collapse simulations to study the growth 
 of the bar mode instability and their impact on the GW emission. The initial 
 condition for their study was based on the configuration at several milliseconds 
 before core bounce in the rapidly rotating 2D models of Zwerger et al. (1996) 
\cite{zweg}. In addition to the configuration, they imposed low mode ($m$ = 3) density
 perturbation and followed the growth of the instability, where $m$ stands for the 
 azimuthal quantum number. They observed the three clumps merged into a bar-like structure due to the growth of the non-axisymmetric instability. In fact, their models are 
 rapid rotators whose values of $T/|W|$ exceeds the critical value beyond $\beta_{\rm bar}$, beyond which 
MacLaurin spheroids become dynamically unstable again triaxial perturbations. However,
 they found that the maximum GW amplitudes were only $\sim 2\%$ different from the 
 2D cases by Zwerger et al \cite{zweg}. 3D SPH simulations of a rapidly rotating model
 with the initial value of $T/|W|_{\rm init}$ of $\sim 3 \%$ 
by Fryer and his collaborators \cite{fryer04a,chris04a} reached a similar conclusion.
  Full GR 3D simulations by Shibata and his collaborators, on the other hand,
 pointed out that the maximum
 GW amplitudes for models with rapid rotation
 ($1 \lesssim T/|W|_{\rm init} \lesssim 3 \%$) can be enhanced by a factor of 10 
  than the ones predicted in the
 Newtonian studies mentioned above \cite{shib05}. They discussed that the enhancement of the 
self-gravity due to the GR effects results in a more efficient spin-up of the core 
 thus leading to the growth of the non-axisymmetric instability. The key question is 
whether such a precollapse rapid rotation ($T/|W|_{\rm init} \gtrsim 1 \%$)
 can be realized or not. 

More recently, Ott et al. (2007) \cite{ott_3d,ott_3db}
and Dimmelmeier et al. (2008)\cite{dimm08} reported a systematic 
 study in which a wide variety of the state-of-the-art stellar evolution
 models\footnote{also by changing a huge parameter space spanned by initial 
rotation rate, degree of differential rotation, and different EOSs.}
 was employed in their 3D full GR and 2D CFC core-collapse simulations.
 Their results 
presented a strong evidence that the postbounce core, even in their extreme models, 
 do not reach values of $\beta$ close to $\beta_{\rm dyn}$ during collapse, bounce 
 and during early postbounce time. The PNS rotation could reach $\beta_{\rm dyn}$
 if it would keep gravitationally contracting with conserving its angular momentum 
 in their cooling phase (lasting on a timescale of $\sim$ minutes).
 However in this case, the threshold of the 
secular bar-mode instability ($\beta_{\rm sec} \sim 0.14$) may be first satisfied
 (e.g., \cite{laishapi}). For more in-depth reviews about this topic 
with their potential GW emission mechanisms, 
see recent reviews of Andersson (2003) \cite{nils1} 
and Kokkotas (2008) \cite{koko}.

\begin{figure}[htpb]
\begin{center}
\epsfxsize=8cm
\epsfbox{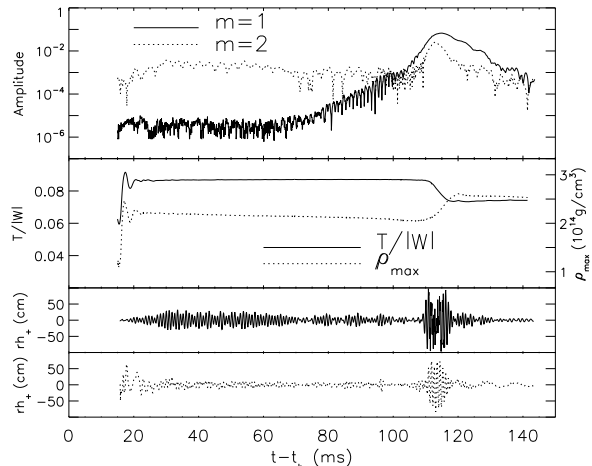}
\caption{Time evolution of various quantities in a 3D model with
 $T/|W|_{\rm init} = 0.2 \%$ calculated
 by Ott et al. (2005) \cite{ott_one_arm}. Time is measured from the epoch of core bounce $t_b$. Top panel shows that the amplitude of the $m = 1$ mode precedes
 that of the $m=2$. Middle panel shows the time evolution of $T/|W|$ and
 the core's maximum density. It is shown from the panel that after the epoch of $t - t_b \sim 100$ ms, when the $m=2$
 mode begins to be amplified, the transfer of the angular momentum
 becomes active which results in the increase of the maximum density and
 the decrease of the $T/|W|$. The bottom panel shows the gravitational
 strain at the distance to the source $r$ as viewed down the rotational
 axis (solid curve) and as viewed along the equatorial plane (dotted curve). One can
 see that the waveform traces the time evolution of the $m=2$ mode. Note in
 the panel that $rh = 100$ cm corresponds to $h \sim 3 \times 10^{-21}$ for a
 galactic supernova. This plot is by courtesy of Ott and the coauthors. }
\label{f14}
\end{center}
\end{figure}

 In contrast to the high $T/|W|$ instabilities mentioned above, recent work,
 some of which has been carried out in idealized setups and assumptions
 \cite{shibata02,saijo03,watts,ou,pablo2007}
 and later also in more self-consistent core-collapse simulations \cite{ott_one_arm,
simon,simon1}, suggest that a differentially rotating PNS can become dynamically unstable
 at much lower $T/|W|$ as low as $\lesssim 0.1$. Despite clear numerical evidence 
 for their existence, the physical origin of the low-$T/|W|$ instability remains 
 unclear. However it has been suggested \cite{watts,saijo06} 
that the instabilities are associated with the 
 existence of corotation points (where the pattern speed of the unstable modes
 matches the local angular velocity) inside the star and are thus likely to be 
a subclass of shear instabilities\footnote{Note that corotation resonance has been long 
 known to the key ingredients in the accretion disk system, such as the
 Papaloizou-Pringle instability \cite{papa}.}.

 Figure \ref{f14} shows various quantities for a non-magnetized 20 $M_{\odot}$ model 
 of Ott et al. (2005) that experiences the low-$T/|W|$ instability. 
They investigated the growth of the non-axisymmetric structure 
until the rather later phases ($\ge 100$ msec) after bounce \cite{ott_one_arm}. 
They found that the growth of the $m=1$ mode, 
the so-called one-armed instability 
\cite{centrella,shibata02,saijo03,saijo06}, 
precedes the growth of the bar-mode ($m=2$) instability (see the top
 panel in Figure \ref{f14}), where $m$ denotes the azimuthal quantum number. 
They pointed out that the initial rotation rate can be as small as 
$T/|W|_{\rm init} = 0.2 \%$  for the sufficient gravitational radiation 
enough to be detected by the future detectors.
This value is indeed much smaller than the one previously 
assumed for igniting the growth of the classical bar-mode instability.   

\begin{figure}[htbp]
    \centering
  \begin{tabular}{cc}
\includegraphics[width=.4\linewidth]{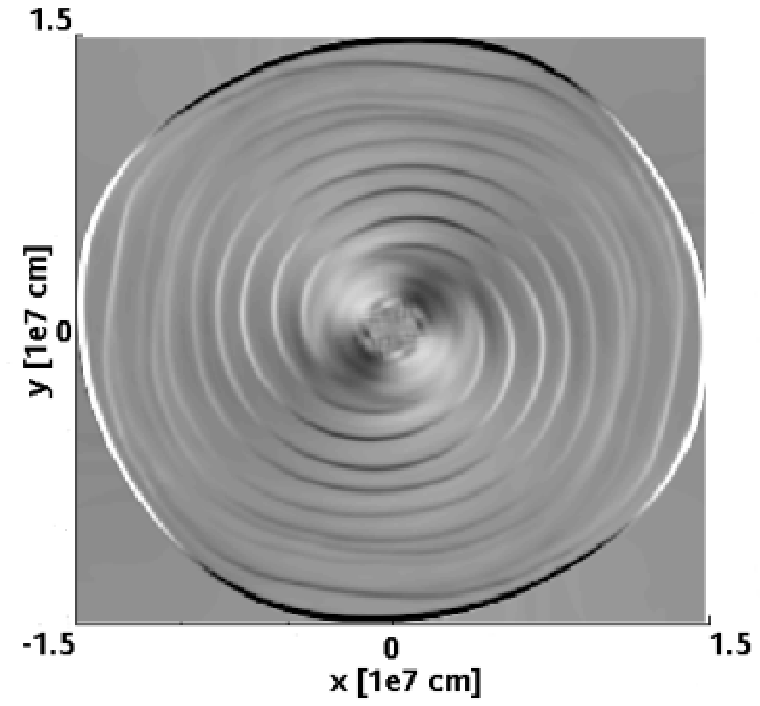} &
\includegraphics[width=.4\linewidth]{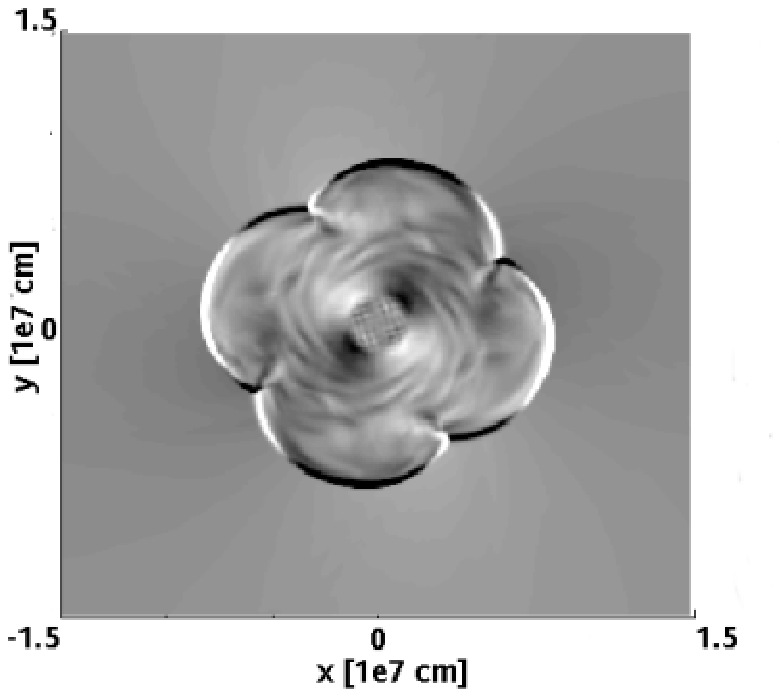}\\
\includegraphics[width=.5\linewidth]{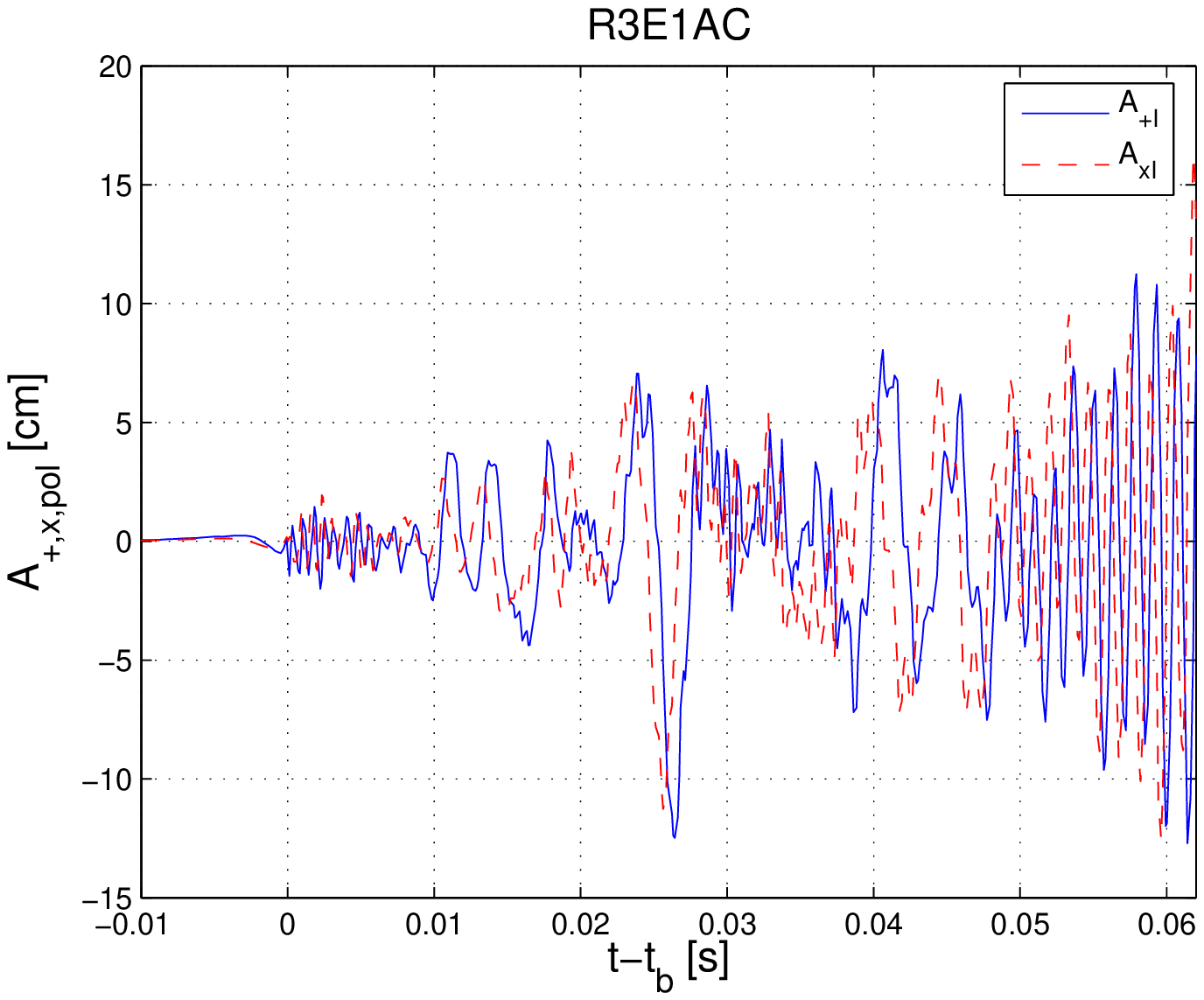} &
\includegraphics[width=.5\linewidth]{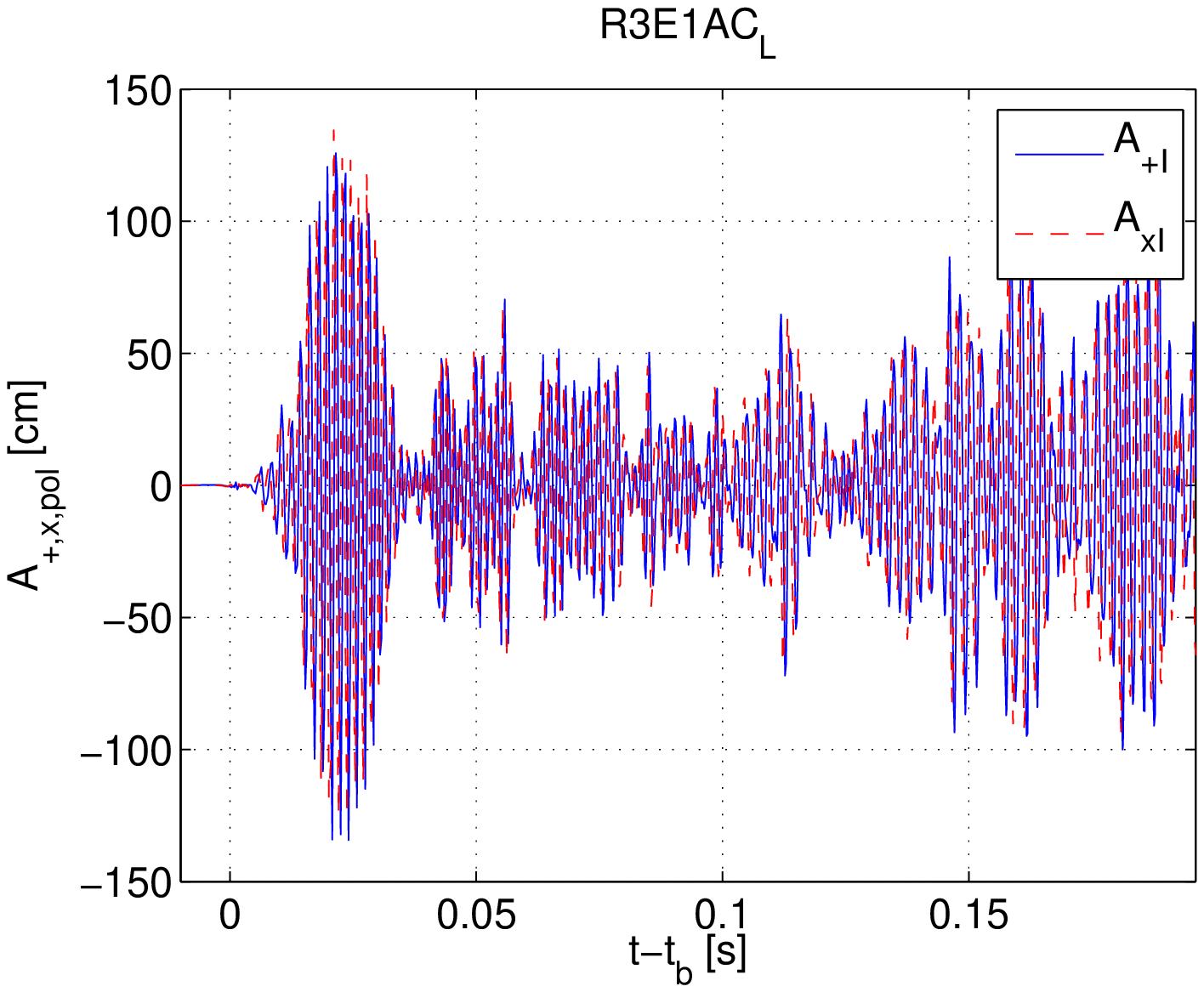} \\
    \end{tabular}
\includegraphics[width=.7\linewidth]{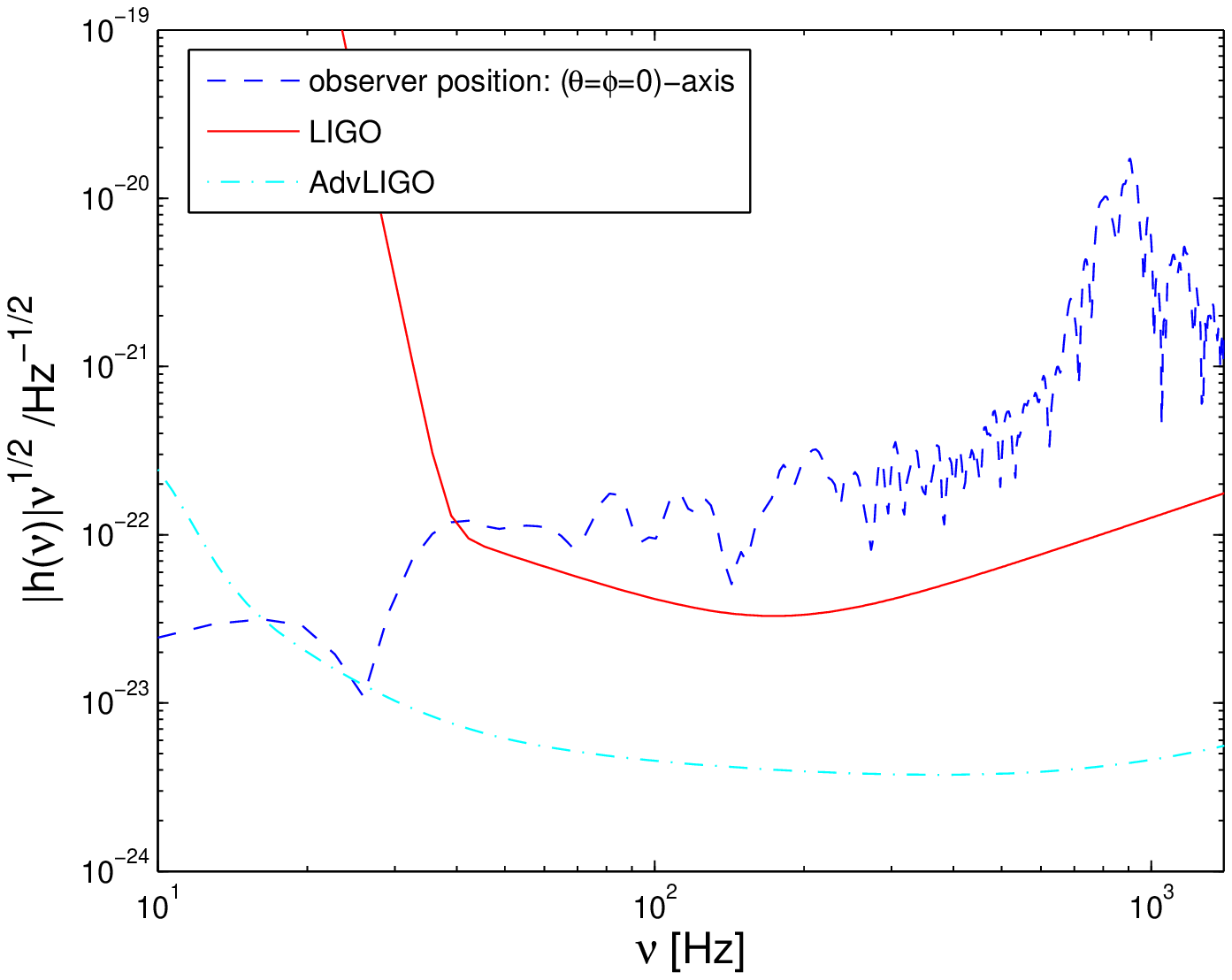}
 \caption{Top panels depict the vorticity distribution in the 
 equatorial plane of the 3D MHD models of R4STAC (left) 
and R4E1FC$_L$ (right) in Scheidegger et al. (2010) \cite{simon1} 
(at $\sim$ 10 ms after bounce). A two-armed $m=2$ pattern behind 
the stalled shock is clearly seen (the side length of the plot is (300km)$^2$)
 either for their 3D model without (top left) or with the deleptonization effect 
(top right). The middle panels display the quadrupole amplitude of the $A_{+}$
 and $A_{\times}$ along the pole for a given 3D MHD model without (left) or with 
 the deleptonization effect (right).
The bottom panel shows spectral energy distributions of the GW signal emitted 
 along the polar axis for their representative 3D model 
experiencing the low-$T/|W|$ instability (model R4STCA)
 at a distance of 10 kpc. These plots 
 are by courtesy of Scheidegger and the coauthors.}
\label{f15}
\end{figure}

More recently, Scheidegger et al. (2008,2010) \cite{simon,simon1} investigated
 this issue by more realistic 3D MHD models that followed consistently the dynamics 
 from core-collapse, through bounce, up to $\sim 200$ ms postbounce in which 
  a treatment for neutrino transport based on a partial implementation of the IDSA scheme
 \cite{idsa}\footnote{see section \ref{s2} for more details of the scheme.} was 
incorporated. Top panels in Figure \ref{f15} show 
the vorticity distribution in the equatorial plane selected from
 their 3D MHD models, in which the major difference originates from the 
 inclusion of the neutrino transport effect (right) or not (left).
 Regardless of 
 the model difference, a two-armed ($m=2$) pattern is clearly seen 
(at $\sim$ 10 ms after bounce for the plot) 
and then the spiral flows develop for more than several hundreds milliseconds later on.
 The middle panels display the quadrupole amplitude of the $A_{+}$
 and $A_{\times}$ along the pole for their 3D MHD model with (right) or without 
the neutrino cooling (left). The most striking feature is that the model
 with neutrino cooling (right) 
shows 5 $\sim$ 10$\times$ bigger maximum GW amplitudes due to the nonaxisymmetric 
dynamics compared to their counterparts that neglect neutrino cooling (left).
 They pointed out that the neutrino cooling in the postbounce phase leads to a 
 more condensed PNS with a shorter dynamical timescale and also with 
 much more matter enclosed in the unstable region (see the compactness of the 
 spiral flows in the top right compared to the top left panel in Figure \ref{f15}),
 leading to much more powerful GW. Another important message from the middle panels is 
 that 
 the gravitational waveforms from the nonaxisymmetric dynamics generally show 
narrow-band and highly quasi-periodic 
signals which persist until the end of simulations. For the periodic signals,
 the effectively measured GW amplitude scales with the number of GW cycles $N$ as 
$h_{\rm eff} \propto h \sqrt{N}$.
 This is the reason why the peak amplitude
 in the GW spectrum (the bottom panel in Figure \ref{f15})  
exceed $10^{-20}$ for the galactic source, making the chance of detection
 quite higher. In fact, the wave amplitude from the 
 short-duration bursts near bounce is generally smaller than $10^{-20}$ for 
the galactic source (e.g., the amplitude near 1 kHz in the bottom panels in Figure 17).
  
For detecting (potentially) the most powerful GW signals, it is therefore 
of crucial importance to understand the properties of the non-axisymmetic instabilities.
 However the numerical difficulty to follow a long-term postbounce evolution in 3D is 
 the main hindrance at present.
 While the angular momentum is continuously brought in the iron core 
from stellar envelope that rotates with higher angular momentum, 
the shear energy in the vicinity of 
 the corotation points will continue to be redistributed or even dissipated by some 
 mechanisms including the MRI. In the mentioned 3D simulations, neutrino heating 
 is not taken into account. Neutrino-driven convection would affect the growth of the 
 non-axisymmetric instabilities as well as through the spiral SASI modes. 
 The general relativity that has been mostly 
treated in a post-Newtonian manner should affect the growth rate.
At first, we could 
 start to address this question by performing parametric explosion models in 3D MHD
 (e.g., sections \ref{exp1} and \ref{exp2}).
 But for a more quantitative discussion, the full  
 3D GRMHD simulation with an appropriate neutrino transport is needed also 
 in this case, towards which the supernova modelers will 
 keep their efforts (as they have done since time immemorial)
 and more recently people in the numerical 
relativity are also joining in the efforts. 

\section{Summary and Concluding remarks}

\begin{table}[htbp]
    \begin{center}
\includegraphics[width=16.0cm]{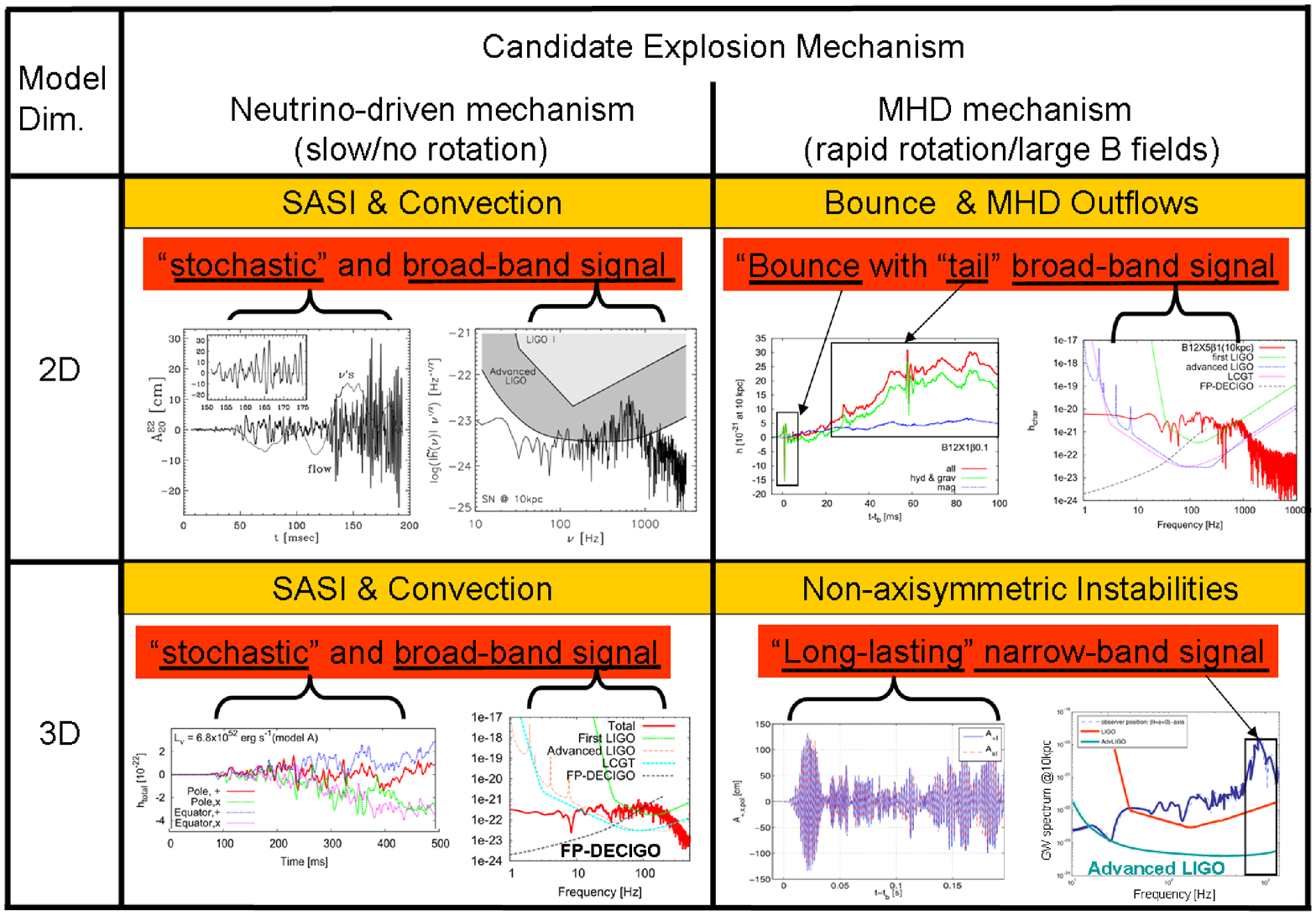}
 \caption{Illustrative summarizing the relation of the potential explosion mechanisms
 (horizontal column) and their emission processes (columns colored by yellow) 
 and GW signatures (columns colored by red) obtained so far in "2D" and "3D" 
simulations (vertical direction). In each column, the waveform (left) and
 the GW spectrum (right) are shown for some representative models, in which 
 top left (2D, neutrino mechanism), top right
 (2D, MHD mechanism), bottom left (3D, neutrino mechanism), and bottom right (3D,
 MHD mechanism) are again taken from Figures \ref{f2}, \ref{f13}, \ref{f8}, and \ref{f15}
 (see each figure and their caption for more details). Figures in the top left and 
bottom right columns are taken from the results by M\"uller et al. (2004) \cite{mueller04}
 and Scheidegger et al. (2010) \cite{simon1} by their courtesy.}
\end{center}
\end{table}

The aim of writing this article was to provide an overview of what we
currently know about the possible GW signatures of core-collapse SNe
 predicted by a number of extensive numerical simulations especially 
 in the last $\sim$15 years. 
In addition to the long-studied 
bounce GW signals,
 multiple new physical inputs have been proposed mostly in the last decade. 
 Among them, we focused on the GW signatures produced by 
prompt convection, non-radial flows inside the PNS and in
 the neutrino-driven bubble, the activity of the SASI, and by asymmetric 
 neutrino emission in the context of the neutrino-driven explosion models in section
 \ref{neutrino}, and by asymmetries associated with the effects of magnetic fields and 
  non-axisymmetric rotational instabilities in section \ref{mhd_gw}.
As mentioned in section \ref{s2}, 
our understanding of the physics to unravel these waveform features
has been progressing in accordance with the sophistication of the numerical 
simulations on which SN modelers have been putting a huge effort for long.

We wish that from Table 3, one might see a very rough illustrative summary of this 
  article that describes a possible association between the potential 
explosion mechanisms (horizontal) and the emission processes with their
 GW signatures (columns colored by yellow and red) obtained in the recent multi-D 
simulations (vertical). At one sight it is a very good news for us that 
 there is a clear correspondence between the explosion mechanisms and 
 their GW signatures. This means that the GW observation could provide 
 an important probe into the potential explosion mechanisms (see also
 Ott (2009) \cite{ott_rev})\footnote{It should be also remembered that 
  galactic core-collapse SN event is a quite rare event happening
 at a rate of one in a few decades. As theoreticians, we sincerely  
 hope to be able to make clear the supernova mechanism and the related 
 issues before observations tell us!}.
Although the criterion that bifurcates between the neutrino-driven 
 mechanism and the MHD mechanism is still gray, rotation and magnetic 
fields in the precollapse cores should hold the key importance.

In the neutrino heating mechanism, the whole story may be summarized as follows. 
If a precollapse iron core has a ``canonical'' rotation rate as predicted by stellar 
evolution calculations ($\Omega_0 \lesssim 0.1$ rad/s), 
the collapse-dynamics before bounce proceeds 
spherically and the structures interior to the PNS are essentially 
spherical at this epoch. The stalling bounce shock gives rise to the GW signals of 
 the prompt convection (see the inset of the top left panel in Table 3, and 
 also Figure 1 in section \ref{neutrino} for more details).
Typically later than $\sim 100$ ms after bounce, convective overturns 
as well as the SASI in the postshock region become much more vigorous with time. 
Since anisotropies of the neutrino flux as well as matter motions 
 in this epoch are governed by the non-linear hydrodynamics, the 
 GWs signatures change also stochastically with time (sections 
\ref{exp1} and \ref{exp2}).
 This feature was already known in the first-principle 
2D simulations (indicated by "stochastic" 
in Table 3, e.g., section \ref{first} for more details). Furthermore 
 recent exploratory 3D simulations revealed that the stochastic nature 
becomes much more
 strong because the explosion geometry changes in all directions (section 
\ref{stochastic}). Note that the stochastic nature could become weaker 
in the presence of rotation as a result of the spiral SASI (section \ref{stochastic}),
 however the first-principle 3D simulations are apparently 
 needed to draw a robust conclusion. Contributed by neutrino GWs dominant 
 in the lower-frequency bands ($\lesssim$ 100 Hz), 
the total GW spectrum tends not to be a single peak but
 rather flat over a broad frequency up to $\sim 1$kHz. 
For detecting these GW signals for a Galactic supernova source 
with a good signal-to-noise ratio, we need next-generation detectors 
such as the advanced LIGO, LCGT, and the Fabry-Perot type DECIGO (e.g., the 
  left panels in Table 3 (right-hand)). As a side remark, 
 neutrino signals at this non-linear phase change also stochastically with time, which 
  could be detectable for a Galactic supernova in a currently running 
 detector such as by IceCube \citep{brandt,marek_gw,lund}. 
 When material behind the stalled shock successfully absorbs enough neutrino energy 
to be gravitationally unbound from the iron core, a powerful neutrino-driven
 explosion with its explosion
 energy exceeding $10^{51}$ erg will be obtained (section \ref{s2}).
 In this case, the GW waveforms would imprint the information
 of the blast morphologies (e.g., section \ref{exp1}).

If a precollapse core rotates enough rapidly (typically initial rotation period
 less than $\sim$ 4 s) with strong magnetic fields (higher than $\sim 10^{11}$ G) imposed
 initially, 
  the MHD mechanism can produce bipolar explosions along the rotational axis predominantly by the field wrapping mechanism (section
 \ref{mhd_gw})\footnote{If the MRI can be sufficiently resolved in global simulations, 
MHD outflows might be produced even for more weakly magnetized cores because the MRI can 
 exponentially amplify the initial magnetic fields to a dynamically important strength 
 within several rotational periods (e.g., collective references in Obergaulinger
 et al. (2009) \cite{MRI}).}. The GW signals
 in energetic MHD explosions exceeding $10^{51}$ erg, are characterized by a 
burst-like bounce signal plus a secularly growing tail (indicated by 
"Bounce" with "tail" signals in the top right panel in Table 3, also 
section \ref{mhd_gw}). Likewise the neutrino GWs in 
 the case of neutrino-driven explosions, the tail component makes the total GW spectrum 
 broad-band. Going beyond 2D, recent 3D MHD simulations show a strong 
 evidence that nonaxisymmetric rotational instabilities can be a source of strong 
 GW emission due to its periodicity and bar-mode character 
(see a sharp peak near $\sim$ 1kHz in the bottom right panel in 
 Table 3). Note that the pole to equator anisotropy of the shock propagation 
in the MHD explosions could lead to 
 a sudden decrease in the SuperKamiokande events through the MSW flavor conversion 
($\sim 2500$ events for a Galactic source). A planned joint analysis of neutrino and GW 
data \cite{leonor} would provide a powerful probe especially into the 
MHD-driven explosions that imprint these peculiar signatures distinct from
 other candidate explosion mechanisms.
Note also that MHD explosions could be a possible $r$-process cite
 because the mass ejection
  from the iron core takes place in much shorter timescales than in 
the delayed neutrino-heating mechanism, which prevents the ejecta from being 
 proton-rich by neutrino capture reactions \cite{nishi06,fujimoto}\footnote{See 
collective references in \cite{wanajo1,wanajo2} for other plausible $r$-process cites.}. 
As mentioned in the final part of section \ref{mhd_gw}, these interesting  
possibilities have been proposed so far by the numerical simulations with a crude 
treatment of neutrino transport, so that they are now awaited to be re-examined. 

 One important notice here is that the explosion energies 
 obtained in some of the first-principle 2D simulations (Table 1 in section \ref{s2}) 
are underpowered by one or two orders of magnitudes to explain 
the canonical supernova kinetic energy ($\sim 10^{51}$ erg).
Moreover, the softer nuclear EOS, such as of the
Lattimer-Swesty (LS) \cite{latt91} EOS with an incompressibility at nuclear
densities, $K$, of 180 MeV, has been employed in all the 2D simulations 
 that succeeded in producing neutrino-driven explosions\footnote{2D explosion was 
obtained in the MPA simulations for a non-rotating 
 11.2 $M_{\odot}$ for the Shen EOS ($K=281$MeV) (H-T. Janka, private 
communication).}. On top
of a striking evidence that favors a stiffer EOS based on the
nuclear experimental data ($K=240 \pm 20$ MeV, \cite{shlo06}), 
the soft EOS may not account
for the recently observed massive neutron star of $\sim 2 M_{\odot}$
\cite{demo10} 
\cite[see the maximum mass for the LS180 EOS in][]{oconnor,kiuc08}. 
  What is then missing furthermore?  To seek the answer is equal to making 
 the GW prediction more accurate. We may get the answer by going to 
 the first-principle 3D simulations or by taking into account new ingredients,
such as exotic physics in the core of the protoneutron star
\cite{sage09,tobi11}, viscous heating by the magnetorotational instability
\cite{thom05,masa11}, or energy dissipation via Alfv\'en waves
\cite{suzu08}. General relativity (GR) so far treated in a very approximate 
 way in multi-D models might help explosions if the advantage of the GR 
(i.e., higher neutrino 
 energies) dominates over the disadvantage (i.e., shorter advection timescale).
 3D effects might help neutrino-driven explosions as have been recently pointed out
 by 3D simulations with a spectral neutrino transport \cite{taki11}.
 Assisted by a growing computer power, all these issues will be tested 
  soon by the forth-coming first-principle simulations. 

In addition to these advances in numerical simulations, 
  the physical understanding of fundamental problems in the supernova theory 
 is progressing at the same pace. What determines the saturation levels of 
 SASI?  A careful analysis on the parasitic instabilities has been 
 reported to answer this question \cite{jerome}. What determines mode couplings between 
the small-scale convection eddies and the large scale SASI modes?  To apply the theory
 of turbulence \cite{murphy11} might put a milestone to address this question. 
The physical understanding of these phenomena should also give us a deep understanding 
 of the GW features from the SASI and convection.
 Although it is also numerically challenging to resolve the small scale MRI modes 
in global simulations, it is very interesting to see their outcome in
 the neutrino-driven mechanism with rapid rotation. 
 Besides the two representative EOSs of LS and Shen, new sets of 
 EOSs have been recently reported \cite{hempel,furusawa,gshen}. Can 
 we extract the information of the different EOSs by the GW observation ?
 All of these questions may not be easy to answer immediately, but they
 should indeed provide us a precious chance to solve them if not for the GW astronomy.

If the neutrino-heating mechanism (or some other mechanisms)
 fails to explode massive stars,
 central PNSs collapse to BHs. Recent GR simulations \cite{ott2011}  
  indicate that the significant GW emission is associated at the moment of
 the BH formation, which can be 
a promising target of the advanced LIGO for a Galactic source. 
As pointed out in \cite{sumi}, disappearing 
 neutrino signals could also tell us the epoch of BH formations for a Galactic
 source. When quarks and pions could appear in a collapsing core,
 one may observe a neutrino burst induced by the recollapse of the PNS,
 which is triggered by the sudden softening of
 the EOS \cite{sage09}. The intervals between bounce and the BH formation 
  depend on the details of exotic physics in the super-dense core
 \cite{naka1,naka2}. All of these observational signatures could be an important probe 
 into the so-called dense QCD region in the QCD phase diagram (e.g., \cite{hatsuda} 
 for recent review) to which lattice 
 calculations are hardly accessible at present. 

Finally, what is about the story if the MHD mechanism fails ? 
 The central cores with significant angular momentum would collapse into a BH. 
Neutrinos emitted from the accretion disk heat matter in the polar funnel region 
 \cite{paz90,mezree} or strong magnetic fields in the
cores of order of $10^{15}$ G play also an active role both for driving 
the magneto-driven jets and for extracting a significant amount of
energy from the BH (e.g.,
\cite{whee00,thom04,uzde07a} and see references therein). 
  This picture, often referred as collapsar \cite{woos93,pacz98,macf99}, has been
 the working hypothesis as a central engine of long-duration gamma-ray bursts for 
these 20 years (see references in \cite{dai98,thom04,uzde07a,bucci,metzger}
 for other candidate mechanism including magnetar models).
 It is indeed a big issue whether the formation
 of outflows predominantly proceeds via the neutrino-heating mechanism 
  or MHD processes (see collective references in \cite{harikae_a,harikae_b,hari} for example). Comparing to various findings illustrated in Table 1, much 
little things are known about the BH-forming supernovae and about 
 the collapsar.
 This may be partly because the requirement for the numerical modeling to this end is 
highly computationally expensive, which necessitates not only GRMHD simulations for 
handling the BH formation, but also multi-angle neutrino transfer for
treating highly anisotropic neutrino radiation from the accretion disks.

To get a unified picture of massive stellar death in view of GW astronomy,
 we need to be able to draw a 
 schematic picture (like Table 3) also in the case of the BH forming supernovae. 
 A documentary film recording our endeavours to make our dream of  
 "GW astronomy of massive stellar core-collapse" come true, 
seems not to come to an end immediately and is becoming even longer by taking into 
account the BH forming supernovae. In addition, a new horizon 
 is now being opening up towards a multi-messenger astronomy by which 
 we could have much more information to decipher the central engines of 
 massive stars from the 
 combined analysis between GWs, neutrinos, and electromagnetic messengers.
  For example, optical observations using a large-aperture telescope such as 
the Subaru telescope with a detailed 
spectropolarimetric technique (e.g., \cite{tanaka}) is expected to 
 give us an indication 
of the explosion geometry, which could be also helpful to constrain the 
asphericity of the central engines. Needless to say, 
it is of crucial importance
 to accurately determine nucleosynthesis in the SN ejecta  
 for modeling supernova light curves (see, \cite{woosh,thielemann} for recent reviews).
 Note that these studies are now able to be updated by using recent results obtained 
 in the first-principle simulations as mentioned in section \ref{s2}.
 Here we would like to lay down our pen hoping 
 that in the near future a great progress will be made in our understanding of 
a number of exciting and unsettled issues raised in this article.

\vspace{5pt}
\noindent
{\bf Acknowledgements:} 
 The author wishes to thank his collaborators, Y. Suwa, T. Takiwaki, 
W. Iwakami-Nakano,
 M. Liebend\"orfer, S. Yamada, N. Ohnishi, K. Sato, M. Hashimoto, S. Harikae, N. Yasutake, N. Nishimura, K. Shaku,  
K. Sumiyoshi, and H. Suzuki. He would like 
 to acknowledge helpful exchanges and/or permission to reprint figures with E. M\"uller, 
H.-T. Janka, M. Shibata, C. Fryer, C.D. Ott, M. Obergaulinger, J. Murphy, S. Scheidegger,
 K. Yakunin, T. Foglizzo, J. Novak, S. Ando, P. Cerd\'a-Dur\'an,
 R. Fern\'andez, E. O'Connor, T. Font, J.M. Iba\~nez, K. Nakamura, and T. Kuroda. 
 It is also pleasure to thank K. Sato for continuing encouragements. 
Finally he would like to thank his family, his wife Ai and two-year-old daughter Yuu.
Numerical computations in \cite{suwa,kotake09,taki_kota,kotake11}
 were carried out in part on XT4 and 
general common use computer system at the center for Computational Astrophysics, CfCA, 
the National Astronomical Observatory of Japan.  This 
study was supported in part by the Grants-in-Aid for the Scientific Research 
from the Ministry of Education, Science and Culture of Japan (Nos. 19540309, 20740150,
  and 23540323) and by HPCI Strategic Program of Japanese MEXT.




\bibliographystyle{elsarticle-num} 
\bibliography{ms}

\begin{thebibliography}{100}
\expandafter\ifx\csname url\endcsname\relax
  \def\url#1{\texttt{#1}}\fi
\expandafter\ifx\csname urlprefix\endcsname\relax\def\urlprefix{URL }\fi
\expandafter\ifx\csname href\endcsname\relax
  \def\href#1#2{#2} \def\path#1{#1}\fi

\bibitem{ando_new}
S.~{Ando}, J.~F. {Beacom}, H.~{Y{\"u}ksel}, {Detection of Neutrinos from
  Supernovae in Nearby Galaxies}, Physical Review Letters 95~(17) (2005)
  171101--+.
\newblock \href {http://arxiv.org/abs/arXiv:astro-ph/0503321}
  {\path{arXiv:arXiv:astro-ph/0503321}}, \href
  {http://dx.doi.org/10.1103/PhysRevLett.95.171101}
  {\path{doi:10.1103/PhysRevLett.95.171101}}.

\bibitem{hirata1987}
K.~{Hirata}, T.~{Kajita}, M.~{Koshiba}, M.~{Nakahata}, Y.~{Oyama}, {Observation
  of a neutrino burst from the supernova SN1987A}, Physical Review Letters 58
  (1987) 1490--1493.
\newblock \href {http://dx.doi.org/10.1103/PhysRevLett.58.1490}
  {\path{doi:10.1103/PhysRevLett.58.1490}}.

\bibitem{bionta1987}
R.~M. {Bionta}, G.~{Blewitt}, C.~B. {Bratton}, D.~{Casper}, A.~{Ciocio},
  {Observation of a neutrino burst in coincidence with supernova 1987A in the
  Large Magellanic Cloud}, Physical Review Letters 58 (1987) 1494--1496.
\newblock \href {http://dx.doi.org/10.1103/PhysRevLett.58.1494}
  {\path{doi:10.1103/PhysRevLett.58.1494}}.

\bibitem{Sato-and-Suzuki}
K.~{Sato}, H.~{Suzuki}, {Analysis of the neutrino burst from supernova 1987A in
  the Large Magellanic Cloud}, Physical Review Letters 58 (1987) 2722--2725.
\newblock \href {http://dx.doi.org/10.1103/PhysRevLett.58.2722}
  {\path{doi:10.1103/PhysRevLett.58.2722}}.

\bibitem{raffelt_review}
G.~G. {Raffelt}, {Physics with supernovae}, Nuclear Physics B Proceedings
  Supplements 110 (2002) 254--267.
\newblock \href {http://arxiv.org/abs/arXiv:hep-ph/0201099}
  {\path{arXiv:arXiv:hep-ph/0201099}}, \href
  {http://dx.doi.org/10.1016/S0920-5632(02)80132-3}
  {\path{doi:10.1016/S0920-5632(02)80132-3}}.

\bibitem{totsuka}
Y.~{Totsuka}, {Neutrino astronomy}, Reports on Progress in Physics 55 (1992)
  377--430.
\newblock \href {http://dx.doi.org/10.1088/0034-4885/55/3/002}
  {\path{doi:10.1088/0034-4885/55/3/002}}.

\bibitem{totani}
T.~{Totani}, K.~{Sato}, H.~E. {Dalhed}, J.~R. {Wilson}, {Future Detection of
  Supernova Neutrino Burst and Explosion Mechanism}, \apj 496 (1998) 216--+.
\newblock \href {http://arxiv.org/abs/arXiv:astro-ph/9710203}
  {\path{arXiv:arXiv:astro-ph/9710203}}, \href
  {http://dx.doi.org/10.1086/305364} {\path{doi:10.1086/305364}}.

\bibitem{beacom99}
J.~F. {Beacom}, {Supernovae and neutrinos}, Nuclear Physics B Proceedings
  Supplements 118 (2003) 307--314.
\newblock \href {http://arxiv.org/abs/arXiv:astro-ph/0209136}
  {\path{arXiv:arXiv:astro-ph/0209136}}, \href
  {http://dx.doi.org/10.1016/S0920-5632(03)01323-9}
  {\path{doi:10.1016/S0920-5632(03)01323-9}}.

\bibitem{raffelt_2010}
G.~G. {Raffelt}, {Physics opportunities with supernova neutrinos}, Progress in
  Particle and Nuclear Physics 64 (2010) 393--399.
\newblock \href {http://dx.doi.org/10.1016/j.ppnp.2009.12.057}
  {\path{doi:10.1016/j.ppnp.2009.12.057}}.

\bibitem{firstligonew}
M.~{Abbott et al.}, {Upper limits from the LIGO and TAMA detectors on the rate
  of gravitational-wave bursts}, \prd 72~(12) (2005) 122004--+.
\newblock \href {http://arxiv.org/abs/arXiv:gr-qc/0507081}
  {\path{arXiv:arXiv:gr-qc/0507081}}, \href
  {http://dx.doi.org/10.1103/PhysRevD.72.122004}
  {\path{doi:10.1103/PhysRevD.72.122004}}.

\bibitem{tamanew}
M.~{Ando}, {the TAMA Collaboration}, {Current status of the TAMA300
  gravitational-wave detector}, Classical and Quantum Gravity 22 (2005) 881--+.
\newblock \href {http://dx.doi.org/10.1088/0264-9381/22/18/S02}
  {\path{doi:10.1088/0264-9381/22/18/S02}}.

\bibitem{hough}
J.~{Hough}, S.~{Rowan}, B.~S. {Sathyaprakash}, {The search for gravitational
  waves}, Journal of Physics B Atomic Molecular Physics 38 (2005) 497--+.
\newblock \href {http://arxiv.org/abs/arXiv:gr-qc/0501007}
  {\path{arXiv:arXiv:gr-qc/0501007}}, \href
  {http://dx.doi.org/10.1088/0953-4075/38/9/004}
  {\path{doi:10.1088/0953-4075/38/9/004}}.

\bibitem{vandel}
M.~{van der Sluys}, {Gravitational waves from compact binaries}, ArXiv
  e-prints\href {http://arxiv.org/abs/1108.1307} {\path{arXiv:1108.1307}}.

\bibitem{harry}
G.~M. {Harry}, {the LIGO Scientific Collaboration}, {Advanced LIGO: the next
  generation of gravitational wave detectors}, Classical and Quantum Gravity
  27~(8) (2010) 084006--+.
\newblock \href {http://dx.doi.org/10.1088/0264-9381/27/8/084006}
  {\path{doi:10.1088/0264-9381/27/8/084006}}.

\bibitem{lcgt}
K.~{Kuroda}, {LCGT Collaboration}, {Status of LCGT}, Classical and Quantum
  Gravity 27~(8) (2010) 084004--+.
\newblock \href {http://dx.doi.org/10.1088/0264-9381/27/8/084004}
  {\path{doi:10.1088/0264-9381/27/8/084004}}.

\bibitem{schutz}
B.~S. {Sathyaprakash}, B.~F. {Schutz}, {Physics, Astrophysics and Cosmology
  with Gravitational Waves}, Living Reviews in Relativity 12 (2009) 2--+.
\newblock \href {http://arxiv.org/abs/0903.0338} {\path{arXiv:0903.0338}}.

\bibitem{faber}
J.~{Faber}, {Status of neutron star-black hole and binary neutron star
  simulations}, Classical and Quantum Gravity 26~(11) (2009) 114004--+.
\newblock \href {http://dx.doi.org/10.1088/0264-9381/26/11/114004}
  {\path{doi:10.1088/0264-9381/26/11/114004}}.

\bibitem{duez}
M.~D. {Duez}, {Numerical relativity confronts compact neutron star binaries: a
  review and status report}, Classical and Quantum Gravity 27~(11) (2010)
  114002--+.
\newblock \href {http://arxiv.org/abs/0912.3529} {\path{arXiv:0912.3529}},
  \href {http://dx.doi.org/10.1088/0264-9381/27/11/114002}
  {\path{doi:10.1088/0264-9381/27/11/114002}}.

\bibitem{nils1}
N.~{Andersson}, {TOPICAL REVIEW: Gravitational waves from instabilities in
  relativistic stars}, Classical and Quantum Gravity 20 (2003) 105--+.
\newblock \href {http://arxiv.org/abs/arXiv:astro-ph/0211057}
  {\path{arXiv:arXiv:astro-ph/0211057}}.

\bibitem{horowitz}
C.~J. {Horowitz}, {Multi-messenger observations of neutron rich matter}, ArXiv
  e-prints\href {http://arxiv.org/abs/1106.1661} {\path{arXiv:1106.1661}}.

\bibitem{kota06}
K.~{Kotake}, K.~{Sato}, K.~{Takahashi}, {Explosion mechanism, neutrino burst
  and gravitational wave in core-collapse supernovae.}, Reports of Progress in
  Physics 69 (2006) 971--1143.
\newblock \href {http://arxiv.org/abs/arXiv:astro-ph/0509456}
  {\path{arXiv:arXiv:astro-ph/0509456}}.

\bibitem{ott_rev}
C.~D. {Ott}, {TOPICAL REVIEW: The gravitational-wave signature of core-collapse
  supernovae}, Classical and Quantum Gravity 26~(6) (2009) 063001--+.
\newblock \href {http://arxiv.org/abs/0809.0695} {\path{arXiv:0809.0695}},
  \href {http://dx.doi.org/10.1088/0264-9381/26/6/063001}
  {\path{doi:10.1088/0264-9381/26/6/063001}}.

\bibitem{fryer11}
C.~L. {Fryer}, K.~C.~B. {New}, {Gravitational Waves from Gravitational
  Collapse}, Living Reviews in Relativity 14 (2011) 1--+.

\bibitem{shap83}
S.~L. {Shapiro}, S.~A. {Teukolsky}, {Black holes, white dwarfs, and neutron
  stars: The physics of compact objects}, Research supported by the National
  Science Foundation.~New York, Wiley-Interscience, 1983, 663 p., 1983.

\bibitem{wang01}
L.~{Wang}, D.~A. {Howell}, P.~{H{\"o}flich}, J.~C. {Wheeler}, {Bipolar
  Supernova Explosions}, Astrophys. J. 550 (2001) 1030--1035.
\newblock \href {http://dx.doi.org/10.1086/319822} {\path{doi:10.1086/319822}}.

\bibitem{wang02}
L.~{Wang}, J.~C. {Wheeler}, P.~{H{\"o}flich}, A.~{Khokhlov}, D.~{Baade},
  D.~{Branch}, P.~{Challis}, A.~V. {Filippenko}, C.~{Fransson}, P.~{Garnavich},
  R.~P. {Kirshner}, P.~{Lundqvist}, R.~{McCray}, N.~{Panagia}, C.~S.~J. {Pun},
  M.~M. {Phillips}, G.~{Sonneborn}, N.~B. {Suntzeff}, {The Axisymmetric Ejecta
  of Supernova 1987A}, Astrophys. J. 579 (2002) 671--677.
\newblock \href {http://dx.doi.org/10.1086/342824} {\path{doi:10.1086/342824}}.

\bibitem{maeda08}
K.~{Maeda}, K.~{Kawabata}, P.~A. {Mazzali}, M.~{Tanaka}, S.~{Valenti},
  K.~{Nomoto}, T.~{Hattori}, J.~{Deng}, E.~{Pian}, S.~{Taubenberger}, M.~{Iye},
  T.~{Matheson}, A.~V. {Filippenko}, K.~{Aoki}, G.~{Kosugi}, Y.~{Ohyama},
  T.~{Sasaki}, T.~{Takata}, {Asphericity in Supernova Explosions from Late-Time
  Spectroscopy}, Science 319 (2008) 1220--.
\newblock \href {http://arxiv.org/abs/0801.1100} {\path{arXiv:0801.1100}},
  \href {http://dx.doi.org/10.1126/science.1149437}
  {\path{doi:10.1126/science.1149437}}.

\bibitem{tanaka}
M.~{Tanaka}, K.~S. {Kawabata}, K.~{Maeda}, M.~{Iye}, T.~{Hattori}, E.~{Pian},
  K.~{Nomoto}, P.~A. {Mazzali}, N.~{Tominaga}, {Spectropolarimetry of the
  Unique Type Ib Supernova 2005bf: Larger Asymmetry Revealed by Later-Phase
  Data}, Astrophys. J. 699 (2009) 1119--1124.
\newblock \href {http://arxiv.org/abs/0906.1062} {\path{arXiv:0906.1062}},
  \href {http://dx.doi.org/10.1088/0004-637X/699/2/1119}
  {\path{doi:10.1088/0004-637X/699/2/1119}}.

\bibitem{87a}
K.~{Kj{\ae}r}, B.~{Leibundgut}, C.~{Fransson}, A.~{Jerkstrand},
  J.~{Spyromilio}, {The 3-D structure of SN 1987A's inner ejecta}, \aap 517
  (2010) A51+.
\newblock \href {http://arxiv.org/abs/1003.5684} {\path{arXiv:1003.5684}},
  \href {http://dx.doi.org/10.1051/0004-6361/201014538}
  {\path{doi:10.1051/0004-6361/201014538}}.

\bibitem{leonard}
D.~C. {Leonard}, A.~V. {Filippenko}, M.~{Ganeshalingam}, F.~J.~D. {Serduke},
  W.~{Li}, B.~J. {Swift}, A.~{Gal-Yam}, R.~J. {Foley}, D.~B. {Fox}, S.~{Park},
  J.~L. {Hoffman}, D.~S. {Wong}, {A non-spherical core in the explosion of
  supernova SN 2004dj}, \nat 440 (2006) 505--507.
\newblock \href {http://arxiv.org/abs/arXiv:astro-ph/0603297}
  {\path{arXiv:arXiv:astro-ph/0603297}}, \href
  {http://dx.doi.org/10.1038/nature04558} {\path{doi:10.1038/nature04558}}.

\bibitem{leon06}
D.~C. {Leonard}, A.~V. {Filippenko}, M.~{Ganeshalingam}, F.~J.~D. {Serduke},
  W.~{Li}, B.~J. {Swift}, A.~{Gal-Yam}, R.~J. {Foley}, D.~B. {Fox}, S.~{Park},
  J.~L. {Hoffman}, D.~S. {Wong}, {A non-spherical core in the explosion of
  supernova SN 2004dj}, \nat 440 (2006) 505--507.
\newblock \href {http://arxiv.org/abs/astro-ph/0603297}
  {\path{arXiv:astro-ph/0603297}}, \href
  {http://dx.doi.org/10.1038/nature04558} {\path{doi:10.1038/nature04558}}.

\bibitem{khok99}
A.~M. {Khokhlov}, P.~A. {H{\"o}flich}, E.~S. {Oran}, J.~C. {Wheeler},
  L.~{Wang}, A.~Y. {Chtchelkanova}, {Jet-induced Explosions of Core Collapse
  Supernovae}, Astrophys. J. Lett. 524 (1999) L107--L110.
\newblock \href {http://dx.doi.org/10.1086/312305} {\path{doi:10.1086/312305}}.

\bibitem{whee00}
J.~C. {Wheeler}, I.~{Yi}, P.~{H{\"o}flich}, L.~{Wang}, {Asymmetric Supernovae,
  Pulsars, Magnetars, and Gamma-Ray Bursts}, Astrophys. J. 537 (2000) 810--823.
\newblock \href {http://arxiv.org/abs/arXiv:astro-ph/9909293}
  {\path{arXiv:arXiv:astro-ph/9909293}}, \href
  {http://dx.doi.org/10.1086/309055} {\path{doi:10.1086/309055}}.

\bibitem{mikheev1986}
S.~P. {Mikheev}, A.~I. {Smirnov}, {Resonant amplification of neutrino
  oscillations in matter and solar-neutrino spectroscopy}, Nuovo Cimento C
  Geophysics Space Physics C 9 (1986) 17--26.
\newblock \href {http://dx.doi.org/10.1007/BF02508049}
  {\path{doi:10.1007/BF02508049}}.

\bibitem{kneller}
H.~{Duan}, J.~P. {Kneller}, {TOPICAL REVIEW: Neutrino flavour transformation in
  supernovae}, Journal of Physics G Nuclear Physics 36~(11) (2009) 113201--+.
\newblock \href {http://arxiv.org/abs/0904.0974} {\path{arXiv:0904.0974}},
  \href {http://dx.doi.org/10.1088/0954-3899/36/11/113201}
  {\path{doi:10.1088/0954-3899/36/11/113201}}.

\bibitem{jank07}
H.-T. {Janka}, K.~{Langanke}, A.~{Marek}, G.~{Mart{\'{\i}}nez-Pinedo},
  B.~{M{\"u}ller}, {Theory of core-collapse supernovae}, \physrep 442 (2007)
  38--74.
\newblock \href {http://arxiv.org/abs/arXiv:astro-ph/0612072}
  {\path{arXiv:arXiv:astro-ph/0612072}}, \href
  {http://dx.doi.org/10.1016/j.physrep.2007.02.002}
  {\path{doi:10.1016/j.physrep.2007.02.002}}.

\bibitem{burr_rev}
A.~{Burrows}, L.~{Dessart}, C.~D. {Ott}, E.~{Livne}, {Multi-dimensional
  explorations in supernova theory}, \physrep 442 (2007) 23--37.
\newblock \href {http://arxiv.org/abs/arXiv:astro-ph/0612460}
  {\path{arXiv:arXiv:astro-ph/0612460}}, \href
  {http://dx.doi.org/10.1016/j.physrep.2007.02.001}
  {\path{doi:10.1016/j.physrep.2007.02.001}}.

\bibitem{eric1}
E.~{Chassande-Mottin}, M.~{Hendry}, P.~J. {Sutton}, S.~{M{\'a}rka},
  {Multimessenger astronomy with the Einstein Telescope}, General Relativity
  and Gravitation 43 (2011) 437--464.
\newblock \href {http://arxiv.org/abs/1004.1964} {\path{arXiv:1004.1964}},
  \href {http://dx.doi.org/10.1007/s10714-010-1019-z}
  {\path{doi:10.1007/s10714-010-1019-z}}.

\bibitem{szabi}
S.~{M{\'a}rka}, {for the LIGO Scientific Collaboration}, {the Virgo
  Collaboration}, {Transient multimessenger astronomy with gravitational
  waves}, Classical and Quantum Gravity 28~(11) (2011) 114013--+.
\newblock \href {http://dx.doi.org/10.1088/0264-9381/28/11/114013}
  {\path{doi:10.1088/0264-9381/28/11/114013}}.

\bibitem{prad}
T.~{Pradier}, {Antares Collaboration}, {The Antares neutrino telescope and
  multi-messenger astronomy}, Classical and Quantum Gravity 27~(19) (2010)
  194004--+.
\newblock \href {http://arxiv.org/abs/1004.5579} {\path{arXiv:1004.5579}},
  \href {http://dx.doi.org/10.1088/0264-9381/27/19/194004}
  {\path{doi:10.1088/0264-9381/27/19/194004}}.

\bibitem{aso}
Y.~{Aso}, Z.~{M{\'a}rka}, C.~{Finley}, J.~{Dwyer}, K.~{Kotake}, S.~{M{\'a}rka},
  {Search method for coincident events from LIGO and IceCube detectors},
  Classical and Quantum Gravity 25~(11) (2008) 114039--+.
\newblock \href {http://arxiv.org/abs/0711.0107} {\path{arXiv:0711.0107}},
  \href {http://dx.doi.org/10.1088/0264-9381/25/11/114039}
  {\path{doi:10.1088/0264-9381/25/11/114039}}.

\bibitem{colgate}
S.~A. {Colgate}, R.~H. {White}, {The Hydrodynamic Behavior of Supernovae
  Explosions}, \apj 143 (1966) 626--+.
\newblock \href {http://dx.doi.org/10.1086/148549} {\path{doi:10.1086/148549}}.

\bibitem{wils85}
J.~R. {Wilson}, {Supernovae and Post-Collapse Behavior}, in: Numerical
  Astrophysics, 1985, pp. 422--+.

\bibitem{bethe85}
H.~A. {Bethe}, J.~R. {Wilson}, {Revival of a stalled supernova shock by
  neutrino heating}, \apj 295 (1985) 14--23.
\newblock \href {http://dx.doi.org/10.1086/163343} {\path{doi:10.1086/163343}}.

\bibitem{bethe}
H.~A. {Bethe}, {Supernova mechanisms}, Reviews of Modern Physics 62 (1990)
  801--866.
\newblock \href {http://dx.doi.org/10.1103/RevModPhys.62.801}
  {\path{doi:10.1103/RevModPhys.62.801}}.

\bibitem{nordhaus}
J.~{Nordhaus}, A.~{Burrows}, A.~{Almgren}, J.~{Bell}, {Dimension as a Key to
  the Neutrino Mechanism of Core-collapse Supernova Explosions}, \apj 720
  (2010) 694--703.
\newblock \href {http://arxiv.org/abs/1006.3792} {\path{arXiv:1006.3792}},
  \href {http://dx.doi.org/10.1088/0004-637X/720/1/694}
  {\path{doi:10.1088/0004-637X/720/1/694}}.

\bibitem{rampp00}
M.~{Rampp}, H.-T. {Janka}, {Spherically Symmetric Simulation with Boltzmann
  Neutrino Transport of Core Collapse and Postbounce Evolution of a 15 Msun
  Star}, Astrophys. J. Lett. 539 (2000) L33--L36.

\bibitem{lieb01}
M.~{Liebend{\" o}rfer}, A.~{Mezzacappa}, F.~{Thielemann}, {Conservative general
  relativistic radiation hydrodynamics in spherical symmetry and comoving
  coordinates}, \prd 63~(10) (2001) 104003--+.

\bibitem{thom03}
T.~A. {Thompson}, A.~{Burrows}, P.~A. {Pinto}, {Shock Breakout in Core-Collapse
  Supernovae and Its Neutrino Signature}, Astrophys. J. 592 (2003) 434--456.

\bibitem{sumi05}
K.~{Sumiyoshi}, S.~{Yamada}, H.~{Suzuki}, H.~{Shen}, S.~{Chiba}, H.~{Toki},
  {Postbounce Evolution of Core-Collapse Supernovae: Long-Term Effects of the
  Equation of State}, Astrophys. J. 629 (2005) 922--932.
\newblock \href {http://arxiv.org/abs/arXiv:astro-ph/0506620}
  {\path{arXiv:arXiv:astro-ph/0506620}}, \href
  {http://dx.doi.org/10.1086/431788} {\path{doi:10.1086/431788}}.

\bibitem{kitaura}
F.~S. {Kitaura}, H.-T. {Janka}, W.~{Hillebrandt}, {Explosions of O-Ne-Mg cores,
  the Crab supernova, and subluminous type II-P supernovae}, \aap 450 (2006)
  345--350.
\newblock \href {http://arxiv.org/abs/arXiv:astro-ph/0512065}
  {\path{arXiv:arXiv:astro-ph/0512065}}, \href
  {http://dx.doi.org/10.1051/0004-6361:20054703}
  {\path{doi:10.1051/0004-6361:20054703}}.

\bibitem{herant}
M.~{Herant}, W.~{Benz}, W.~R. {Hix}, C.~L. {Fryer}, S.~A. {Colgate}, {Inside
  the supernova: A powerful convective engine}, \apj 435 (1994) 339--361.
\newblock \href {http://arxiv.org/abs/arXiv:astro-ph/9404024}
  {\path{arXiv:arXiv:astro-ph/9404024}}, \href
  {http://dx.doi.org/10.1086/174817} {\path{doi:10.1086/174817}}.

\bibitem{burr95}
A.~{Burrows}, J.~{Hayes}, B.~A. {Fryxell}, {On the Nature of Core-Collapse
  Supernova Explosions}, \apj 450 (1995) 830--+.
\newblock \href {http://arxiv.org/abs/arXiv:astro-ph/9506061}
  {\path{arXiv:arXiv:astro-ph/9506061}}, \href
  {http://dx.doi.org/10.1086/176188} {\path{doi:10.1086/176188}}.

\bibitem{jankamueller96}
H.~{Janka}, E.~{M\"uller}, {Neutrino heating, convection, and the mechanism of
  Type-II supernova explosions.}, \aap 306 (1996) 167--+.

\bibitem{frye02}
C.~L. {Fryer}, D.~E. {Holz}, S.~A. {Hughes}, {Gravitational Wave Emission from
  Core Collapse of Massive Stars}, Astrophys. J. 565 (2002) 430--446.

\bibitem{fryer04a}
C.~L. {Fryer}, {Neutron Star Kicks from Asymmetric Collapse}, Astrophys. J.
  Lett. 601 (2004) L175--L178.

\bibitem{blon03}
J.~M. {Blondin}, A.~{Mezzacappa}, C.~{DeMarino}, {Stability of Standing
  Accretion Shocks, with an Eye toward Core-Collapse Supernovae}, Astrophys. J.
  584 (2003) 971--980.
\newblock \href {http://arxiv.org/abs/astro-ph/0210634}
  {\path{arXiv:astro-ph/0210634}}, \href {http://dx.doi.org/10.1086/345812}
  {\path{doi:10.1086/345812}}.

\bibitem{sche04}
L.~{Scheck}, T.~{Plewa}, H.-T. {Janka}, K.~{Kifonidis}, E.~{M{\"u}ller},
  {Pulsar Recoil by Large-Scale Anisotropies in Supernova Explosions}, Physical
  Review Letters 92~(1) (2004) 011103--+.
\newblock \href {http://arxiv.org/abs/astro-ph/0307352}
  {\path{arXiv:astro-ph/0307352}}.

\bibitem{scheck06}
L.~{Scheck}, K.~{Kifonidis}, H.~{Janka}, E.~{M{\"u}ller}, {Multidimensional
  supernova simulations with approximative neutrino transport. I. Neutron star
  kicks and the anisotropy of neutrino-driven explosions in two spatial
  dimensions}, \aap 457 (2006) 963--986.
\newblock \href {http://arxiv.org/abs/arXiv:astro-ph/0601302}
  {\path{arXiv:arXiv:astro-ph/0601302}}, \href
  {http://dx.doi.org/10.1051/0004-6361:20064855}
  {\path{doi:10.1051/0004-6361:20064855}}.

\bibitem{Ohni05}
N.~{Ohnishi}, K.~{Kotake}, S.~{Yamada}, {Numerical Analysis of Standing
  Accretion Shock Instability with Neutrino Heating in Supernova Cores},
  Astrophys. J. 641 (2006) 1018--1028.
\newblock \href {http://arxiv.org/abs/astro-ph/0509765}
  {\path{arXiv:astro-ph/0509765}}, \href {http://dx.doi.org/10.1086/500554}
  {\path{doi:10.1086/500554}}.

\bibitem{ohnishi07}
N.~{Ohnishi}, K.~{Kotake}, S.~{Yamada}, {Inelastic Neutrino-Helium Scatterings
  and Standing Accretion Shock Instability in Core-Collapse Supernovae}, \apj
  667 (2007) 375--381.
\newblock \href {http://arxiv.org/abs/arXiv:astro-ph/0606187}
  {\path{arXiv:arXiv:astro-ph/0606187}}, \href
  {http://dx.doi.org/10.1086/520755} {\path{doi:10.1086/520755}}.

\bibitem{thierry}
T.~{Foglizzo}, L.~{Scheck}, H.-T. {Janka}, {Neutrino-driven Convection versus
  Advection in Core-Collapse Supernovae}, \apj 652 (2006) 1436--1450.
\newblock \href {http://arxiv.org/abs/arXiv:astro-ph/0507636}
  {\path{arXiv:arXiv:astro-ph/0507636}}, \href
  {http://dx.doi.org/10.1086/508443} {\path{doi:10.1086/508443}}.

\bibitem{murphy08}
J.~W. {Murphy}, A.~{Burrows}, {Criteria for Core-Collapse Supernova Explosions
  by the Neutrino Mechanism}, \apj 688 (2008) 1159--1175.
\newblock \href {http://arxiv.org/abs/0805.3345} {\path{arXiv:0805.3345}},
  \href {http://dx.doi.org/10.1086/592214} {\path{doi:10.1086/592214}}.

\bibitem{rodrigo09}
R.~{Fern{\'a}ndez}, C.~{Thompson}, {Stability of a Spherical Accretion Shock
  with Nuclear Dissociation}, \apj 697 (2009) 1827--1841.
\newblock \href {http://arxiv.org/abs/0811.2795} {\path{arXiv:0811.2795}},
  \href {http://dx.doi.org/10.1088/0004-637X/697/2/1827}
  {\path{doi:10.1088/0004-637X/697/2/1827}}.

\bibitem{rodrigo09_2}
R.~{Fern{\'a}ndez}, C.~{Thompson}, {Dynamics of a Spherical Accretion Shock
  with Neutrino Heating and Alpha-Particle Recombination}, \apj 703 (2009)
  1464--1485.
\newblock \href {http://arxiv.org/abs/0812.4574} {\path{arXiv:0812.4574}},
  \href {http://dx.doi.org/10.1088/0004-637X/703/2/1464}
  {\path{doi:10.1088/0004-637X/703/2/1464}}.

\bibitem{iwakami1}
W.~{Iwakami}, K.~{Kotake}, N.~{Ohnishi}, S.~{Yamada}, K.~{Sawada},
  {Three-Dimensional Simulations of Standing Accretion Shock Instability in
  Core-Collapse Supernovae}, Astrophys. J. 678 (2008) 1207--1222.
\newblock \href {http://arxiv.org/abs/0710.2191} {\path{arXiv:0710.2191}},
  \href {http://dx.doi.org/10.1086/533582} {\path{doi:10.1086/533582}}.

\bibitem{iwakami2}
W.~{Iwakami}, K.~{Kotake}, N.~{Ohnishi}, S.~{Yamada}, K.~{Sawada}, {Effects of
  Rotation on Standing Accretion Shock Instability in Nonlinear Phase for
  Core-Collapse Supernovae}, Astrophys. J. 700 (2009) 232--242.
\newblock \href {http://arxiv.org/abs/0811.0651} {\path{arXiv:0811.0651}},
  \href {http://dx.doi.org/10.1088/0004-637X/700/1/232}
  {\path{doi:10.1088/0004-637X/700/1/232}}.

\bibitem{fern}
R.~{Fern{\'a}ndez}, {The Spiral Modes of the Standing Accretion Shock
  Instability}, \apj 725 (2010) 1563--1580.
\newblock \href {http://arxiv.org/abs/1003.1730} {\path{arXiv:1003.1730}},
  \href {http://dx.doi.org/10.1088/0004-637X/725/2/1563}
  {\path{doi:10.1088/0004-637X/725/2/1563}}.

\bibitem{jank01}
H.-T. {Janka}, {Conditions for shock revival by neutrino heating in
  core-collapse supernovae}, \aap 368 (2001) 527--560.
\newblock \href {http://arxiv.org/abs/arXiv:astro-ph/0008432}
  {\path{arXiv:arXiv:astro-ph/0008432}}, \href
  {http://dx.doi.org/10.1051/0004-6361:20010012}
  {\path{doi:10.1051/0004-6361:20010012}}.

\bibitem{annop1}
A.~{Wongwathanarat}, N.~J. {Hammer}, E.~{M{\"u}ller}, {An axis-free overset
  grid in spherical polar coordinates for simulating 3D self-gravitating
  flows}, \aap 514 (2010) A48+.
\newblock \href {http://arxiv.org/abs/1003.1633} {\path{arXiv:1003.1633}},
  \href {http://dx.doi.org/10.1051/0004-6361/200913435}
  {\path{doi:10.1051/0004-6361/200913435}}.

\bibitem{nomo88}
K.~{Nomoto}, M.~{Mashimoto}, {Presupernova evolution of massive stars.},
  \physrep 163 (1988) 13--36.
\newblock \href {http://dx.doi.org/10.1016/0370-1573(88)90032-4}
  {\path{doi:10.1016/0370-1573(88)90032-4}}.

\bibitem{woos02}
S.~E. {Woosley}, A.~{Heger}, T.~A. {Weaver}, {The evolution and explosion of
  massive stars}, Reviews of Modern Physics 74 (2002) 1015--1071.
\newblock \href {http://dx.doi.org/10.1103/RevModPhys.74.1015}
  {\path{doi:10.1103/RevModPhys.74.1015}}.

\bibitem{WW95}
S.~E. {Woosley}, T.~A. {Weaver}, {The Evolution and Explosion of Massive Stars.
  II. Explosive Hydrodynamics and Nucleosynthesis}, \apjs 101 (1995) 181--+.
\newblock \href {http://dx.doi.org/10.1086/192237} {\path{doi:10.1086/192237}}.

\bibitem{janka08}
H.-T. {Janka}, B.~{M{\"u}ller}, F.~S. {Kitaura}, R.~{Buras}, {Dynamics of shock
  propagation and nucleosynthesis conditions in O-Ne-Mg core supernovae}, \aap
  485 (2008) 199--208.
\newblock \href {http://arxiv.org/abs/0712.4237} {\path{arXiv:0712.4237}},
  \href {http://dx.doi.org/10.1051/0004-6361:20079334}
  {\path{doi:10.1051/0004-6361:20079334}}.

\bibitem{burr06}
A.~{Burrows}, E.~{Livne}, L.~{Dessart}, C.~D. {Ott}, J.~{Murphy}, {A New
  Mechanism for Core-Collapse Supernova Explosions}, Astrophys. J. 640 (2006)
  878--890.
\newblock \href {http://arxiv.org/abs/astro-ph/0510687}
  {\path{arXiv:astro-ph/0510687}}, \href {http://dx.doi.org/10.1086/500174}
  {\path{doi:10.1086/500174}}.

\bibitem{sage09}
I.~{Sagert}, T.~{Fischer}, M.~{Hempel}, G.~{Pagliara}, J.~{Schaffner-Bielich},
  A.~{Mezzacappa}, F.~{Thielemann}, M.~{Liebend{\"o}rfer}, {Signals of the QCD
  Phase Transition in Core-Collapse Supernovae}, Physical Review Letters
  102~(8) (2009) 081101--+.
\newblock \href {http://arxiv.org/abs/0809.4225} {\path{arXiv:0809.4225}},
  \href {http://dx.doi.org/10.1103/PhysRevLett.102.081101}
  {\path{doi:10.1103/PhysRevLett.102.081101}}.

\bibitem{buras06}
R.~{Buras}, H.-T. {Janka}, M.~{Rampp}, K.~{Kifonidis}, {Two-dimensional
  hydrodynamic core-collapse supernova simulations with spectral neutrino
  transport. II. Models for different progenitor stars}, \aap 457 (2006)
  281--308.
\newblock \href {http://arxiv.org/abs/arXiv:astro-ph/0512189}
  {\path{arXiv:arXiv:astro-ph/0512189}}, \href
  {http://dx.doi.org/10.1051/0004-6361:20054654}
  {\path{doi:10.1051/0004-6361:20054654}}.

\bibitem{burrows2}
A.~{Burrows}, E.~{Livne}, L.~{Dessart}, C.~D. {Ott}, J.~{Murphy}, {Features of
  the Acoustic Mechanism of Core-Collapse Supernova Explosions}, Astrophys. J.
  655 (2007) 416--433.
\newblock \href {http://arxiv.org/abs/arXiv:astro-ph/0610175}
  {\path{arXiv:arXiv:astro-ph/0610175}}, \href
  {http://dx.doi.org/10.1086/509773} {\path{doi:10.1086/509773}}.

\bibitem{taki11}
T.~{Takiwaki}, K.~{Kotake}, Y.~{Suwa}, {Three-dimensional Hydrodynamic
  Core-Collapse Supernova Simulations for an \$11.2
  M\_$\{$$\backslash$odot$\}$\$ Star with Spectral Neutrino Transport}, ArXiv
  e-prints\href {http://arxiv.org/abs/1108.3989} {\path{arXiv:1108.3989}}.

\bibitem{bruenn}
S.~W. {Bruenn}, A.~{Mezzacappa}, W.~R. {Hix}, J.~M. {Blondin}, P.~{Marronetti},
  O.~E.~B. {Messer}, C.~J. {Dirk}, S.~{Yoshida}, {2D and 3D Core-Collapse
  Supernovae Simulation Results Obtained with the CHIMERA Code}, ArXiv
  e-prints\href {http://arxiv.org/abs/1002.4914} {\path{arXiv:1002.4914}}.

\bibitem{suwa}
Y.~{Suwa}, K.~{Kotake}, T.~{Takiwaki}, S.~C. {Whitehouse},
  M.~{Liebend{\"o}rfer}, K.~{Sato}, {Explosion Geometry of a Rotating 13Msun
  Star Driven by the SASI-Aided Neutrino-Heating Supernova Mechanism}, \pasj 62
  (2010) L49+.
\newblock \href {http://arxiv.org/abs/0912.1157} {\path{arXiv:0912.1157}}.

\bibitem{marek}
A.~{Marek}, H.-T. {Janka}, {Delayed Neutrino-Driven Supernova Explosions Aided
  by the Standing Accretion-Shock Instability}, Astrophys. J. 694 (2009)
  664--696.
\newblock \href {http://arxiv.org/abs/0708.3372} {\path{arXiv:0708.3372}},
  \href {http://dx.doi.org/10.1088/0004-637X/694/1/664}
  {\path{doi:10.1088/0004-637X/694/1/664}}.

\bibitem{cardall}
C.~Y. {Cardall}, {Supernova neutrino challenges}, Nuclear Physics B Proceedings
  Supplements 145 (2005) 295--300.
\newblock \href {http://arxiv.org/abs/arXiv:astro-ph/0502232}
  {\path{arXiv:arXiv:astro-ph/0502232}}, \href
  {http://dx.doi.org/10.1016/j.nuclphysbps.2005.04.026}
  {\path{doi:10.1016/j.nuclphysbps.2005.04.026}}.

\bibitem{hege05}
A.~{Heger}, S.~E. {Woosley}, H.~C. {Spruit}, {Presupernova Evolution of
  Differentially Rotating Massive Stars Including Magnetic Fields}, Astrophys.
  J. 626 (2005) 350--363.
\newblock \href {http://arxiv.org/abs/arXiv:astro-ph/0409422}
  {\path{arXiv:arXiv:astro-ph/0409422}}, \href
  {http://dx.doi.org/10.1086/429868} {\path{doi:10.1086/429868}}.

\bibitem{buras1}
R.~{Buras}, M.~{Rampp}, H.-T. {Janka}, K.~{Kifonidis}, {Two-dimensional
  hydrodynamic core-collapse supernova simulations with spectral neutrino
  transport. I. Numerical method and results for a 15 M{\.o} star}, \aap 447
  (2006) 1049--1092.
\newblock \href {http://arxiv.org/abs/arXiv:astro-ph/0507135}
  {\path{arXiv:arXiv:astro-ph/0507135}}, \href
  {http://dx.doi.org/10.1051/0004-6361:20053783}
  {\path{doi:10.1051/0004-6361:20053783}}.

\bibitem{ott_multi}
C.~D. {Ott}, A.~{Burrows}, L.~{Dessart}, E.~{Livne}, {Two-Dimensional
  Multiangle, Multigroup Neutrino Radiation-Hydrodynamic Simulations of
  Postbounce Supernova Cores}, \apj 685 (2008) 1069--1088.
\newblock \href {http://arxiv.org/abs/0804.0239} {\path{arXiv:0804.0239}},
  \href {http://dx.doi.org/10.1086/591440} {\path{doi:10.1086/591440}}.

\bibitem{brandt}
T.~D. {Brandt}, A.~{Burrows}, C.~D. {Ott}, E.~{Livne}, {Results from
  Core-collapse Simulations with Multi-dimensional, Multi-angle Neutrino
  Transport}, \apj 728 (2011) 8--+.
\newblock \href {http://arxiv.org/abs/1009.4654} {\path{arXiv:1009.4654}},
  \href {http://dx.doi.org/10.1088/0004-637X/728/1/8}
  {\path{doi:10.1088/0004-637X/728/1/8}}.

\bibitem{bruenn85}
S.~W. {Bruenn}, {Stellar core collapse - Numerical model and infall epoch},
  \apjs 58 (1985) 771--841.
\newblock \href {http://dx.doi.org/10.1086/191056} {\path{doi:10.1086/191056}}.

\bibitem{yakunin}
K.~N. {Yakunin}, P.~{Marronetti}, A.~{Mezzacappa}, S.~W. {Bruenn}, C.~{Lee},
  M.~A. {Chertkow}, W.~R. {Hix}, J.~M. {Blondin}, E.~J. {Lentz}, O.~E. {Bronson
  Messer}, S.~{Yoshida}, {Gravitational waves from core collapse supernovae},
  Classical and Quantum Gravity 27~(19) (2010) 194005--+.
\newblock \href {http://arxiv.org/abs/1005.0779} {\path{arXiv:1005.0779}},
  \href {http://dx.doi.org/10.1088/0264-9381/27/19/194005}
  {\path{doi:10.1088/0264-9381/27/19/194005}}.

\bibitem{idsa}
M.~{Liebend{\"o}rfer}, S.~C. {Whitehouse}, T.~{Fischer}, {The Isotropic
  Diffusion Source Approximation for Supernova Neutrino Transport}, \apj 698
  (2009) 1174--1190.
\newblock \href {http://arxiv.org/abs/0711.2929} {\path{arXiv:0711.2929}},
  \href {http://dx.doi.org/10.1088/0004-637X/698/2/1174}
  {\path{doi:10.1088/0004-637X/698/2/1174}}.

\bibitem{lieb05}
M.~{Liebend{\"o}rfer}, M.~{Rampp}, H.-T. {Janka}, A.~{Mezzacappa}, {Supernova
  Simulations with Boltzmann Neutrino Transport: A Comparison of Methods},
  Astrophys. J. 620 (2005) 840--860.
\newblock \href {http://arxiv.org/abs/arXiv:astro-ph/0310662}
  {\path{arXiv:arXiv:astro-ph/0310662}}, \href
  {http://dx.doi.org/10.1086/427203} {\path{doi:10.1086/427203}}.

\bibitem{hari}
S.~{Harikae}, K.~{Kotake}, T.~{Takiwaki}, Y.-i. {Sekiguchi}, {A General
  Relativistic Ray-tracing Method for Estimating the Energy and Momentum
  Deposition by Neutrino Pair Annihilation in Collapsars}, \apj 720 (2010)
  614--625.
\newblock \href {http://dx.doi.org/10.1088/0004-637X/720/1/614}
  {\path{doi:10.1088/0004-637X/720/1/614}}.

\bibitem{mueller}
B.~{M{\"u}ller}, H.-T. {Janka}, H.~{Dimmelmeier}, {A New Multi-dimensional
  General Relativistic Neutrino Hydrodynamic Code for Core-collapse Supernovae.
  I. Method and Code Tests in Spherical Symmetry}, \apjs 189 (2010) 104--133.
\newblock \href {http://arxiv.org/abs/1001.4841} {\path{arXiv:1001.4841}},
  \href {http://dx.doi.org/10.1088/0067-0049/189/1/104}
  {\path{doi:10.1088/0067-0049/189/1/104}}.

\bibitem{shibata11}
M.~{Shibata}, K.~{Kiuchi}, Y.-i. {Sekiguchi}, Y.~{Suwa}, {Truncated Moment
  Formalism for Radiation Hydrodynamics in Numerical Relativity}, ArXiv
  e-prints\href {http://arxiv.org/abs/1104.3937} {\path{arXiv:1104.3937}}.

\bibitem{hempel}
M.~{Hempel}, J.~{Schaffner-Bielich}, {A statistical model for a complete
  supernova equation of state}, Nuclear Physics A 837 (2010) 210--254.
\newblock \href {http://arxiv.org/abs/0911.4073} {\path{arXiv:0911.4073}},
  \href {http://dx.doi.org/10.1016/j.nuclphysa.2010.02.010}
  {\path{doi:10.1016/j.nuclphysa.2010.02.010}}.

\bibitem{furusawa}
S.~{Furusawa}, S.~{Yamada}, K.~{Sumiyoshi}, H.~{Suzuki}, {A new baryonic
  equation of state at sub-nuclear densities for core-collapse simulations},
  ArXiv e-prints\href {http://arxiv.org/abs/1103.6129}
  {\path{arXiv:1103.6129}}.

\bibitem{gshen}
G.~{Shen}, C.~J. {Horowitz}, E.~{O'Connor}, {Second relativistic mean field and
  virial equation of state for astrophysical simulations}, \prc 83~(6) (2011)
  065808--+.
\newblock \href {http://arxiv.org/abs/1103.5174} {\path{arXiv:1103.5174}},
  \href {http://dx.doi.org/10.1103/PhysRevC.83.065808}
  {\path{doi:10.1103/PhysRevC.83.065808}}.

\bibitem{ewald82}
E.~{M\"uller}, {Gravitational radiation from collapsing rotating stellar
  cores}, \aap 114 (1982) 53--59.

\bibitem{mm}
R.~{M\"{o}nchmeyer}, G.~{Schaefer}, E.~{Mueller}, R.~E. {Kates}, {Gravitational
  waves from the collapse of rotating stellar cores}, Astron. Astrophys. 246
  (1991) 417--440.

\bibitem{yama95}
S.~{Yamada}, K.~{Sato}, {Gravitational Radiation from Rotational Collapse of a
  Supernova Core}, \apj 450 (1995) 245--+.
\newblock \href {http://dx.doi.org/10.1086/176135} {\path{doi:10.1086/176135}}.

\bibitem{zwer97}
T.~{Zwerger}, E.~{M\"uller}, {Dynamics and gravitational wave signature of
  axisymmetric rotational core collapse.}, Astron. Astrophys. 320 (1997)
  209--227.

\bibitem{kotakegw}
K.~{Kotake}, S.~{Yamada}, K.~{Sato}, {Gravitational radiation from axisymmetric
  rotational core collapse}, \prd 68~(4) (2003) 044023--+.

\bibitem{kota04a}
K.~{Kotake}, S.~{Yamada}, K.~{Sato}, K.~{Sumiyoshi}, H.~{Ono}, H.~{Suzuki},
  {Gravitational radiation from rotational core collapse: Effects of magnetic
  fields and realistic equations of state}, \prd 69~(12) (2004) 124004--+.

\bibitem{shibaseki}
M.~{Shibata}, Y.-I. {Sekiguchi}, {Gravitational waves from axisymmetric
  rotating stellar core collapse to a neutron star in full general relativity},
  \prd 69~(8) (2004) 084024--+.
\newblock \href {http://arxiv.org/abs/arXiv:gr-qc/0402040}
  {\path{arXiv:arXiv:gr-qc/0402040}}, \href
  {http://dx.doi.org/10.1103/PhysRevD.69.084024}
  {\path{doi:10.1103/PhysRevD.69.084024}}.

\bibitem{ott}
C.~D. {Ott}, A.~{Burrows}, E.~{Livne}, R.~{Walder}, {Gravitational Waves from
  Axisymmetric, Rotating Stellar Core Collapse}, Astrophys. J. 600 (2004)
  834--864.
\newblock \href {http://arxiv.org/abs/arXiv:astro-ph/0307472}
  {\path{arXiv:arXiv:astro-ph/0307472}}, \href
  {http://dx.doi.org/10.1086/379822} {\path{doi:10.1086/379822}}.

\bibitem{ott_prl}
C.~D. {Ott}, H.~{Dimmelmeier}, A.~{Marek}, H.~{Janka}, I.~{Hawke}, B.~{Zink},
  E.~{Schnetter}, {3D Collapse of Rotating Stellar Iron Cores in General
  Relativity Including Deleptonization and a Nuclear Equation of State},
  Physical Review Letters 98~(26) (2007) 261101--+.
\newblock \href {http://arxiv.org/abs/arXiv:astro-ph/0609819}
  {\path{arXiv:arXiv:astro-ph/0609819}}, \href
  {http://dx.doi.org/10.1103/PhysRevLett.98.261101}
  {\path{doi:10.1103/PhysRevLett.98.261101}}.

\bibitem{ott_2007}
C.~D. {Ott}, H.~{Dimmelmeier}, A.~{Marek}, H.~{Janka}, B.~{Zink}, I.~{Hawke},
  E.~{Schnetter}, {Rotating collapse of stellar iron cores in general
  relativity}, Classical and Quantum Gravity 24 (2007) 139--+.
\newblock \href {http://arxiv.org/abs/arXiv:astro-ph/0612638}
  {\path{arXiv:arXiv:astro-ph/0612638}}, \href
  {http://dx.doi.org/10.1088/0264-9381/24/12/S10}
  {\path{doi:10.1088/0264-9381/24/12/S10}}.

\bibitem{dimm02}
H.~{Dimmelmeier}, J.~A. {Font}, E.~{M{\" u}ller}, {Relativistic simulations of
  rotational core collapse II. Collapse dynamics and gravitational radiation},
  Astron. Astrophys. 393 (2002) 523--542.

\bibitem{dimmelprl}
H.~{Dimmelmeier}, C.~D. {Ott}, H.-T. {Janka}, A.~{Marek}, E.~{M{\"u}ller},
  {Generic Gravitational-Wave Signals from the Collapse of Rotating Stellar
  Cores}, Physical Review Letters 98~(25) (2007) 251101--+.
\newblock \href {http://arxiv.org/abs/arXiv:astro-ph/0702305}
  {\path{arXiv:arXiv:astro-ph/0702305}}, \href
  {http://dx.doi.org/10.1103/PhysRevLett.98.251101}
  {\path{doi:10.1103/PhysRevLett.98.251101}}.

\bibitem{dimm08}
H.~{Dimmelmeier}, C.~D. {Ott}, A.~{Marek}, H.~{Janka}, {Gravitational wave
  burst signal from core collapse of rotating stars}, \prd 78~(6) (2008)
  064056--+.
\newblock \href {http://arxiv.org/abs/0806.4953} {\path{arXiv:0806.4953}},
  \href {http://dx.doi.org/10.1103/PhysRevD.78.064056}
  {\path{doi:10.1103/PhysRevD.78.064056}}.

\bibitem{simon}
S.~{Scheidegger}, T.~{Fischer}, S.~C. {Whitehouse}, M.~{Liebend{\"o}rfer},
  {Gravitational waves from 3D MHD core collapse simulations}, Astron.
  Astrophys. 490 (2008) 231--241.
\newblock \href {http://arxiv.org/abs/0709.0168} {\path{arXiv:0709.0168}},
  \href {http://dx.doi.org/10.1051/0004-6361:20078577}
  {\path{doi:10.1051/0004-6361:20078577}}.

\bibitem{simon1}
S.~{Scheidegger}, R.~{K{\"a}ppeli}, S.~C. {Whitehouse}, T.~{Fischer},
  M.~{Liebend{\"o}rfer}, {The influence of model parameters on the prediction
  of gravitational wave signals from stellar core collapse}, \aap 514 (2010)
  A51+.
\newblock \href {http://dx.doi.org/10.1051/0004-6361/200913220}
  {\path{doi:10.1051/0004-6361/200913220}}.

\bibitem{matthias_dep}
M.~{Liebend{\"o}rfer}, {A Simple Parameterization of the Consequences of
  Deleptonization for Simulations of Stellar Core Collapse}, \apj 633 (2005)
  1042--1051.
\newblock \href {http://arxiv.org/abs/arXiv:astro-ph/0504072}
  {\path{arXiv:arXiv:astro-ph/0504072}}, \href
  {http://dx.doi.org/10.1086/466517} {\path{doi:10.1086/466517}}.

\bibitem{woos06}
S.~E. {Woosley}, A.~{Heger}, {The Progenitor Stars of Gamma-Ray Bursts},
  Astrophys. J. 637 (2006) 914--921.
\newblock \href {http://arxiv.org/abs/arXiv:astro-ph/0508175}
  {\path{arXiv:arXiv:astro-ph/0508175}}, \href
  {http://dx.doi.org/10.1086/498500} {\path{doi:10.1086/498500}}.

\bibitem{yoon}
S.-C. {Yoon}, N.~{Langer}, {Evolution of rapidly rotating metal-poor massive
  stars towards gamma-ray bursts}, Astron. Astrophys. 443 (2005) 643--648.
\newblock \href {http://arxiv.org/abs/arXiv:astro-ph/0508242}
  {\path{arXiv:arXiv:astro-ph/0508242}}, \href
  {http://dx.doi.org/10.1051/0004-6361:20054030}
  {\path{doi:10.1051/0004-6361:20054030}}.

\bibitem{ott_birth}
C.~D. {Ott}, A.~{Burrows}, T.~A. {Thompson}, E.~{Livne}, R.~{Walder}, {The Spin
  Periods and Rotational Profiles of Neutron Stars at Birth}, Astrophys. J.,
  Suppl. Ser. 164 (2006) 130--155.
\newblock \href {http://arxiv.org/abs/arXiv:astro-ph/0508462}
  {\path{arXiv:arXiv:astro-ph/0508462}}, \href
  {http://dx.doi.org/10.1086/500832} {\path{doi:10.1086/500832}}.

\bibitem{muyan97}
E.~{M\"{u}ler}, H.-T. {Janka}, {Gravitational radiation from convective
  instabilities in Type II supernova explosions.}, Astron. Astrophys. 317
  (1997) 140--163.

\bibitem{burohey}
A.~{Burrows}, J.~{Hayes}, {Pulsar Recoil and Gravitational Radiation Due to
  Asymmetrical Stellar Collapse and Explosion}, Physical Review Letters 76
  (1996) 352--355.
\newblock \href {http://arxiv.org/abs/arXiv:astro-ph/9511106}
  {\path{arXiv:arXiv:astro-ph/9511106}}, \href
  {http://dx.doi.org/10.1103/PhysRevLett.76.352}
  {\path{doi:10.1103/PhysRevLett.76.352}}.

\bibitem{fryersingle}
C.~L. {Fryer}, {Neutron Star Kicks from Asymmetric Collapse}, Astrophys. J.
  Lett. 601 (2004) L175--L178.
\newblock \href {http://arxiv.org/abs/arXiv:astro-ph/0312265}
  {\path{arXiv:arXiv:astro-ph/0312265}}, \href
  {http://dx.doi.org/10.1086/382044} {\path{doi:10.1086/382044}}.

\bibitem{mueller04}
E.~{M{\"u}ller}, M.~{Rampp}, R.~{Buras}, H.-T. {Janka}, D.~H. {Shoemaker},
  {Toward Gravitational Wave Signals from Realistic Core-Collapse Supernova
  Models}, Astrophys. J. 603 (2004) 221--230.
\newblock \href {http://arxiv.org/abs/arXiv:astro-ph/0309833}
  {\path{arXiv:arXiv:astro-ph/0309833}}, \href
  {http://dx.doi.org/10.1086/381360} {\path{doi:10.1086/381360}}.

\bibitem{rampp}
M.~{Rampp}, E.~{Mueller}, M.~{Ruffert}, {Simulations of non-axisymmetric
  rotational core collapse}, \aap 332 (1998) 969--983.
\newblock \href {http://arxiv.org/abs/arXiv:astro-ph/9711122}
  {\path{arXiv:arXiv:astro-ph/9711122}}.

\bibitem{ott_3d}
C.~D. {Ott}, H.~{Dimmelmeier}, A.~{Marek}, H.-T. {Janka}, I.~{Hawke},
  B.~{Zink}, E.~{Schnetter}, {3D Collapse of Rotating Stellar Iron Cores in
  General Relativity Including Deleptonization and a Nuclear Equation of
  State}, Physical Review Letters 98~(26) (2007) 261101--+.
\newblock \href {http://arxiv.org/abs/arXiv:astro-ph/0609819}
  {\path{arXiv:arXiv:astro-ph/0609819}}, \href
  {http://dx.doi.org/10.1103/PhysRevLett.98.261101}
  {\path{doi:10.1103/PhysRevLett.98.261101}}.

\bibitem{ott_new}
C.~D. {Ott}, A.~{Burrows}, L.~{Dessart}, E.~{Livne}, {A New Mechanism for
  Gravitational-Wave Emission in Core-Collapse Supernovae}, Physical Review
  Letters 96~(20) (2006) 201102--+.
\newblock \href {http://arxiv.org/abs/arXiv:astro-ph/0605493}
  {\path{arXiv:arXiv:astro-ph/0605493}}, \href
  {http://dx.doi.org/10.1103/PhysRevLett.96.201102}
  {\path{doi:10.1103/PhysRevLett.96.201102}}.

\bibitem{nils}
N.~{Andersson}, V.~{Ferrari}, D.~I. {Jones}, K.~D. {Kokkotas}, B.~{Krishnan},
  J.~S. {Read}, L.~{Rezzolla}, B.~{Zink}, {Gravitational waves from neutron
  stars: promises and challenges}, General Relativity and Gravitation (2010)
  156--+\href {http://arxiv.org/abs/0912.0384} {\path{arXiv:0912.0384}}, \href
  {http://dx.doi.org/10.1007/s10714-010-1059-4}
  {\path{doi:10.1007/s10714-010-1059-4}}.

\bibitem{kotake07}
K.~{Kotake}, N.~{Ohnishi}, S.~{Yamada}, {Gravitational Radiation from Standing
  Accretion Shock Instability in Core-Collapse Supernovae}, Astrophys. J. 655
  (2007) 406--415.
\newblock \href {http://arxiv.org/abs/arXiv:astro-ph/0607224}
  {\path{arXiv:arXiv:astro-ph/0607224}}, \href
  {http://dx.doi.org/10.1086/509320} {\path{doi:10.1086/509320}}.

\bibitem{kotake_ray}
K.~{Kotake}, W.~{Iwakami}, N.~{Ohnishi}, S.~{Yamada}, {Ray-Tracing Analysis of
  Anisotropic Neutrino Radiation for Estimating Gravitational Waves in
  Core-Collapse Supernovae}, Astrophys. J. 704 (2009) 951--963.
\newblock \href {http://arxiv.org/abs/0909.3622} {\path{arXiv:0909.3622}},
  \href {http://dx.doi.org/10.1088/0004-637X/704/2/951}
  {\path{doi:10.1088/0004-637X/704/2/951}}.

\bibitem{kotake09}
K.~{Kotake}, W.~{Iwakami}, N.~{Ohnishi}, S.~{Yamada}, {Stochastic Nature of
  Gravitational Waves from Supernova Explosions with Standing Accretion Shock
  Instability}, Astrophys. J. Lett. 697 (2009) L133--L136.
\newblock \href {http://arxiv.org/abs/0904.4300} {\path{arXiv:0904.4300}},
  \href {http://dx.doi.org/10.1088/0004-637X/697/2/L133}
  {\path{doi:10.1088/0004-637X/697/2/L133}}.

\bibitem{marek_gw}
A.~{Marek}, H.-T. {Janka}, E.~{M{\"u}ller}, {Equation-of-state dependent
  features in shock-oscillation modulated neutrino and gravitational-wave
  signals from supernovae}, Astron. Astrophys. 496 (2009) 475--494.
\newblock \href {http://arxiv.org/abs/0808.4136} {\path{arXiv:0808.4136}},
  \href {http://dx.doi.org/10.1051/0004-6361/200810883}
  {\path{doi:10.1051/0004-6361/200810883}}.

\bibitem{murphy}
J.~W. {Murphy}, C.~D. {Ott}, A.~{Burrows}, {A Model for Gravitational Wave
  Emission from Neutrino-Driven Core-Collapse Supernovae}, ArXiv e-prints\href
  {http://arxiv.org/abs/0907.4762} {\path{arXiv:0907.4762}}.

\bibitem{arnett}
W.~D. {Arnett}, C.~{Meakin}, {Toward Realistic Progenitors of Core-collapse
  Supernovae}, \apj 733 (2011) 78--+.
\newblock \href {http://arxiv.org/abs/1101.5646} {\path{arXiv:1101.5646}},
  \href {http://dx.doi.org/10.1088/0004-637X/733/2/78}
  {\path{doi:10.1088/0004-637X/733/2/78}}.

\bibitem{ewald11}
E.~{M{\"u}ller}, H.~. {Janka}, A.~{Wongwathanarat}, {Parametrized 3D models of
  neutrino-driven supernova explosions: Neutrino emission asymmetries and
  gravitational-wave signals}, ArXiv e-prints\href
  {http://arxiv.org/abs/1106.6301} {\path{arXiv:1106.6301}}.

\bibitem{epstein}
R.~{Epstein}, {The generation of gravitational radiation by escaping supernova
  neutrinos}, \apj 223 (1978) 1037--1045.
\newblock \href {http://dx.doi.org/10.1086/156337} {\path{doi:10.1086/156337}}.

\bibitem{turner1979}
M.~S. {Turner}, R.~V. {Wagoner}, {Gravitational radiation from slowly-rotating
  'supernovae' - Preliminary results}, in: {L.~L.~Smarr} (Ed.), Sources of
  Gravitational Radiation, 1979, pp. 383--407.

\bibitem{favata}
M.~{Favata}, {The gravitational-wave memory effect}, Classical and Quantum
  Gravity 27~(8) (2010) 084036--+.
\newblock \href {http://arxiv.org/abs/1003.3486} {\path{arXiv:1003.3486}},
  \href {http://dx.doi.org/10.1088/0264-9381/27/8/084036}
  {\path{doi:10.1088/0264-9381/27/8/084036}}.

\bibitem{hiramatsu}
T.~{Hiramatsu}, K.~{Kotake}, H.~{Kudoh}, A.~{Taruya}, {Gravitational wave
  background from neutrino-driven gamma-ray bursts}, \mnras 364 (2005)
  1063--1068.
\newblock \href {http://arxiv.org/abs/arXiv:astro-ph/0509787}
  {\path{arXiv:arXiv:astro-ph/0509787}}, \href
  {http://dx.doi.org/10.1111/j.1365-2966.2005.09643.x}
  {\path{doi:10.1111/j.1365-2966.2005.09643.x}}.

\bibitem{suwa09}
Y.~{Suwa}, K.~{Murase}, {Probing the central engine of long gamma-ray bursts
  and hypernovae with gravitational waves and neutrinos}, \prd 80~(12) (2009)
  123008--+.
\newblock \href {http://arxiv.org/abs/0906.3833} {\path{arXiv:0906.3833}},
  \href {http://dx.doi.org/10.1103/PhysRevD.80.123008}
  {\path{doi:10.1103/PhysRevD.80.123008}}.

\bibitem{suwa07b}
Y.~{Suwa}, T.~{Takiwaki}, K.~{Kotake}, K.~{Sato}, {Gravitational Wave
  Background from Population III Stars}, Astrophys. J. Lett. 665 (2007)
  L43--L46.
\newblock \href {http://arxiv.org/abs/arXiv:0706.3495}
  {\path{arXiv:arXiv:0706.3495}}, \href {http://dx.doi.org/10.1086/521078}
  {\path{doi:10.1086/521078}}.

\bibitem{kota03a}
K.~{Kotake}, S.~{Yamada}, K.~{Sato}, {Anisotropic Neutrino Radiation in
  Rotational Core Collapse}, Astrophys. J. 595 (2003) 304--316.
\newblock \href {http://dx.doi.org/10.1086/377196} {\path{doi:10.1086/377196}}.

\bibitem{jamu}
H.~{Janka}, R.~{M\"onchmeyer}, {Anisotropic neutrino emission from rotating
  protoneutron stars}, \aap 209 (1989) L5--L8.

\bibitem{walder}
R.~{Walder}, A.~{Burrows}, C.~D. {Ott}, E.~{Livne}, I.~{Lichtenstadt},
  M.~{Jarrah}, {Anisotropies in the Neutrino Fluxes and Heating Profiles in
  Two-dimensional, Time-dependent, Multigroup Radiation Hydrodynamics
  Simulations of Rotating Core-Collapse Supernovae}, \apj 626 (2005) 317--332.
\newblock \href {http://arxiv.org/abs/arXiv:astro-ph/0412187}
  {\path{arXiv:arXiv:astro-ph/0412187}}, \href
  {http://dx.doi.org/10.1086/429816} {\path{doi:10.1086/429816}}.

\bibitem{dessart_07}
L.~{Dessart}, A.~{Burrows}, E.~{Livne}, C.~D. {Ott}, {Magnetically Driven
  Explosions of Rapidly Rotating White Dwarfs Following Accretion-Induced
  Collapse}, \apj 669 (2007) 585--599.
\newblock \href {http://arxiv.org/abs/0705.3678} {\path{arXiv:0705.3678}},
  \href {http://dx.doi.org/10.1086/521701} {\path{doi:10.1086/521701}}.

\bibitem{suwa_in_prep}
Y.~{Suwa}, K.~{Kotake}, T.~{Takiwaki}, {The Gravitational-wave signatures in a
  rotating vs. non-rotating 13Msun star driven by the SASI-aided
  neutrino-heating supernova mechanism}, in preparation.

\bibitem{annop2}
A.~{Wongwathanarat}, N.~J. {Hammer}, E.~{M{\"u}ller}, {An axis-free overset
  grid in spherical polar coordinates for simulating 3D self-gravitating
  flows}, \aap 514 (2010) A48+.
\newblock \href {http://arxiv.org/abs/1003.1633} {\path{arXiv:1003.1633}},
  \href {http://dx.doi.org/10.1051/0004-6361/200913435}
  {\path{doi:10.1051/0004-6361/200913435}}.

\bibitem{chris04}
C.~L. {Fryer}, D.~E. {Holz}, S.~A. {Hughes}, {Gravitational Waves from Stellar
  Collapse: Correlations to Explosion Asymmetries}, \apj 609 (2004) 288--300.
\newblock \href {http://arxiv.org/abs/arXiv:astro-ph/0403188}
  {\path{arXiv:arXiv:astro-ph/0403188}}, \href
  {http://dx.doi.org/10.1086/421040} {\path{doi:10.1086/421040}}.

\bibitem{kotake11}
K.~{Kotake}, W.~{Iwakami Nakano}, N.~{Ohnishi}, {Effects of Rotation on
  Stochasticity of Gravitational Waves in Nonlinear Phase of Core-Collapse
  Supernovae}, ArXiv e-prints\href {http://arxiv.org/abs/1106.0544}
  {\path{arXiv:1106.0544}}.

\bibitem{annop}
A.~{Wongwathanarat}, H.~{Janka}, E.~{Mueller}, {Hydrodynamical Neutron Star
  Kicks in Three Dimensions}, ArXiv e-prints\href
  {http://arxiv.org/abs/1010.0167} {\path{arXiv:1010.0167}}.

\bibitem{firstligo}
K.~S. {Thorne}, {Gravitational Waves}, in: {E.~W.~Kolb \& R.~D.~Peccei} (Ed.),
  Particle and Nuclear Astrophysics and Cosmology in the Next Millenium, 1995,
  pp. 160--+.

\bibitem{fpdecigo}
S.~e.~a. {Kawamura}, {The Japanese space gravitational wave antenna DECIGO},
  Classical and Quantum Gravity 23 (2006) 125--+.
\newblock \href {http://dx.doi.org/10.1088/0264-9381/23/8/S17}
  {\path{doi:10.1088/0264-9381/23/8/S17}}.

\bibitem{leblanc}
J.~M. {LeBlanc}, J.~R. {Wilson}, {A Numerical Example of the Collapse of a
  Rotating Magnetized Star}, Astrophys. J. 161 (1970) 541--+.

\bibitem{bisno76}
G.~S. {Bisnovatyi-Kogan}, I.~P. {Popov}, A.~A. {Samokhin}, {The
  magnetohydrodynamic rotational model of supernova explosion}, \apss 41 (1976)
  287--320.
\newblock \href {http://dx.doi.org/10.1007/BF00646184}
  {\path{doi:10.1007/BF00646184}}.

\bibitem{muller79}
E.~{M\"uller}, W.~{Hillebrandt}, {A magnetohydrodynamical supernova model},
  \aap 80 (1979) 147--154.

\bibitem{symb84}
E.~M.~D. {Symbalisty}, {Magnetorotational iron core collapse}, Astrophys. J.
  285 (1984) 729--746.
\newblock \href {http://dx.doi.org/10.1086/162551} {\path{doi:10.1086/162551}}.

\bibitem{arde00}
N.~V. {Ardeljan}, G.~S. {Bisnovatyi-Kogan}, S.~G. {Moiseenko}, {Nonstationary
  magnetorotational processes in a rotating magnetized cloud}, Astron.
  Astrophys. 355 (2000) 1181--1190.
\newblock \href {http://arxiv.org/abs/arXiv:astro-ph/9912513}
  {\path{arXiv:arXiv:astro-ph/9912513}}.

\bibitem{yama04}
S.~{Yamada}, H.~{Sawai}, {Numerical Study on the Rotational Collapse of
  Strongly Magnetized Cores of Massive Stars}, Astrophys. J. 608 (2004)
  907--924.
\newblock \href {http://dx.doi.org/10.1086/420760} {\path{doi:10.1086/420760}}.

\bibitem{kota04b}
K.~{Kotake}, H.~{Sawai}, S.~{Yamada}, K.~{Sato}, {Magnetorotational Effects on
  Anisotropic Neutrino Emission and Convection in Core-Collapse Supernovae},
  Astrophys. J. 608 (2004) 391--404.
\newblock \href {http://dx.doi.org/10.1086/392530} {\path{doi:10.1086/392530}}.

\bibitem{kotake05}
K.~{Kotake}, S.~{Yamada}, K.~{Sato}, {North-South Neutrino Heating Asymmetry in
  Strongly Magnetized and Rotating Stellar Cores}, \apj 618 (2005) 474--484.
\newblock \href {http://arxiv.org/abs/arXiv:astro-ph/0409244}
  {\path{arXiv:arXiv:astro-ph/0409244}}, \href
  {http://dx.doi.org/10.1086/425911} {\path{doi:10.1086/425911}}.

\bibitem{sawa05}
H.~{Sawai}, K.~{Kotake}, S.~{Yamada}, {Core-Collapse Supernovae with Nonuniform
  Magnetic Fields}, Astrophys. J. 631 (2005) 446--455.
\newblock \href {http://dx.doi.org/10.1086/432529} {\path{doi:10.1086/432529}}.

\bibitem{ober06a}
M.~{Obergaulinger}, M.~A. {Aloy}, H.~{Dimmelmeier}, E.~{M{\"u}ller},
  {Axisymmetric simulations of magnetorotational core collapse: approximate
  inclusion of general relativistic effects}, Astron. Astrophys. 457 (2006)
  209--222.
\newblock \href {http://arxiv.org/abs/arXiv:astro-ph/0602187}
  {\path{arXiv:arXiv:astro-ph/0602187}}, \href
  {http://dx.doi.org/10.1051/0004-6361:20064982}
  {\path{doi:10.1051/0004-6361:20064982}}.

\bibitem{ober06b}
M.~{Obergaulinger}, M.~A. {Aloy}, E.~{M{\"u}ller}, {Axisymmetric simulations of
  magneto-rotational core collapse: dynamics and gravitational wave signal},
  Astron. Astrophys. 450 (2006) 1107--1134.
\newblock \href {http://arxiv.org/abs/arXiv:astro-ph/0510184}
  {\path{arXiv:arXiv:astro-ph/0510184}}, \href
  {http://dx.doi.org/10.1051/0004-6361:20054306}
  {\path{doi:10.1051/0004-6361:20054306}}.

\bibitem{burr07}
A.~{Burrows}, L.~{Dessart}, E.~{Livne}, C.~D. {Ott}, J.~{Murphy}, {Simulations
  of Magnetically Driven Supernova and Hypernova Explosions in the Context of
  Rapid Rotation}, Astrophys. J. 664 (2007) 416--434.
\newblock \href {http://arxiv.org/abs/arXiv:astro-ph/0702539}
  {\path{arXiv:arXiv:astro-ph/0702539}}, \href
  {http://dx.doi.org/10.1086/519161} {\path{doi:10.1086/519161}}.

\bibitem{cerd07}
P.~{Cerd{\'a}-Dur{\'a}n}, J.~A. {Font}, H.~{Dimmelmeier}, {General relativistic
  simulations of magneto-rotational core collapse with microphysics}, ArXiv
  Astrophysics e-prints\href {http://arxiv.org/abs/astro-ph/0703597}
  {\path{arXiv:astro-ph/0703597}}.

\bibitem{suwa07a}
Y.~{Suwa}, T.~{Takiwaki}, K.~{Kotake}, K.~{Sato}, {Magnetorotational Collapse
  of Population III Stars}, \pasj 59 (2007) 771--785.

\bibitem{taki04}
T.~{Takiwaki}, K.~{Kotake}, S.~{Nagataki}, K.~{Sato}, {Magneto-driven Shock
  Waves in Core-Collapse Supernovae}, \apj 616 (2004) 1086--1094.
\newblock \href {http://arxiv.org/abs/arXiv:astro-ph/0408388}
  {\path{arXiv:arXiv:astro-ph/0408388}}, \href
  {http://dx.doi.org/10.1086/424993} {\path{doi:10.1086/424993}}.

\bibitem{taki09}
T.~{Takiwaki}, K.~{Kotake}, K.~{Sato}, {Special Relativistic Simulations of
  Magnetically Dominated Jets in Collapsing Massive Stars}, Astrophys. J. 691
  (2009) 1360--1379.
\newblock \href {http://arxiv.org/abs/0712.1949} {\path{arXiv:0712.1949}},
  \href {http://dx.doi.org/10.1088/0004-637X/691/2/1360}
  {\path{doi:10.1088/0004-637X/691/2/1360}}.

\bibitem{scheid}
S.~{Scheidegger}, R.~{Kaeppeli}, S.~C. {Whitehouse}, T.~{Fischer},
  M.~{Liebendoerfer}, {The influence of model parameters on the prediction of
  gravitational wave signals from stellar core collapse}, ArXiv e-prints\href
  {http://arxiv.org/abs/1001.1570} {\path{arXiv:1001.1570}}.

\bibitem{martin11}
M.~{Obergaulinger}, H.-T. {Janka}, {Magnetic field amplification in collapsing,
  non-rotating stellar cores}, ArXiv e-prints\href
  {http://arxiv.org/abs/1101.1198} {\path{arXiv:1101.1198}}.

\bibitem{dunc92}
R.~C. {Duncan}, C.~{Thompson}, {Formation of very strongly magnetized neutron
  stars - Implications for gamma-ray bursts}, Astrophys. J. Lett. 392 (1992)
  L9--L13.
\newblock \href {http://dx.doi.org/10.1086/186413} {\path{doi:10.1086/186413}}.

\bibitem{latt07}
J.~M. {Lattimer}, M.~{Prakash}, {Neutron star observations: Prognosis for
  equation of state constraints}, \physrep 442 (2007) 109--165.
\newblock \href {http://arxiv.org/abs/arXiv:astro-ph/0612440}
  {\path{arXiv:arXiv:astro-ph/0612440}}, \href
  {http://dx.doi.org/10.1016/j.physrep.2007.02.003}
  {\path{doi:10.1016/j.physrep.2007.02.003}}.

\bibitem{meier}
D.~L. {Meier}, R.~I. {Epstein}, W.~D. {Arnett}, D.~N. {Schramm},
  {Magnetohydrodynamic phenomena in collapsing stellar cores}, Astrophys. J.
  204 (1976) 869--878.

\bibitem{balb98}
S.~A. {Balbus}, J.~F. {Hawley}, {Instability, turbulence, and enhanced
  transport in accretion disks}, Reviews of Modern Physics 70 (1998) 1--53.

\bibitem{MRI}
M.~{Obergaulinger}, P.~{Cerd{\'a}-Dur{\'a}n}, E.~{M{\"u}ller}, M.~A. {Aloy},
  {Semi-global simulations of the magneto-rotational instability in core
  collapse supernovae}, \aap 498 (2009) 241--271.
\newblock \href {http://arxiv.org/abs/0811.1652} {\path{arXiv:0811.1652}},
  \href {http://dx.doi.org/10.1051/0004-6361/200811323}
  {\path{doi:10.1051/0004-6361/200811323}}.

\bibitem{woos_blom}
S.~E. {Woosley}, J.~S. {Bloom}, {The Supernova Gamma-Ray Burst Connection},
  \araa 44 (2006) 507--556.
\newblock \href {http://arxiv.org/abs/arXiv:astro-ph/0609142}
  {\path{arXiv:arXiv:astro-ph/0609142}}, \href
  {http://dx.doi.org/10.1146/annurev.astro.43.072103.150558}
  {\path{doi:10.1146/annurev.astro.43.072103.150558}}.

\bibitem{shib06}
M.~{Shibata}, Y.~T. {Liu}, S.~L. {Shapiro}, B.~C. {Stephens},
  {Magnetorotational collapse of massive stellar cores to neutron stars:
  Simulations in full general relativity}, \prd 74~(10) (2006) 104026--+.
\newblock \href {http://arxiv.org/abs/arXiv:astro-ph/0610840}
  {\path{arXiv:arXiv:astro-ph/0610840}}, \href
  {http://dx.doi.org/10.1103/PhysRevD.74.104026}
  {\path{doi:10.1103/PhysRevD.74.104026}}.

\bibitem{taki_kota}
T.~{Takiwaki}, K.~{Kotake}, {Gravitational-Wave Signatures in
  Magnetically-driven Supernova Explosions}, ArXiv e-prints, ApJ in press\href
  {http://arxiv.org/abs/1004.2896} {\path{arXiv:1004.2896}}.

\bibitem{thorne}
K.~S. {Thorne}, {Multipole expansions of gravitational radiation}, Reviews of
  Modern Physics 52 (1980) 299--340.
\newblock \href {http://dx.doi.org/10.1103/RevModPhys.52.299}
  {\path{doi:10.1103/RevModPhys.52.299}}.

\bibitem{kudoh}
H.~{Kudoh}, A.~{Taruya}, T.~{Hiramatsu}, Y.~{Himemoto}, {Detecting a
  gravitational-wave background with next-generation space interferometers},
  \prd 73~(6) (2006) 064006--+.
\newblock \href {http://arxiv.org/abs/arXiv:gr-qc/0511145}
  {\path{arXiv:arXiv:gr-qc/0511145}}, \href
  {http://dx.doi.org/10.1103/PhysRevD.73.064006}
  {\path{doi:10.1103/PhysRevD.73.064006}}.

\bibitem{chandra69}
S.~{Chandrasekhar}, {Ellipsoidal figures of equilibrium}, 1969.

\bibitem{tass78}
J.-L. {Tassoul}, {Theory of rotating stars}, Princeton Series in Astrophysics,
  Princeton: University Press, 1978, 1978.

\bibitem{ramp98}
M.~{Rampp}, E.~{Mueller}, M.~{Ruffert}, {Simulations of non-axisymmetric
  rotational core collapse}, Astron. Astrophys. 332 (1998) 969--983.

\bibitem{zweg}
T.~{Zwerger}, E.~{Mueller}, {Dynamics and gravitational wave signature of
  axisymmetric rotational core collapse.}, Astron. Astrophys. 320 (1997)
  209--227.

\bibitem{chris04a}
C.~L. {Fryer}, M.~S. {Warren}, {The Collapse of Rotating Massive Stars in Three
  Dimensions}, \apj 601 (2004) 391--404.
\newblock \href {http://arxiv.org/abs/arXiv:astro-ph/0309539}
  {\path{arXiv:arXiv:astro-ph/0309539}}, \href
  {http://dx.doi.org/10.1086/380193} {\path{doi:10.1086/380193}}.

\bibitem{shib05}
M.~{Shibata}, Y.-I. {Sekiguchi}, {Three-dimensional simulations of stellar core
  collapse in full general relativity: Nonaxisymmetric dynamical
  instabilities}, \prd 71~(2) (2005) 024014.
\newblock \href {http://arxiv.org/abs/arXiv:astro-ph/0412243}
  {\path{arXiv:arXiv:astro-ph/0412243}}, \href
  {http://dx.doi.org/10.1103/PhysRevD.71.024014}
  {\path{doi:10.1103/PhysRevD.71.024014}}.

\bibitem{ott_3db}
C.~D. {Ott}, H.~{Dimmelmeier}, A.~{Marek}, H.-T. {Janka}, B.~{Zink},
  I.~{Hawke}, E.~{Schnetter}, {Rotating collapse of stellar iron cores in
  general relativity}, Classical and Quantum Gravity 24 (2007) 139--+.
\newblock \href {http://arxiv.org/abs/arXiv:astro-ph/0612638}
  {\path{arXiv:arXiv:astro-ph/0612638}}, \href
  {http://dx.doi.org/10.1088/0264-9381/24/12/S10}
  {\path{doi:10.1088/0264-9381/24/12/S10}}.

\bibitem{laishapi}
D.~{Lai}, S.~L. {Shapiro}, {Gravitational radiation from rapidly rotating
  nascent neutron stars}, \apj 442 (1995) 259--272.
\newblock \href {http://arxiv.org/abs/arXiv:astro-ph/9408053}
  {\path{arXiv:arXiv:astro-ph/9408053}}, \href
  {http://dx.doi.org/10.1086/175438} {\path{doi:10.1086/175438}}.

\bibitem{koko}
K.~D. {Kokkotas}, {Gravitational Wave Astronomy}, ArXiv e-prints\href
  {http://arxiv.org/abs/0809.1602} {\path{arXiv:0809.1602}}.

\bibitem{ott_one_arm}
C.~D. {Ott}, S.~{Ou}, J.~E. {Tohline}, A.~{Burrows}, {One-armed Spiral
  Instability in a Low-T/|W| Postbounce Supernova Core}, \apjl 625 (2005)
  L119--L122.
\newblock \href {http://arxiv.org/abs/arXiv:astro-ph/0503187}
  {\path{arXiv:arXiv:astro-ph/0503187}}, \href
  {http://dx.doi.org/10.1086/431305} {\path{doi:10.1086/431305}}.

\bibitem{shibata02}
M.~{Shibata}, S.~{Karino}, Y.~{Eriguchi}, {Dynamical instability of
  differentially rotating stars}, \mnras 334 (2002) L27--L31.
\newblock \href {http://arxiv.org/abs/arXiv:gr-qc/0206002}
  {\path{arXiv:arXiv:gr-qc/0206002}}, \href
  {http://dx.doi.org/10.1046/j.1365-8711.2002.05724.x}
  {\path{doi:10.1046/j.1365-8711.2002.05724.x}}.

\bibitem{saijo03}
M.~{Saijo}, T.~W. {Baumgarte}, S.~L. {Shapiro}, {One-armed Spiral Instability
  in Differentially Rotating Stars}, \apj 595 (2003) 352--364.
\newblock \href {http://arxiv.org/abs/arXiv:astro-ph/0302436}
  {\path{arXiv:arXiv:astro-ph/0302436}}, \href
  {http://dx.doi.org/10.1086/377334} {\path{doi:10.1086/377334}}.

\bibitem{watts}
A.~L. {Watts}, N.~{Andersson}, D.~I. {Jones}, {The Nature of Low T/|W|
  Dynamical Instabilities in Differentially Rotating Stars}, \apjl 618 (2005)
  L37--L40.
\newblock \href {http://arxiv.org/abs/arXiv:astro-ph/0309554}
  {\path{arXiv:arXiv:astro-ph/0309554}}, \href
  {http://dx.doi.org/10.1086/427653} {\path{doi:10.1086/427653}}.

\bibitem{ou}
S.~{Ou}, J.~E. {Tohline}, {Unexpected Dynamical Instabilities in Differentially
  Rotating Neutron Stars}, \apj 651 (2006) 1068--1078.
\newblock \href {http://arxiv.org/abs/arXiv:astro-ph/0604099}
  {\path{arXiv:arXiv:astro-ph/0604099}}, \href
  {http://dx.doi.org/10.1086/507597} {\path{doi:10.1086/507597}}.

\bibitem{pablo2007}
P.~{Cerd{\'a}-Dur{\'a}n}, V.~{Quilis}, J.~A. {Font}, {AMR simulations of the
  low T/|W| bar-mode instability of neutron stars}, Computer Physics
  Communications 177 (2007) 288--297.
\newblock \href {http://arxiv.org/abs/0704.0356} {\path{arXiv:0704.0356}},
  \href {http://dx.doi.org/10.1016/j.cpc.2007.04.001}
  {\path{doi:10.1016/j.cpc.2007.04.001}}.

\bibitem{saijo06}
M.~{Saijo}, S.~{Yoshida}, {Low T/|W| dynamical instability in differentially
  rotating stars: diagnosis with canonical angular momentum}, \mnras 368 (2006)
  1429--1442.
\newblock \href {http://arxiv.org/abs/arXiv:astro-ph/0505543}
  {\path{arXiv:arXiv:astro-ph/0505543}}, \href
  {http://dx.doi.org/10.1111/j.1365-2966.2006.10229.x}
  {\path{doi:10.1111/j.1365-2966.2006.10229.x}}.

\bibitem{papa}
J.~C.~B. {Papaloizou}, J.~E. {Pringle}, {The dynamical stability of
  differentially rotating discs. II}, \mnras 213 (1985) 799--820.

\bibitem{centrella}
J.~M. {Centrella}, K.~C.~B. {New}, L.~L. {Lowe}, J.~D. {Brown}, {Dynamical
  Rotational Instability at Low T/W}, \apjl 550 (2001) L193--L196.
\newblock \href {http://arxiv.org/abs/arXiv:astro-ph/0010574}
  {\path{arXiv:arXiv:astro-ph/0010574}}, \href
  {http://dx.doi.org/10.1086/319634} {\path{doi:10.1086/319634}}.

\bibitem{lund}
T.~{Lund}, A.~{Marek}, C.~{Lunardini}, H.-T. {Janka}, G.~{Raffelt}, {Fast time
  variations of supernova neutrino fluxes and their detectability}, \prd 82~(6)
  (2010) 063007--+.
\newblock \href {http://arxiv.org/abs/1006.1889} {\path{arXiv:1006.1889}},
  \href {http://dx.doi.org/10.1103/PhysRevD.82.063007}
  {\path{doi:10.1103/PhysRevD.82.063007}}.

\bibitem{leonor}
I.~{Leonor}, L.~{Cadonati}, E.~{Coccia}, S.~{D'Antonio}, A.~{Di Credico},
  V.~{Fafone}, R.~{Frey}, W.~{Fulgione}, E.~{Katsavounidis}, C.~D. {Ott},
  G.~{Pagliaroli}, K.~{Scholberg}, E.~{Thrane}, F.~{Vissani}, {Searching for
  prompt signatures of nearby core-collapse supernovae by a joint analysis of
  neutrino and gravitational-wave data}, ArXiv e-prints\href
  {http://arxiv.org/abs/1002.1511} {\path{arXiv:1002.1511}}.

\bibitem{nishi06}
S.~{Nishimura}, K.~{Kotake}, M.-a. {Hashimoto}, S.~{Yamada}, N.~{Nishimura},
  S.~{Fujimoto}, K.~{Sato}, {r-Process Nucleosynthesis in Magnetohydrodynamic
  Jet Explosions of Core-Collapse Supernovae}, \apj 642 (2006) 410--419.
\newblock \href {http://arxiv.org/abs/arXiv:astro-ph/0504100}
  {\path{arXiv:arXiv:astro-ph/0504100}}, \href
  {http://dx.doi.org/10.1086/500786} {\path{doi:10.1086/500786}}.

\bibitem{fujimoto}
S.-i. {Fujimoto}, M.-a. {Hashimoto}, K.~{Kotake}, S.~{Yamada}, {Heavy-Element
  Nucleosynthesis in a Collapsar}, \apj 656 (2007) 382--392.
\newblock \href {http://arxiv.org/abs/arXiv:astro-ph/0602460}
  {\path{arXiv:arXiv:astro-ph/0602460}}, \href
  {http://dx.doi.org/10.1086/509908} {\path{doi:10.1086/509908}}.

\bibitem{wanajo1}
S.~{Wanajo}, H.-T. {Janka}, S.~{Kubono}, {Uncertainties in the
  {$\nu$}p-process: Supernova Dynamics Versus Nuclear Physics}, \apj 729 (2011)
  46--+.
\newblock \href {http://arxiv.org/abs/1004.4487} {\path{arXiv:1004.4487}},
  \href {http://dx.doi.org/10.1088/0004-637X/729/1/46}
  {\path{doi:10.1088/0004-637X/729/1/46}}.

\bibitem{wanajo2}
S.~{Wanajo}, H.-T. {Janka}, {The r-process in the neutrino-driven wind from a
  black-hole torus}, ArXiv e-prints\href {http://arxiv.org/abs/1106.6142}
  {\path{arXiv:1106.6142}}.

\bibitem{latt91}
J.~M. {Lattimer}, F.~D. {Swesty}, {A generalized equation of state for hot,
  dense matter}, Nuclear Physics A 535 (1991) 331--376.
\newblock \href {http://dx.doi.org/10.1016/0375-9474(91)90452-C}
  {\path{doi:10.1016/0375-9474(91)90452-C}}.

\bibitem{shlo06}
S.~{Shlomo}, V.~M. {Kolomietz}, G.~{Col{\`o}}, {Deducing the nuclear-matter
  incompressibility coefficient from data on isoscalar compression modes},
  European Physical Journal A 30 (2006) 23--30.
\newblock \href {http://dx.doi.org/10.1140/epja/i2006-10100-3}
  {\path{doi:10.1140/epja/i2006-10100-3}}.

\bibitem{demo10}
P.~B. {Demorest}, T.~{Pennucci}, S.~M. {Ransom}, M.~S.~E. {Roberts}, J.~W.~T.
  {Hessels}, {A two-solar-mass neutron star measured using Shapiro delay}, \nat
  467 (2010) 1081--1083.
\newblock \href {http://arxiv.org/abs/1010.5788} {\path{arXiv:1010.5788}},
  \href {http://dx.doi.org/10.1038/nature09466}
  {\path{doi:10.1038/nature09466}}.

\bibitem{oconnor}
E.~{O'Connor}, C.~D. {Ott}, {Black Hole Formation in Failing Core-Collapse
  Supernovae}, \apj 730 (2011) 70--+.
\newblock \href {http://arxiv.org/abs/1010.5550} {\path{arXiv:1010.5550}},
  \href {http://dx.doi.org/10.1088/0004-637X/730/2/70}
  {\path{doi:10.1088/0004-637X/730/2/70}}.

\bibitem{kiuc08}
K.~{Kiuchi}, K.~{Kotake}, {Equilibrium configurations of strongly magnetized
  neutron stars with realistic equations of state}, \mnras 385 (2008)
  1327--1347.
\newblock \href {http://arxiv.org/abs/0708.3597} {\path{arXiv:0708.3597}},
  \href {http://dx.doi.org/10.1111/j.1365-2966.2007.12791.x}
  {\path{doi:10.1111/j.1365-2966.2007.12791.x}}.

\bibitem{tobi11}
T.~{Fischer}, I.~{Sagert}, G.~{Pagliara}, M.~{Hempel}, J.~{Schaffner-Bielich},
  T.~{Rauscher}, F.-K. {Thielemann}, R.~{K{\"a}ppeli},
  G.~{Mart{\'{\i}}nez-Pinedo}, M.~{Liebend{\"o}rfer}, {Core-collapse Supernova
  Explosions Triggered by a Quark-Hadron Phase Transition During the Early
  Post-bounce Phase}, \apjs 194 (2011) 39--+.
\newblock \href {http://arxiv.org/abs/1011.3409} {\path{arXiv:1011.3409}},
  \href {http://dx.doi.org/10.1088/0067-0049/194/2/39}
  {\path{doi:10.1088/0067-0049/194/2/39}}.

\bibitem{thom05}
T.~A. {Thompson}, E.~{Quataert}, A.~{Burrows}, {Viscosity and Rotation in
  Core-Collapse Supernovae}, \apj 620 (2005) 861--877.
\newblock \href {http://arxiv.org/abs/arXiv:astro-ph/0403224}
  {\path{arXiv:arXiv:astro-ph/0403224}}, \href
  {http://dx.doi.org/10.1086/427177} {\path{doi:10.1086/427177}}.

\bibitem{masa11}
Y.~{Masada}, T.~{Takiwaki}, K.~{Kotake}, submitted to \apj.

\bibitem{suzu08}
T.~K. {Suzuki}, K.~{Sumiyoshi}, S.~{Yamada}, {Alfv{\'e}n Wave-Driven Supernova
  Explosion}, \apj 678 (2008) 1200--1206.
\newblock \href {http://arxiv.org/abs/0707.4345} {\path{arXiv:0707.4345}},
  \href {http://dx.doi.org/10.1086/533515} {\path{doi:10.1086/533515}}.

\bibitem{jerome}
J.~{Guilet}, J.~{Sato}, T.~{Foglizzo}, {The Saturation of SASI by Parasitic
  Instabilities}, \apj 713 (2010) 1350--1362.
\newblock \href {http://arxiv.org/abs/0910.3953} {\path{arXiv:0910.3953}},
  \href {http://dx.doi.org/10.1088/0004-637X/713/2/1350}
  {\path{doi:10.1088/0004-637X/713/2/1350}}.

\bibitem{murphy11}
J.~W. {Murphy}, C.~{Meakin}, {A Global Turbulence Model for Neutrino-Driven
  Convection in Core-Collapse Supernovae}, ArXiv e-prints\href
  {http://arxiv.org/abs/1106.5496} {\path{arXiv:1106.5496}}.

\bibitem{ott2011}
C.~D. {Ott}, C.~{Reisswig}, E.~{Schnetter}, E.~{O'Connor}, U.~{Sperhake},
  F.~{L{\"o}ffler}, P.~{Diener}, E.~{Abdikamalov}, I.~{Hawke}, A.~{Burrows},
  {Dynamics and Gravitational Wave Signature of Collapsar Formation}, Physical
  Review Letters 106~(16) (2011) 161103--+.
\newblock \href {http://arxiv.org/abs/1012.1853} {\path{arXiv:1012.1853}},
  \href {http://dx.doi.org/10.1103/PhysRevLett.106.161103}
  {\path{doi:10.1103/PhysRevLett.106.161103}}.

\bibitem{sumi}
K.~{Sumiyoshi}, S.~{Yamada}, H.~{Suzuki}, S.~{Chiba}, {Neutrino Signals from
  the Formation of a Black Hole: A Probe of the Equation of State of Dense
  Matter}, Physical Review Letters 97~(9) (2006) 091101--+.
\newblock \href {http://arxiv.org/abs/arXiv:astro-ph/0608509}
  {\path{arXiv:arXiv:astro-ph/0608509}}, \href
  {http://dx.doi.org/10.1103/PhysRevLett.97.091101}
  {\path{doi:10.1103/PhysRevLett.97.091101}}.

\bibitem{naka1}
K.~{Nakazato}, K.~{Sumiyoshi}, H.~{Suzuki}, S.~{Yamada}, {Exploring hadron
  physics in black hole formations: A new promising target of neutrino
  astronomy}, \prd 81~(8) (2010) 083009--+.
\newblock \href {http://arxiv.org/abs/1004.0291} {\path{arXiv:1004.0291}},
  \href {http://dx.doi.org/10.1103/PhysRevD.81.083009}
  {\path{doi:10.1103/PhysRevD.81.083009}}.

\bibitem{naka2}
K.~{Nakazato}, K.~{Sumiyoshi}, S.~{Yamada}, {Impact of Quarks and Pions on
  Dynamics and Neutrino Signal of Black Hole Formation in Non-rotating Stellar
  Core Collapse}, \apj 721 (2010) 1284--1294.
\newblock \href {http://arxiv.org/abs/1001.5084} {\path{arXiv:1001.5084}},
  \href {http://dx.doi.org/10.1088/0004-637X/721/2/1284}
  {\path{doi:10.1088/0004-637X/721/2/1284}}.

\bibitem{hatsuda}
K.~{Fukushima}, T.~{Hatsuda}, {The phase diagram of dense QCD}, Reports on
  Progress in Physics 74~(1) (2011) 014001--+.
\newblock \href {http://arxiv.org/abs/1005.4814} {\path{arXiv:1005.4814}},
  \href {http://dx.doi.org/10.1088/0034-4885/74/1/014001}
  {\path{doi:10.1088/0034-4885/74/1/014001}}.

\bibitem{paz90}
B.~{Paczynski}, {Super-Eddington winds from neutron stars}, \apj 363 (1990)
  218--226.
\newblock \href {http://dx.doi.org/10.1086/169332} {\path{doi:10.1086/169332}}.

\bibitem{mezree}
P.~{Meszaros}, M.~J. {Rees}, {High-entropy fireballs and jets in gamma-ray
  burst sources}, \mnras 257 (1992) 29P--31P.

\bibitem{thom04}
T.~A. {Thompson}, P.~{Chang}, E.~{Quataert}, {Magnetar Spin-Down,
  Hyperenergetic Supernovae, and Gamma-Ray Bursts}, Astrophys. J. 611 (2004)
  380--393.
\newblock \href {http://arxiv.org/abs/arXiv:astro-ph/0401555}
  {\path{arXiv:arXiv:astro-ph/0401555}}, \href
  {http://dx.doi.org/10.1086/421969} {\path{doi:10.1086/421969}}.

\bibitem{uzde07a}
D.~A. {Uzdensky}, A.~I. {MacFadyen}, {Magnetar-Driven Magnetic Tower as a Model
  for Gamma-Ray Bursts and Asymmetric Supernovae}, Astrophys. J. 669 (2007)
  546--560.
\newblock \href {http://arxiv.org/abs/arXiv:astro-ph/0609047}
  {\path{arXiv:arXiv:astro-ph/0609047}}, \href
  {http://dx.doi.org/10.1086/521322} {\path{doi:10.1086/521322}}.

\bibitem{woos93}
S.~E. {Woosley}, {Gamma-ray bursts from stellar mass accretion disks around
  black holes}, \apj 405 (1993) 273--277.
\newblock \href {http://dx.doi.org/10.1086/172359} {\path{doi:10.1086/172359}}.

\bibitem{pacz98}
B.~{Paczynski}, {Are Gamma-Ray Bursts in Star-Forming Regions?}, \apjl 494
  (1998) L45+.
\newblock \href {http://arxiv.org/abs/arXiv:astro-ph/9710086}
  {\path{arXiv:arXiv:astro-ph/9710086}}, \href
  {http://dx.doi.org/10.1086/311148} {\path{doi:10.1086/311148}}.

\bibitem{macf99}
A.~I. {MacFadyen}, S.~E. {Woosley}, {Collapsars: Gamma-Ray Bursts and
  Explosions in ``Failed Supernovae''}, Astrophys. J. 524 (1999) 262--289.
\newblock \href {http://arxiv.org/abs/arXiv:astro-ph/9810274}
  {\path{arXiv:arXiv:astro-ph/9810274}}, \href
  {http://dx.doi.org/10.1086/307790} {\path{doi:10.1086/307790}}.

\bibitem{dai98}
Z.~G. {Dai}, T.~{Lu}, {Gamma-ray burst afterglows and evolution of postburst
  fireballs with energy injection from strongly magnetic millisecond pulsars},
  \aap 333 (1998) L87--L90.
\newblock \href {http://arxiv.org/abs/arXiv:astro-ph/9810402}
  {\path{arXiv:arXiv:astro-ph/9810402}}.

\bibitem{bucci}
N.~{Bucciantini}, E.~{Quataert}, J.~{Arons}, B.~D. {Metzger}, T.~A. {Thompson},
  {Relativistic jets and long-duration gamma-ray bursts from the birth of
  magnetars}, \mnras (2007) L126+\href {http://arxiv.org/abs/arXiv:0707.2100}
  {\path{arXiv:arXiv:0707.2100}}, \href
  {http://dx.doi.org/10.1111/j.1745-3933.2007.00403.x}
  {\path{doi:10.1111/j.1745-3933.2007.00403.x}}.

\bibitem{metzger}
B.~D. {Metzger}, D.~{Giannios}, T.~A. {Thompson}, N.~{Bucciantini},
  E.~{Quataert}, {The protomagnetar model for gamma-ray bursts}, \mnras 413
  (2011) 2031--2056.
\newblock \href {http://arxiv.org/abs/1012.0001} {\path{arXiv:1012.0001}},
  \href {http://dx.doi.org/10.1111/j.1365-2966.2011.18280.x}
  {\path{doi:10.1111/j.1365-2966.2011.18280.x}}.

\bibitem{harikae_a}
S.~{Harikae}, T.~{Takiwaki}, K.~{Kotake}, {Long-Term Evolution of Slowly
  Rotating Collapsar in Special Relativistic Magnetohydrodynamics}, Astrophys.
  J. 704 (2009) 354--371.
\newblock \href {http://arxiv.org/abs/0905.2006} {\path{arXiv:0905.2006}},
  \href {http://dx.doi.org/10.1088/0004-637X/704/1/354}
  {\path{doi:10.1088/0004-637X/704/1/354}}.

\bibitem{harikae_b}
S.~{Harikae}, K.~{Kotake}, T.~{Takiwaki}, {Neutrino Pair Annihilation in
  Collapsars: A Ray-tracing Method in Special Relativity}, Astrophys. J. 713
  (2010) 304--317.
\newblock \href {http://arxiv.org/abs/0912.2590} {\path{arXiv:0912.2590}},
  \href {http://dx.doi.org/10.1088/0004-637X/713/1/304}
  {\path{doi:10.1088/0004-637X/713/1/304}}.

\bibitem{woosh}
S.~E. {Woosley}, A.~{Heger}, {Nucleosynthesis and remnants in massive stars of
  solar metallicity}, \physrep 442 (2007) 269--283.
\newblock \href {http://arxiv.org/abs/arXiv:astro-ph/0702176}
  {\path{arXiv:arXiv:astro-ph/0702176}}, \href
  {http://dx.doi.org/10.1016/j.physrep.2007.02.009}
  {\path{doi:10.1016/j.physrep.2007.02.009}}.

\bibitem{thielemann}
F.-K. {Thielemann}, R.~{Hirschi}, M.~{Liebend{\"o}rfer}, R.~{Diehl}, {Massive
  Stars and Their Supernovae}, in: {R.~Diehl, D.~H.~Hartmann, \& N.~Prantzos}
  (Ed.), Lecture Notes in Physics, Berlin Springer Verlag, Vol. 812 of Lecture
  Notes in Physics, Berlin Springer Verlag, 2011, pp. 153--232.
\newblock \href {http://arxiv.org/abs/1008.2144} {\path{arXiv:1008.2144}}.

\end{thebibliography}







\end{document}